\documentclass{elsart}

\usepackage[square,comma]{natbib}
\usepackage{graphicx}
\usepackage{pxfonts}

\usepackage{amssymb}

\journal{}

\journal{}

\begin{document}

\thispagestyle{empty}
\begin{Large}
\textbf{DEUTSCHES ELEKTRONEN-SYNCHROTRON}

\textbf{\large{in der HELMHOLTZ-GEMEINSCHAFT}\\}
\end{Large}

DESY 08-118

August 2008

\begin{eqnarray}
\nonumber &&\cr \nonumber && \cr \nonumber &&\cr
\end{eqnarray}
\begin{eqnarray}
\nonumber
\end{eqnarray}
\begin{center}
\begin{Large}
\textbf{Theory of Edge Radiation}
\end{Large}
\begin{eqnarray}
\nonumber &&\cr \nonumber && \cr
\end{eqnarray}

\begin{large}
Gianluca Geloni, Vitali Kocharyan, Evgeni Saldin, Evgeni
Schneidmiller and Mikhail Yurkov
\end{large}
\textsl{\\Deutsches Elektronen-Synchrotron DESY, Hamburg}
\begin{eqnarray}
\nonumber
\end{eqnarray}
\begin{eqnarray}
\nonumber
\end{eqnarray}
\begin{eqnarray}
\nonumber
\end{eqnarray}
ISSN 0418-9833
\begin{eqnarray}
\nonumber
\end{eqnarray}
\begin{large}
\textbf{NOTKESTRASSE 85 - 22607 HAMBURG}
\end{large}
\end{center}
\clearpage
\newpage

\begin{frontmatter}



\title{Theory of edge radiation}


\author{Gianluca Geloni,}
\author{Vitali Kocharyan,}
\author{Evgeni Saldin,}
\author{Evgeni Schneidmiller}
\author{and Mikhail Yurkov}

\address{Deutsches Elektronen-Synchrotron (DESY), Hamburg,
Germany}

\begin{abstract}

We formulate a complete theory of Edge Radiation based on a novel
method relying on Fourier Optics techniques. Similar types of
radiation like Transition Undulator Radiation are addressed in the
framework of the same formalism. Special attention is payed in
discussing the validity of approximations upon which the theory is
built. Our study makes consistent use of both similarity
techniques and comparisons with numerical results from simulation.
We discuss both near and far zone. Physical understanding of many
asymptotes is discussed. Based on the solution of the field
equation with a tensor Green's function technique, we also discuss
an analytical model to describe the presence of a vacuum chamber.
In particular, explicit calculations for a circular vacuum chamber
are reported. Finally, we consider the use of Edge Radiation as a
tool for electron beam diagnostics. We discuss Coherent Edge
Radiation, Extraction of Edge Radiation by a mirror, and other
issues becoming important at high electron energy and long
radiation wavelength. Based on this work we also study the impact
of Edge Radiation on XFEL setups and we discuss recent results.
\end{abstract}


\end{frontmatter}


\clearpage

\section{\label{sec:intro} Introduction}

Synchrotron Radiation (SR) sources from bending magnets are
brilliant, and cover the continuous spectral range from microwaves
to X-rays. However, in order to optimally meet the needs of basic
research with SR, it is desirable to provide specific radiation
characteristics, which cannot be obtained from bending magnets,
but require special magnetic setups, called insertion devices.
These are installed along the particle beam path between two
bending magnets, and introduce no net beam deflection. Therefore,
they can be incorporated in a given beamline without changing its
geometry. Undulators are a typical example of such devices,
generating specific radiation characteristics in the short
wavelength range.

The history of SR utilization in the long wavelength region (from
micrometer to millimeter) is more recent than that in the short
wavelength range. Long wavelength SR sources may have a strong
potential for infrared spectroscopy or imaging techniques. In
fact, they are some order of magnitude brighter than standard
thermal sources in the same spectral range.

\begin{figure}
\begin{center}
\includegraphics*[width=100mm]{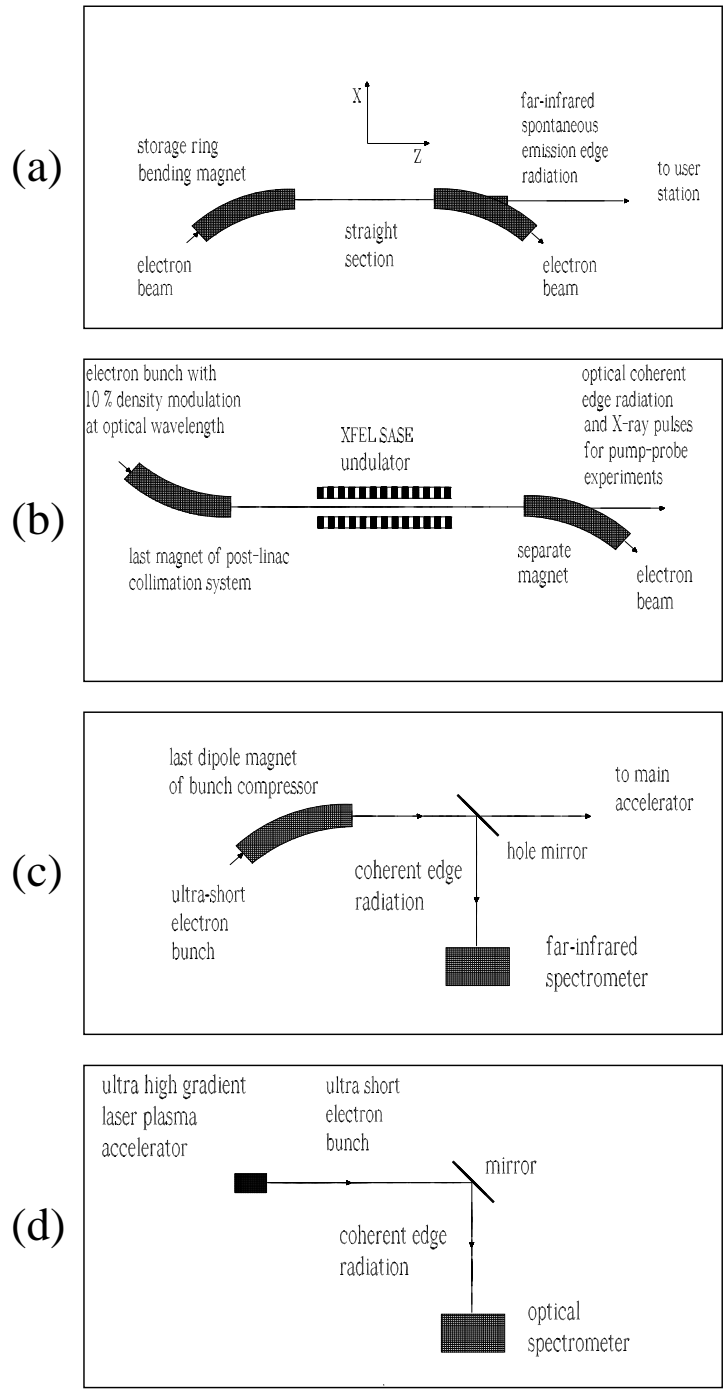}
\caption{\label{geome} Four main types of edge radiation setups:
(a) Far-infrared beamline for synchrotron radiation source using
edge radiation. (b) Arrival-time monitor for XFEL source using
optical coherent edge radiation. (c)  Electron bunch length
monitor for XFEL using  far-infrared coherent  edge radiation. (d)
Ultra-short electron bunch diagnostic for laser-plasma accelerator
facility using optical coherent edge radiation.  }
\end{center}
\end{figure}

Large angles are required to extract long wavelength SR from
bending magnets, because the "natural" opening angle in this case
increases up to several tens milliradians in the far-infrared
range. However, the situation changes dramatically if a straight
section is introduced between two bends, like in Fig.
\ref{geome}(a). Long-wavelength radiation emitted by relativistic
electrons in this setup is called Edge Radiation (ER), and
presents a significantly smaller opening angle than standard SR
from bends (see, among others, \cite{NIK1}-\cite{BOS2}). In other
words, in the long wavelength region (compared to the critical
bending-magnet radiation wavelength) a simple straight section
between bends can play the role of a kind of insertion device.

ER and bending magnet radiation have equivalent brightness. In
fact, the physical process of ER emission is not different from
that of radiation emission from a single bend. However, radiation
from the setup in Fig. \ref{geome}(a) exhibits special features.
Due to a narrower opening angle of ER over SR from bends, as the
wavelength gets longer ER yields significant advantages in terms
of simplicity of the photon beamline.

ER theory is a part of the more general SR theory, very much like
Undulator Radiation (UR) theory is a part of SR theory. Similarly
to the UR case, also for ER the knowledge of the applicability
region of the far-field formulas and corrections for near-field
effects are of practical importance. In most practical cases, the
distance between ER source and observer (i.e. the first optical
element of the photon beamline) are comparable or even much
smaller than the length of the straight section, which plays the
role of the length of the insertion device for ER.

In this paper we developed a theory of near-field ER based on
Fourier Optics (FO) techniques. These techniques can be exploited
without limitations for ER setups, because the paraxial
approximation can always be applied in the case of electrons in
ultra-relativistic motion \cite{OURF}. The use of the paraxial
approximation allows reconstruction of the field in the near-zone
from the knowledge of the far-field data. The solvability of the
inverse problem for the field allows characterization of any ER
setup, starting from the far-zone field, in terms of virtual
sources. These sources exhibit a plane wavefront, and can be
pictured as waists of laser-like beams. Using this kind of
description we develop our theory in close relation with
laser-beam optics. In particular, usual FO can be exploited to
characterize the field at any distance, providing a tool for
designing and analyzing ER setups.

It is the purpose of this article to discuss the principles of
production and properties for all applications of ER. First, we
treat the relatively simple case of ER from a setup composed by
straight section and two bending magnets at its ends (see Fig.
\ref{geome}(a)). We begin calculating an analytical expression for
ER from a single electron in the far-zone. Then, we characterize
the near-zone with the help of the virtual-source technique. Two
alternative techniques for the field propagation are given, based
on a single virtual source located in the middle of the ER setup,
and based on two virtual sources located at its edges.

Based on this study-case we turn to analyze a more complicated
setup, consisting of an undulator preceded and followed by two
straight sections and two bends (see Fig. \ref{geome}(b)). ER from
this kind of setup is commonly known as Transition Undulator
Radiation (TUR). The first study on TUR appeared more than a
decade ago in \cite{KIM1}. In that work it was pointed out for the
first time that, since an electron entering or leaving an
undulator experiences a sudden change in longitudinal velocity,
highly collimated radiation with broadband spectrum, similar to
transition radiation, had to be expected in the low-frequency
region in addition to the usual UR. Reference \cite{KIM1}
constitutes a theoretical basis for many other studies. Here we
remind only a few \cite{BOS4,KINC,CAST,BOS6,ROY2}, dealing both
with theoretical and experimental issues. More recently, TUR has
been given consideration in the framework of X-ray Free-Electron
Laser (XFEL) projects like \cite{XFEL,SLAC,SCSS}. For example, an
arrival-time monitor for XFELs using infrared coherent ER from a
setup similar to that in Fig. \ref{geome}(b) has been proposed in
\cite{OURM}, which should be used for pump-probe experiments with
femtosecond-scale resolution. In view of these applications, there
is a need to extend the characterization of TUR to the near-zone,
and to the coherent case. From this viewpoint, specification of
what precedes and follows the undulator is of fundamental
importance. As has been recognized for TUR many years ago
\cite{BOS4}, if this information is not known, any discussion
about the intensity distribution of TUR is meaningless. According
to our approach, the two straight sections and the undulator in
the setup in Fig. \ref{geome}(b) will be associated to virtual
sources with plane wavefronts. The field from the setup can then
be described, in the near as well as in the far-zone, as a
superposition of laser-like beams, radiating at the same
wavelength and separated by different phase shifts.

Our study makes consistent use of both dimensional analysis and
comparisons with outcomes from numerical simulation. All
simulations in this paper are performed with the help of the
computer code SRW \cite{CHUB}.

There are, however, situations when existing computer codes cannot
predict the radiation characteristics. One of these is the case
when perturbations of the long-wavelength radiation by vacuum
chambers are present, which may potentially affect the performance
of ER setups. Since the diffraction size of the THz radiation
exceeds the vacuum chamber dimensions, characterization of
far-infrared ER must be performed accounting for the presence of a
waveguide. In order to deal with this situation we developed a
theory of ER in a waveguide. The task to be solved differs from
the unbounded-space case only in the formulation of boundary
conditions. The paraxial approximation applies as in the
unbounded-space case. Only, on perfectly conducting walls the
electric field must be orthogonal to the vacuum chamber surface.
As in the unbounded-space case, one can use the Green's function
approach to solve the field equations. The presence of different
boundary conditions complicates the solution of the paraxial
equation for the field, which can nevertheless be explicitly found
by accounting for the tensorial nature of the Green's function.
Here we take advantage of a mode expansion approach to calculate
ER emission in the metallic waveguide structure. We solve the
field equations with a tensor Green's function technique, and we
extract figure of merits describing in a simple way the influence
of the vacuum chamber on the radiation pulse as a function of the
problem parameters. We put particular emphasis on a vacuum chamber
with circular cross-section, which is natural for future
linac-based sources (XFELs and Energy Recovery Linacs (ERLs)
\cite{EDGA}).

Finally, we address the long-standing interest of electron beam
characterization for linac-based sources and laser-plasma
accelerators. One possibility to perform electron-beam diagnostics
in these kind of facilities is to use coherent ER. This is an
attractive tool, because it can provide valuable and detailed
information on the electron beam. By detecting coherent ER, 3D
distributions, divergence, micro-bunching may be measured in
principle (See Figs. \ref{geome}(c) and (d)). Usually, electrons
in accelerators are highly collimated and monochromatic. In this
case, coherent ER can be used for longitudinal and transverse
beam-size monitoring. In contrast to this, electrons generated in
laser-plasma interactions have different properties compared with
those in conventional accelerators. Namely, in this case,
electrons have both divergence angle and energy distribution. We
address applications of coherent ER by studying, first, the
relatively simple case when only the influence of longitudinal and
transverse structure factor of the electron bunch is accounted
for. In particular, we give particular attention to the case when
microbunching at optical wavelengths is imprinted onto an electron
bunch, and we analyze the more complicated case of a bunch
produced in a laser-plasma accelerator, i.e. accounting for
divergence angle distribution of the beam. Energy distribution can
be easily accounted for, based on this analysis. The problem of
extraction of ER by a mirror, strictly related to diagnostics
applications, is included too.

\section{\label{sec:param} General relations for edge radiation phenomena}

\subsection{\label{subs:num} Physical discussion of some numerical
experiment}

This Section constitutes an attempt to introduce ER theory to
readers in as intuitive and simple a fashion as possible by
simulating the angular spectral flux as a function of observation
angles for the geometry in \ref{geome}(a). For this purpose we
take advantage of the code SRW \cite{CHUB}, which provides a
numerical solution of Maxwell's equations.

The origin of a Cartesian coordinate system is placed at the
center of straight section. The $z$-axis is in the direction of
straight section and electron motion is in $xz$ plane. Parameters
of the problem are the radiation wavelength $\lambda$, the radius
of the bend $R$, the relativistic Lorentz factor $\gamma$, the
length of the straight section $L$ and, additionally, the position
of the observation plane down the beamline, $z$. We work in the
far zone. In this Section it is operatively defined as a region
where $z$ is large enough, so that the simulated angular spectral
flux does not show dependence on $z$ anymore.

\begin{figure}
\begin{center}
\includegraphics*[width=140mm]{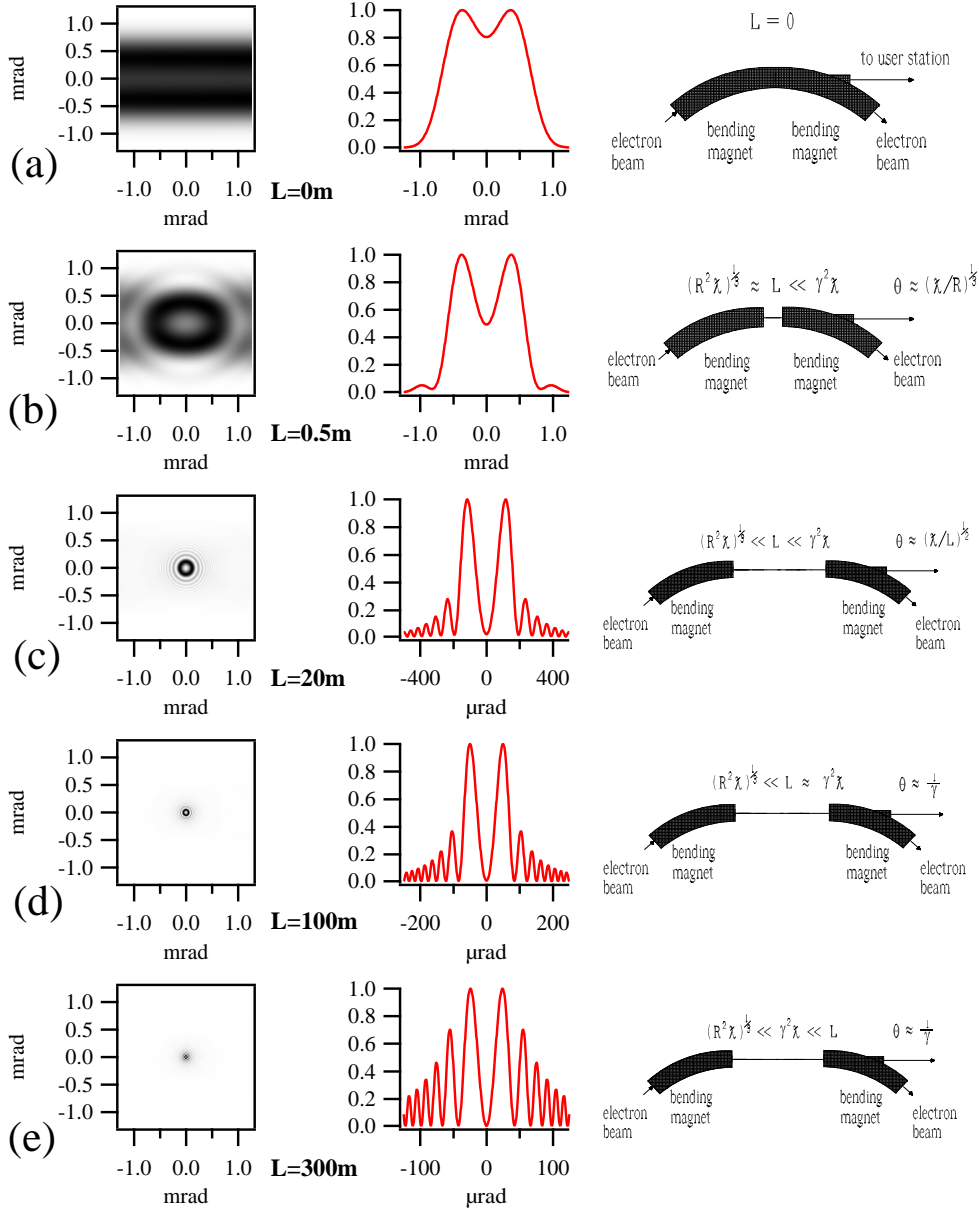}
\caption{\label{FIGV1} Illustrative calculations of the effect of
bending magnet separation on the directivity diagram of the
radiation. The bending magnet radius $R=400$ m, the relativistic
factor $\gamma = 3.42 \cdot 10^4$, and the wavelength of
interested $\lambda = 400$ nm are fixed, while the straight
section length varies from $L=0$ up to $L \gg \gamma^2 \lambdabar
\simeq 100$ m. In this setup (as well as in all others in this
paper) $\lambda \gg \lambda_c \simeq 0.1 \AA$. Case (a) is a
bending magnet setup. Case (b) is a complex setup, where the
radiation beam divergence is practically the same as in (a). Case
(c) illustrates an ER setup. Bending magnet separation
dramatically lowers the radiation beam divergence. (d) Optimal
bending magnet separation. The straight section length $L \simeq
\gamma^2 \lambdabar$ corresponds to a radiation beam divergence
$\theta \simeq 1/\gamma$. (e) Further increase of $L$ only leads
to the appearance of finer structures in the radiation profile. 2D
plots on the left show the angular spectral flux as a function of
the horizontal and vertical angles $\theta_x$ and $\theta_y$ for
various lengths of the straight section. Middle plots are obtained
cutting the 2D angular spectral flux profile at $x=0$. Right plots
show a schematic of the considered layout. }
\end{center}
\end{figure}

ER carries advantages over bending magnet radiation in the limit
for $\lambdabar/\lambdabar_c \gg 1$, where $\lambdabar_c \sim
R/\gamma^3$ (here $\lambdabar=\lambda/(2\pi)$ is the reduced
wavelength) is the critical wavelength of bending magnet
radiation. We will work, therefore, in this limit. We set $\gamma
= 3.42 \cdot 10^4$ ($17.5$ GeV), $R = 400$ m, which are typical
values for XFELs. Note that in this case $\lambda_c \simeq 0.1
\AA$. Here we take $\lambda = 400$ nm. We begin with the case
$L=0$ (bending magnet), and we increase the straight section
length (see Fig. \ref{FIGV1}). As one can see from the figure,
radiation becomes more collimated, up to about $L \simeq \gamma^2
\lambdabar \simeq 100$ m (case (d)), where the collimation angle
reaches $1/\gamma \sim 30 \mu$rad. Further increase of $L$ only
leads to the appearance of finer structures in the radiation
profile. It is important to remark that the total number of
photons in the $\pm 1$ mrad window shown in Fig. \ref{FIGV1}(a) is
roughly the same in the $\pm 100 \mu$rad window in Fig.
\ref{FIGV1}(d). It is clear that the length of the straight
section $L$ is strongly related with the collimation of the
radiation.

\subsection{\label{sum:simi} Similarity techniques}

To study ER further we apply similarity techniques. Similarity is
a special symmetry where a change in scale of independent
variables can be compensated by a similarity transformation of
other variables. This is a familiar concept in hydrodynamics,
where the cardinal example is given by the Reynolds number.
Similarity allows one to reduce the number of parameters to a few
dimensionless ones that are directly linked to the physics of the
process, and that control it in full. Such parameters are found by
analysis of the underlying equations characterizing the system
under study. In this Section we limit ourselves to list them, to
show their correctness with the help of the code SRW, and to
describe their physical meaning. This allows one to obtain general
properties of the ER process. A comprehensive theory of ER will be
presented in the following Sections.

For the setup in Fig. \ref{geome}(a), two dimensionless parameters
controlling the radiation characteristics can be extracted from
Maxwell's equations. In the next Section we will show how these
parameters can be derived. Here we limit ourselves to write them:

\begin{eqnarray}
\delta \equiv \frac{\sqrt[3]{R^2 \lambdabar}}{L}~,~~~~\phi \equiv
\frac{L}{\gamma^2 \lambdabar} ~. \label{param}
\end{eqnarray}
The detector is supposed to be far away from the source so that
the above-given definition of far-zone holds.

The most important general statement concerning ER is that all
possible situations correspond to different values of the two
dimensionless parameters $\delta$ and $\phi$.

\begin{figure}
\begin{center}
\includegraphics*[width=100mm]{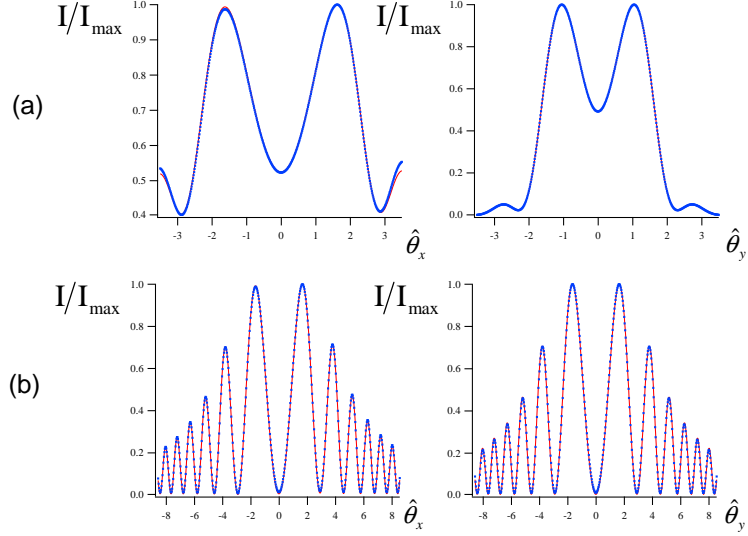}
\caption{\label{simil2} Verification of similarity techniques.
Left and right plots show the normalized angular spectral flux as
a function of the horizontal and vertical angles $\hat{\theta}_x$
and $\hat{\theta}_y$ respectively (at $\hat{\theta}_y =0$ and
$\hat{\theta}_x = 0$ respectively). (a) Case $\delta \simeq 0.43$
and $\phi \simeq 6.7 \cdot 10^{-3}$. Solid curve is the result of
SRW calculations with $L = 0.5$ m, $R = 400$ m, $\lambda = 400 $
nm at $17.5$ GeV. Dotted curve is the result for $L = 1$ m, $R =
800$ m, $\lambda = 800$ nm at $17.5$ GeV. (b) Case $\delta \ll 1$
and $\phi \simeq 4$. Solid curve is the result of SRW calculations
with $L = 300$ m, $R = 400$ m, $\lambda = 400 $ nm, at $17.5$ GeV
(corresponding to $\delta \simeq 7 \cdot 10^{-4}$). Dotted curve
is the result for $L = 150$ m, $R = 400$ m, $\lambda = 800$ nm at
$8.5$ GeV (corresponding to $\delta \simeq 2 \cdot 10^{-3}$). }
\end{center}
\end{figure}

\begin{figure}
\begin{center}
\includegraphics*[width=100mm]{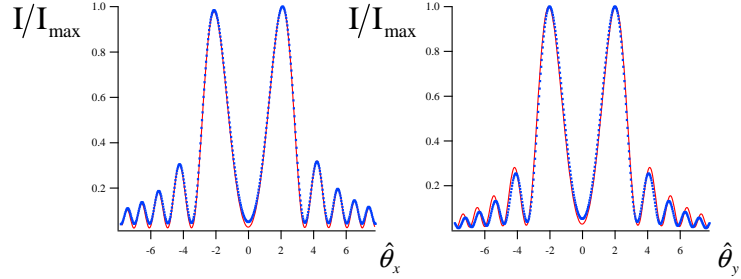}
\caption{\label{simil3} Illustration of self-similarity
techniques. Left and right plots show the normalized angular
spectral flux as a function of the horizontal and vertical angles
$\hat{\theta}_x$ and $\hat{\theta}_y$ respectively (at
$\hat{\theta}_y =0$ and $\hat{\theta}_x = 0$ respectively). The
angular spectral flux profile asymptotically approaches the
self-similar form ${I}/{I}_\mathrm{max} = F(\hat{\theta}_x,
\hat{\theta}_y)$ for $\delta \ll 1$ and $\phi \ll 1$. Solid curve
is the result of SRW calculations with $\delta \simeq 0.02$ and
$\phi \simeq 0.13$. Dotted curve is refers to the case $\delta =
0.01$ and $\phi = 0.27$ instead. }
\end{center}
\end{figure}

Note that the working limit $\lambdabar/\lambdabar_c \gg 1$ means
$\phi \cdot \delta \ll 1$ in terms of dimensionless parameters.
For any two cases characterized by the same values of $\delta$ and
$\phi$ the angular spectral flux from set up in Fig.
\ref{geome}(a) will "look" the same in terms of angles scaled to
$\sqrt{\lambdabar/L}$, i.e. $\hat{\theta} =
\theta/\sqrt{\lambdabar/L}$. In other words, data for different
sets of problem parameters corresponding to the same values of
$\delta$ and $\phi$ reduce to a single curve when properly
normalized.   We tested the scaling properties of ER by running
numerical simulations with the first principle computer code SRW.
We used two different sets of dimensional parameters corresponding
to the same case in terms of parameters $\delta$ and $\phi$, and
we checked that the angular spectral fluxes normalized to their
maximal values are identical. Results are presented in Fig.
\ref{simil2} and Fig. \ref{simil3}, where the normalized angular
spectral flux is indicated with ${I}/{I}_\mathrm{max}$.


When $\delta \sim 1$, the presence of the bending magnet radiation
strongly influences the radiation profile (see Fig.
\ref{simil2}(a)). When $\delta$ decreases up to $\delta \ll 1$,
one can neglect bending magnet contributions (see Fig.
\ref{simil2}(b)): what is left in this case is ER. These
situations are realized, for example, if one works at fixed
$\lambdabar$, $\gamma$ and $R$ while increasing the length $L$ as
in the case of Fig. \ref{FIGV1}. It follows that $\delta$ is
responsible for the relative weight of ER and bending magnet
radiation contributions in the radiation profile. Since we are
interested in ER emission, it is natural to consider more in
detail the limit for $\delta \ll 1$. In this case, results are
independent on the actual value of $\delta$, and the only
parameter left is $\phi$. This fact can be seen from Fig.
\ref{simil2}(b), where the two sets of dimensional parameters
refer to two different value of $\delta \ll 1$. We will name this
situation the sharp-edge asymptote.

In the limit for $\phi \ll 1$, the opening angle of the radiation
is independent of the actual value of $\phi$ too. In this case we
obtain the universal plot shown in Fig. \ref{simil3}, and one
talks about a self-similar behavior of the angular spectral flux
profile, which asymptotically approaches the self-similar form
${I}/{I}_\mathrm{max} = F(\hat{\theta}_x, \hat{\theta}_y)$. Note
that the separation distance $L$ between the bends dramatically
lowers the radiation beam divergence, but the characteristic angle
of emission is still larger than $1/\gamma$. In fact, radiation
peaks at ${\theta} \simeq 2.2 \sqrt{\lambdabar/L}$. When $\phi$
increases, radiation becomes better and better collimated, up to
angles $\theta \sim 1/\gamma$. This happens for values $\phi
\simeq 1$. Radiation has reached the best possible collimation
angle and further increase of $\phi$ (see Fig. \ref{simil2}(b))
only modifies fine structures in the radiation profile.

\subsection{\label{sub:quali} Qualitative description}

It is possible to present intuitive arguments to explain why all
problem parameters ($R$, $\gamma$, $L$ and $\lambdabar$) are
effectively grouped in $\delta$ and $\phi$.

To this purpose let us consider first the parameter $\delta$. By
definition, $1/\delta$ is a measure of the straight section length
$L$ in units of a characteristic length $\sqrt[3]{R^2
\lambdabar}$.

\begin{figure}
\begin{center}
\includegraphics*[width=100mm]{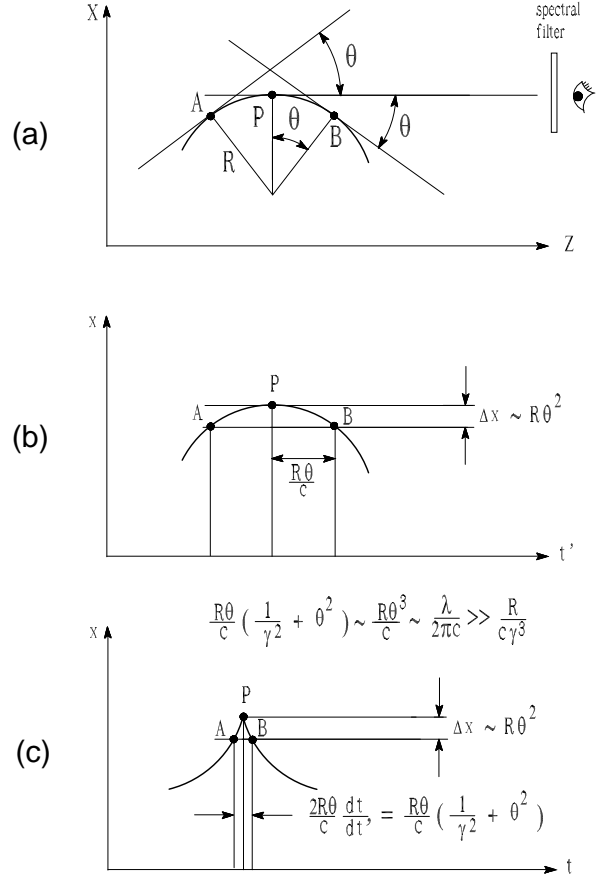}
\caption{\label{forle} Geometry for SR from a bending magnet.}
\end{center}
\end{figure}
To explain the meaning of the quantity $\sqrt[3]{R^2 \lambdabar}$,
following \cite{KIMM} we consider Fig. \ref{forle}(a), and we
focus on the region of parameters $\lambdabar \ll R$ and $\gamma^2
\gg 1$. A posteriori, this region of parameters will turn out to
correspond to an angular dimension along the trajectory  $2 \theta
\ll 1$ within the bending magnet. Radiation from an electron
passing through the setup is observed through a spectral filter by
a fixed observer positioned on the tangent to the bend at point
$P$. Electromagnetic sources propagate through the system, as a
function of time, as shown in Fig. \ref{forle}(b). However,
electromagnetic signals emitted at time $t'$ at a given position
$x(t')$ arrives at the observer position at a different time $t$,
due to finite speed of light. As a result, the observer in Fig.
\ref{forle}(a) sees the electromagnetic source motion as a
function of $t$. What one needs to know, in order to calculate the
electric field, is the apparent motion $x(t)$ shown in Fig.
\ref{forle}(c), which is a hypocycloid, and not the real motion
$x(t')$. In fact, the electric field at the observation point is
proportional to the second derivative of the x-coordinate with
respect to the retarded time $t$, because the observer sees
everything as delayed. We discuss the case when the source is
heading towards the observer. Using the fact that $\theta \ll 1$,
one obtains the well-known relation ${dt}/{dt'}={1}/{2}\cdot
\left({1}/{\gamma^2}+\theta^2\right)$. The observer sees a
time-compressed motion of the sources, which go from point $A$ to
point $B$ in an apparent time corresponding to an apparent
distance $2R \theta dt/(dt')$. Let us assume (this assumption will
be justified in a moment) $\theta^2 > 1/\gamma^2$. In this case
one has $2R \theta dt/(dt') \simeq R\theta^3$. Obviously one can
distinguish between radiation emitted at point $A$ and radiation
emitted at point $B$ only when $R \theta^3 \gg \lambdabar$, i.e.
for $\theta \gg (\lambdabar/R)^{1/3}$. This means that, as
concerns the radiative process, we cannot distinguish between
point $A$ and $B$ on the bend such that $R \theta \lesssim (R^2
\lambdabar)^{1/3}$. It does not make sense at all to talk about
the position where electromagnetic signals are emitted within
$L_{fb} = (R^2 \lambdabar)^{1/3}$ (here we are assuming that the
bend is longer than $L_{fb}$). This characteristic length is
called the formation length for the bend. The formation length can
also be considered as a longitudinal size of a single-electron
source. Note that a single electron always produces
diffraction-limited radiation $d \cdot \Delta \theta \sim
\lambdabar$, $d$ being the transverse size and $\Delta \theta$ the
divergence of the source. Since $d \sim L_{fb} \Delta \theta$, it
follows that the divergence angle $\Delta \theta$ is strictly
related to $L_{fb}$ and $\lambdabar$: $\theta \sim
\sqrt{{\lambdabar}/{L_f}}$. One may check that, using $L_{fb} \sim
\sqrt[3]{R^2 \lambdabar_c}$, one obtains $\theta \sim
\sqrt[3]{\lambdabar/R}$; in particular, at $\lambdabar \sim
\lambdabar_c \sim R/\gamma^3$ one obtains $\theta \sim 1/\gamma$,
as is well-known for bending magnet radiation.

Let us now consider the case of a straight section of length $L$
inserted between the two halves of a bend. Since we cannot
distinguish between points within $L_{fb}$, the case $L=0$ is
obviously indistinguishable from the case $L \ll L_f$. Significant
deviations from the bending magnet case are to be expected when $L
\gtrsim L_{fb}$, i.e. when $\delta \lesssim 1$. This hints to the
fact that $\delta$ is responsible for the relative weight of ER
and bending magnet radiation contributions in the radiation
profile.

Let us now discuss the parameter $\phi$. By definition, $\phi$ is
a measure of the straight section length $L$ in units of a
characteristic length $\gamma^2 \lambdabar$.  One can still use
the same reasoning considered for the bend to define a region of
the trajectory where it does not make sense to distinguish between
different points. In the case of a straight section of length $L$
connecting $A$ and $B$, $dt/dt'=1/(2 \gamma^2)$. It follows that
the apparent distance $AB$ is equal to $L/(2 \gamma^2)$. Since it
does not make sense to distinguish between points within the
apparent electron trajectory such that $L/(2\gamma^2) \lesssim
\lambdabar$, one obtains a critical length of interest $\sim
\gamma^2 \lambdabar$. This hints to the fact that for values $\phi
\simeq 1$ radiation has reached the best collimation angle.

Note that for ultrarelativistic systems in general, the formation
length is always much longer than the radiation wavelength. This
is related with a large compression factor $dt/dt'$. For
comparison, in the case of non-relativistic motion the compression
factor $dt/dt' \simeq 1$, and the formation length is simply of
order of the radiation wavelength. The counterintuitive result
follows, that for ultrarelativistic systems one cannot localize
sources of radiation within a macroscopic part of the trajectory.

\section{\label{sec:theor} Paraxial approximation}

In the next two Sections we present a complete theory of ER. All
electrodynamical theories are based on the presence of small or
large parameters.

In general, the theory of Synchrotron Radiation (SR) is based on
the exploitation, for ultra-relativistic particles, of the small
parameter $\gamma^{-2}$. By this, Maxwell's equations are reduced
to much simpler equations with the help of paraxial approximation.
ER theory constitutes a particular case of SR theory, based on the
extra small-parameter $\delta$.

Here and everywhere else in this paper we will make consistent use
of Gaussian units.

In this Section we deal with the paraxial approximation of
Maxwell's equations. We will treat both near and far zone cases,
with special attention to the applicability region of equations
describing ER in different regions of the parameter space.

Whatever the method used to present results, one needs to solve
Maxwell's equations in unbounded space. We introduce a cartesian
coordinate system, where a point in space is identified by a
longitudinal coordinate $z$ and transverse position $\vec{r}$.
Accounting for electromagnetic sources, i.e. in a region of space
where current and charge densities are present, the following
equation for the field in the space-frequency domain holds in all
generality:

\begin{equation}
c^2 \nabla^2 \vec{\bar{E}} + \omega^2 \vec{\bar{E}} = 4 \pi c^2
\vec{\nabla} \bar{\rho} - 4 \pi i \omega \vec{\bar{j}}~,
\label{trdisoo}
\end{equation}
where $\bar{\rho}(z,\vec{r},\omega)$ and
$\vec{\bar{j}}(z,\vec{r},\omega)$ are the Fourier
transforms\footnote{We explicitly write the definitions of the
Fourier transform and inverse transform of a function $f(t)$ in
agreement with the notations used in this paper. The Fourier
transform and inverse transform pair reads:

\begin{eqnarray}
\nonumber \tilde{f}(\omega) = \int \d t ~f(t) \exp\left[i \omega t
\right]~;~~{f}(t) = \frac{1}{2\pi} \int \d \omega
~\tilde{f}(\omega) \exp\left[-i \omega t \right] . \label{spatgft}
\end{eqnarray}
} of the charge density $\rho(z,\vec{r},t)$ and of the current
density $\vec{j}(z,\vec{r},t)$. Eq. (\ref{trdisoo}) is the
well-known Helmholtz equation. Here $\vec{\bar{E}}$ indicates the
Fourier transform of the electric field in the space-time domain.

A system of electromagnetic sources in the space-time can be
conveniently described by $\rho(z,\vec{r},t)$ and
$\vec{j}(z,\vec{r},t)$. Considering a single electron and using
the Dirac delta distribution, we can write

\begin{equation}
\rho\left(z,\vec{r},t\right) = -e
\delta\left(\vec{r}-\vec{r}_0(t)\right)\delta(z-z_0(t)) =
-\frac{e}{v_z(z)}
\delta\left(\vec{r}-\vec{r}_0(z)\right)\delta\left(\frac{s(z)}{v}-t\right)\label{charge}
\end{equation}
\begin{equation}
\vec{j}\left(z,\vec{r},t\right) =  \vec{v}(t)
\rho\left(z,\vec{r},t\right) ~, \label{current}
\end{equation}
where $(z_0(t),\vec{r}_0(t))$ and $\vec{v}(t)$ are, respectively,
position and velocity of the particle at a given time $t$ in a
fixed reference frame, $v_z$ is the longitudinal velocity of the
electron , and $(-e)$ is the electron charge. Additionally, we
defined the curvilinear abscissa $s(z)=v t(z)$, where $v =
\left|\vec{v}(t(z))\right|$ is a constant. In the space-frequency
domain the electromagnetic sources transform to:

\begin{equation}
\bar{\rho}(\vec{r},z,\omega) = -{e\over{v_z(z)}}
\delta\left(\vec{r}-\vec{r}_0(z)\right) \exp\left[\frac{i \omega
s(z)}{v}\right] \label{charge2tr}
\end{equation}
and

\begin{equation}
\vec{\bar{j}}(\vec{r},z,\omega) = \vec{v}(z)
\bar{\rho}(\vec{r},z,\omega) \label{curr2tr}
\end{equation}
Since we will only be interested in the transverse components of
the field, from now on we will consider the transverse field
envelope $\vec{\widetilde{E}}$, a 2D vector defined in the
space-frequency domain as $\vec{\widetilde{E}} =
\vec{\bar{E}}_\bot \exp[-i \omega z/c]$, the symbol "$\bot$"
indicating projection on the transverse plane. By substitution in
Helmholtz equation we obtain

\begin{eqnarray}
\left({\nabla}^2 + {2 i \omega \over{c}} {\partial\over{\partial
z}}\right) \vec{\widetilde{E}} &=& {4 \pi e\over{v_z(z)}}
\exp\left[{i \omega
\left({s(z)\over{v}}-{z\over{c}}\right)}\right]
\left[{i\omega\over{c^2}}\vec{v}_\bot(z) -\vec{\nabla}_\bot\right]
\delta\left(\vec{r}-\vec{r}_0(z)\right),\label{incipit}
\end{eqnarray}
where, according to our notation, $\vec{\nabla}_\bot$ indicates a
gradient with respect to transverse coordinates only, and
$\vec{v}_\bot$ is the transverse velocity of the electron.  Eq.
(\ref{incipit}) is still fully general and may be solved in any
fixed reference system $(x,y,z)$ of choice with the help of an
appropriate Green's function.

When the longitudinal velocity of the electron, $v_z$, is close to
the speed of light $c$, one has $\gamma_z^2 \gg 1$, where
${\gamma}_z^{-2}=1-v_z^2/c^2$. The Fourier components of the
source are then almost synchronized with the electromagnetic wave
travelling at the speed of light. Note that this synchronization
is the reason for the time compression factor described in Section
\ref{sub:quali}. In this case, the phase $\omega
({s(z)/{v}}-{z/{c}})$ is a slow function of $z$ compared to the
wavelength. For example, in the particular case of motion on a
straight section, one has $s(z) = z/v_z$, so that $\omega
({s(z)/{v}}-{z/{c}}) = \omega z/(2\gamma_z^2 c)$, and if
$\gamma_z^2 \gg 1$ such phase grows slowly in $z$ with respect to
the wavelength. For a more generic motion, one similarly obtains:

\begin{equation}
\omega \left({s(z_2)-s(z_1)\over{v}}-{z_2-z_1\over{c}}\right) =
\int_{z_1}^{z_2} d \bar{z} \frac{\omega}{2 \gamma_z^2(\bar{z})
c}~, \label{moregen}
\end{equation}
Mathematically, the phase in Eq. (\ref{moregen}) enters in the
Green's function solution of Eq. (\ref{incipit}) as a factor in
the integrand. As we integrate along $z$, the factor
$\omega(s(z)/v - z/c)$ leads to an oscillatory behavior of the
integrand over a certain integration range in $z$. Such range can
be identified with the value of $z_2-z_1$ for which the right hand
side of Eq. (\ref{moregen}) is of order unity, and it is naturally
defined as the radiation formation length $L_f$ of the system at
frequency $\omega$. Of course there exist some freedom in the
choice of such definition: "order of unity" is not a precise
number, and reflects the fact that there is no abrupt threshold
between "oscillatory" and "non-oscillatory" behavior of the
integrand in the solution of Eq. (\ref{incipit}). It is easy to
see by inspection of Eq. (\ref{moregen}) that if $v_z$ is sensibly
smaller than $c$, but still of order $c$, i.e. $v_z\sim c$ but
$1/\gamma_z^2 \sim 1$, then $L_f \sim \lambdabar$. On the
contrary, when $v_z$ is very close to $c$, i.e. $1/\gamma_z^2 \ll
1$, the right hand side of Eq. (\ref{moregen}) is of order unity
for $L_f = z_2-z_1 \gg \lambdabar$. When the radiation formation
length is much longer than $\lambdabar$, $\vec{\widetilde{E}}$
does not vary much along $z$ on the scale of $\lambdabar$, that is
$\mid
\partial_z \widetilde{E}_{x,y}\mid \ll \omega/c \mid
\widetilde{E}_{x,y}\mid$. Therefore, the second order derivative
with respect to $z$ in the $\nabla^2$ operator on the left hand
side of Eq. (\ref{incipit}) is negligible with respect to the
first order derivative, and Eq. (\ref{incipit}) can be simplified
to

\begin{eqnarray}
\left({\nabla_\bot}^2 + {2 i \omega \over{c}}
{\partial\over{\partial z}}\right) \vec{\widetilde{E}} &=& {4 \pi
e\over{c}} \exp\left[{i \omega
\left({s(z)\over{v}}-{z\over{c}}\right)}\right]
\left[{i\omega\over{c^2}}\vec{v}_\bot(z)-\vec{\nabla}_\bot\right]
\delta\left(\vec{r}-\vec{r}_0(z)\right)~,\cr && \label{incipit2}
\end{eqnarray}
where, as said before, we consider transverse components of
$\vec{\widetilde{E}}$, and we substituted $v_z(z)$ with $c$, based
on the fact that $1/\gamma_z^2 \ll 1$. Eq. (\ref{incipit2}) is
Maxwell's equation in paraxial approximation. Eq. (\ref{incipit}),
which is an elliptic partial differential equation, has thus been
transformed into Eq. (\ref{incipit2}), that is of parabolic type.
Note that the applicability of the paraxial approximation depends
on the ultra-relativistic assumption $\gamma^2 \gg 1$ but not on
the choice of the $z$ axis. If, for a certain choice of the
longitudinal $z$ direction, part of the trajectory is such that
$\gamma_z^2 \sim 1$, the formation length is very short ($L_f \sim
\lambdabar$), and the radiated field is practically zero. As a
result, Eq. (\ref{incipit2}) can always be applied, i.e. the
paraxial approximation can always be applied, whenever $\gamma^2
\gg 1$.

Complementarily, it should also be remarked here that the status
of the paraxial equation Eq. (\ref{incipit2}) in Synchrotron
Radiation theory is different from that of the paraxial equation
in Physical Optics. In the latter case, the paraxial approximation
is satisfied only by small observation angles. For example, one
may think of a setup where a thermal source is studied by an
observer positioned at a long distance from the source and behind
a limiting aperture. Only if a small-angle acceptance is
considered the paraxial approximation can be applied. On the
contrary, due to the ultra-relativistic nature of the emitting
electrons, contributions to the SR field from parts of the
trajectory with formation length $L_f \gg \lambdabar$ (the only
non-negligible) are highly collimated. As a result, the paraxial
equation can be applied at any angle of interest, because it
practically returns zero field at angles where it should not be
applied.

Finally, since the characteristic scale of variation of
$\vec{\widetilde{E}}$ is much larger than $\lambdabar$, the
paraxial approximation is valid up to distances of the observer
from the electromagnetic sources of order $\lambdabar$.

The Green's function for Eq. (\ref{incipit2}), namely the solution
corresponding to the unit point source can explicitly be written
in an unbounded region as

\begin{eqnarray}
G(\vec{r}-\vec{r'},z-z') = -{1\over{4\pi |z-z'|}} \exp\left[
i\omega{\mid \vec{r}-\vec{r'}\mid^2\over{2c |z-z'|}}\right]
\label{green}~,
\end{eqnarray}
Note that when $z-z' < 0$ the paraxial approximation does not
hold, and the paraxial wave equation Eq. (\ref{incipit2}) should
be substituted, in the space-frequency domain, by the more general
Helmholtz equation. However, the radiation formation length for $z
- z'<0$ is very short with respect to the case $z - z' >0$, i.e.
there is effectively no radiation for observer positions $z-z'
<0$. As a result, in this paper we will consider only $z - z'> 0$.
The reason why $|z-z'|$ is present in Eq. (\ref{green}) (while
$z-z'>0$ always) is that, mathematically, the Green's function $G$
is actually related to the \textit{operator} on the left hand side
of Eq. (\ref{incipit2}), and not to the whole equation. For
example, when dealing with wavefront propagation, one must
consider the homogeneous version of Eq. (\ref{incipit2}), and the
same Green's function in Eq. (\ref{green}) can be used, as we will
see, to propagate the electric field. In this case, propagation
can be performed in the backward direction as well, i.e. for $z -
z'<0$.

Note that Eq. (\ref{green}) automatically include information
about the boundary condition for the field. In the present case,
since we are dealing with unbounded space, the field vanishes at
large distance from the sources. Due to this fact, the Green's
function in Eq. (\ref{green}) is a scalar function, while in
general it admits tensorial values. Using the definition of
Green's function, and carrying out integration over transverse
coordinates we obtain

\begin{eqnarray}
\vec{\widetilde{E}} &=& \frac{4\pi e}{c} \int_{-\infty}^{z} dz'
\left\{ \frac{i\omega}{c^2} \vec{v}_\bot(z')
G\left(\vec{r}-\vec{r}_0(z'), z-z'
\right)+\left[\vec{\nabla}'_\bot G\left(\vec{r}-\vec{r'},
z-z'\right) \right]_{\vec{r'}=\vec{r}_0(z')}\right\}\cr &&\times
\exp\left[{i \omega
\left({s(z)\over{v}}-{z\over{c}}\right)}\right]~,\label{efielGfree}
\end{eqnarray}
where$\vec{\nabla}'_\bot $ indicates derivative with respect to
$\vec{r'}$. Explicit substitution of Eq. (\ref{green}) yields the
following result

\begin{eqnarray}
\vec{\widetilde{E}}(z, \vec{r}) &=& -\frac{i \omega e}{c^2}
\int_{-\infty}^{z} dz' \frac{1}{z-z'}
\left[\frac{\vec{v}_\bot(z')}{c}
-{\vec{r}-\vec{r}_0(z')\over{z-z'}}\right]\cr &&\times
\exp\left\{i\omega\left[{\mid \vec{r}-\vec{r'} \mid^2\over{2c
(z-z')}}+ \int_{0}^{z'} d \bar{z} \frac{1}{2 \gamma_z^2(\bar{z})
c}\right] \right\} ~, \label{generalfin2}
\end{eqnarray}
where we the choice of integration limits in $d\bar{z}$ indicate
that the electron arrives at position $z=0$ at time $t_a=0$. Eq.
(\ref{generalfin2}) is valid at any observation position $z$ such
that the paraxial approximation is valid, i.e. up to distances
between the observer and the electromagnetic sources comparable
with the radiation wavelength. One may recognize two terms in Eq.
(\ref{generalfin2}). The first in $\vec{v}_\bot(z')$ can be traced
back to the current term on the right hand side of Eq.
(\ref{incipit}), while the second, in $\vec{r}-\vec{r}_0(z')$,
corresponds to the gradient term on the right hand side of Eq.
(\ref{incipit}).

Eq. (\ref{generalfin2}) is used as starting point for numerical
codes like SRW. The only approximation used is the paraxial
approximation. This rules out the possibility of using SRW to
study the region of applicability of the paraxial approximation.
However, once the paraxial approximation is granted for valid,
SRW, or Eq. (\ref{generalfin2}), can be used to investigate the
applicability of ER theory, which is built within the constraints
of the paraxial approximation. Note that the evaluation of the
field begins with the knowledge of the trajectory followed by the
electron, which is completely generic. In other words, one needs
to know the electromagnetic sources to evaluate the field at any
position $z$ down the beamline.

Alternatively, the knowledge of the far-zone field distribution,
i.e. a limit of Eq. (\ref{generalfin2}), allows one to specify an
algorithm to reconstruct the field in the near zone up to
distances of the observer from the sources much larger than
$\lambdabar$. An important characteristic of this algorithm is
that it works within the region of applicability of the paraxial
approximation, $\gamma^2 \gg 1$ only. Such algorithm was developed
in \cite{OURF} to deal with SR problems in full generality, and
will be used in this paper to develop our ER theory. It follows
three major steps.

\paragraph*{I.} The first step is the characterization of ER emission in the
far zone. From Eq. (\ref{generalfin2}) follows directly:

\begin{eqnarray}
\vec{\widetilde{{E}}}(z, \vec{\theta})&& = -{i \omega
e\over{c^2}z} \int_{-\infty}^{z} dz'
\left({\vec{v}_\bot(z')\over{c}} -{\vec{\theta}}\right) \cr &&
\times{\exp{\left[i \omega \int_{0}^{z'} \frac{d \bar{z}}{2 c
{\gamma}_z^2(\bar{z}) }+ \frac{i \omega}{{2 c}}\left( z
~\theta^2-2 \vec{\theta}\cdot\vec{ {r}}_0(z')+{z'
\theta^2}\right)\right]}} ~, \label{generalfin}
\end{eqnarray}
where $\vec{\theta}= \vec{r}/z$ defines the observation direction,
and $\theta \equiv |\vec{\theta}|$. Note that the concept of
formation length is strictly related to the concept of observation
angle of interest. In fact, given a certain formation length
$L_f$, and substituting it into the phase in $z' \theta^2$ in Eq.
(\ref{generalfin}), one sees that the integrand starts to be
highly oscillatory for angles $\theta \simeq
\sqrt{\lambdabar/L_f}$. Eq. (\ref{generalfin}) is taken as the
starting point for our algorithm.

\paragraph*{II.} The second step consists in interpreting the far-zone field in Eq.
(\ref{generalfin}) as a laser-like beam, generated by one or more
virtual sources. These sources are not present in reality, because
they are located at positions inside the magnetic setup, but they
produce the same field as that of the real system. Hence the
denomination "virtual".  A virtual source is similar, in many
aspects, to the waist of a laser beam and, in our case, exhibits a
plane wavefront. It is then completely specified, for any given
polarization component, by a real-valued amplitude distribution of
the field, located at a fixed longitudinal position.

Suppose we know the field at a given plane at $z$, and we want to
calculate the field at another plane at $z_s$. In paraxial
approximation and in free space, the homogeneous version of Eq.
(\ref{incipit2}) holds for the complex envelope
$\vec{\widetilde{E}}$ of the Fourier transform of the electric
field along a fixed polarization component, that is $[{
{\nabla}_\bot}^2 + ({2 i \omega}/{c}) {\partial_z}]
\vec{\widetilde{E}} = 0$. One has to solve this equation with a
given initial condition at ${z}$, which defines a Cauchy problem.
We obtain

\begin{equation}
\vec{ \widetilde{E}}( {z}_s,\vec {r}) = - \frac{2 i \omega}{c}
\int d \vec{ {r}'}~ \vec{\widetilde{E}}(z,\vec{r'})
G\left(\vec{r}-\vec{r'},z_s-z\right)~, \label{fieldpropback}
\end{equation}
where the integral is performed over the transverse plane and the
Green's function $G$ in unbounded space is given in Eq.
(\ref{green}). Similarly as before, it is important to remark that
since $\vec{\widetilde{E}}$ is a slowly-varying function with
respect to the wavelength, one cannot resolve the evolution of the
field on a longitudinal scale of order of the wavelength within
the accuracy of the paraxial approximation. In order to do so, the
paraxial equation should be replaced by the more general Helmholtz
equation. Let us now consider the limit $z \longrightarrow
\infty$, with finite ratio $\vec{r'}/z$. In this case, the
exponential function in Eq. (\ref{fieldpropback}) can be expanded
giving

\begin{eqnarray}
\vec{ \widetilde{E}}(z_s,\vec{r}) &=& \frac{i \omega}{2 \pi c {z}}
\int d \vec{ {r}'}~ \vec{\widetilde{E}}(z,\vec{r'})
\exp{\left[-\frac{i \omega}{{2 c {z}}}\left(  r^2-2 \vec{
{r}}\cdot\vec{ {r}'}+\frac{z_s r^2}{z}\right)\right]}~.
\label{fieldpropback2}
\end{eqnarray}
Letting $\vec{\theta} = \vec{r'}/z$ we have

\begin{eqnarray}
\vec{ \widetilde{E}}\left( z_s,\vec{\theta} \right)&=& \frac{i
\omega {z}}{2 \pi c} \int d\vec{\theta}\exp{\left[-\frac{i \omega
\theta^2}{2 c}(z+z_s)\right]}\vec{
\widetilde{E}}\left(z,\vec{\theta}\right) \exp\left[\frac {i
\omega}{c} \vec{r} \cdot \vec{\theta}\right] ~ ,
\label{virfiemody}
\end{eqnarray}
where the transverse vector $\vec{r}$ defines a transverse
position on the virtual source plane at $z=z_s$. Eq.
(\ref{virfiemody}) allows to calculate the field at the virtual
source once the field in the far zone is known. The specification
of the virtual source amounts to the specification of an initial
condition for the electric field, that is then propagated at any
distance. From this viewpoint, identification of the position
$z=z_s$ with the virtual source position is possible independently
of the choice of $z_s$. In other words, like in laser physics, the
SR field can be propagated starting from any point $z_s$. However,
there are choices that are more convenient than others, exactly
like in laser physics the waist plane is privileged with respect
to others. The most convenient choice of $z_s$ is the one that
allows maximal simplification of the phase contained in the
far-zone field ${\widetilde{E}}(\vec{\theta})$ with the quadratic
phase factor in ${\theta}^2$ in Eq. (\ref{virfiemody}). In
practical situations of interest it is possible to choose $z_s$ in
such a way that the field at the virtual source exhibits a plane
wavefront, exactly as for the waist of a laser beam. Finally, in
some cases, it is convenient to consider the far zone field
${\widetilde{E}}(\vec{\theta})$ as a superposition of different
contributions. In this way, more than one virtual source can be
identified and treated independently, provided that different
contributions are finally summed together.

\paragraph*{III.} The third, and final step, consists in the
propagation of the field from the virtual sources in paraxial
approximation. Each source $\vec{\widetilde{E}}( z_s,\vec{r} )$
generates the field

\begin{eqnarray}
\vec{\widetilde{E}}({z},\vec {r}) = \frac{i \omega}{2 \pi c( {z}-
{z_s})} \int d \vec{{r}'}~ \vec{\widetilde{E}}(z_s,\vec{r'})
\exp{\left[\frac{i \omega \left|{\vec{ {r}}}-\vec{
{r}'}\right|^2}{2 c ( {z}- {z_s})}\right]}~,
\label{fieldpropbackx}
\end{eqnarray}
as follows directly from Eq. (\ref{fieldpropback}).

Here it should be stressed that the inverse field problem, which
relies on far-field data only, cannot be solved without
application of the paraxial approximation.

If the paraxial approximation were not applicable, we should have
solved the homogeneous version of Eq. (\ref{trdisoo}). Boundary
conditions would have been constituted by the knowledge of the
field on a open surface (for example, a transverse plane) and
additionally, by Rayleigh-Sommerfeld radiation condition at
infinity, separately for all polarization components of the field.
However, this would not have been enough to reconstruct the field
at any position in space. In order to do so, we would have had to
specify the sources. In fact, the boundary conditions specified
above allow one to solve the direct transmission problem, but not
the inverse one.

Since, however, the paraxial approximation is applicable, the
inverse field problem turns out to have unique and stable
solution. A conservative estimate of the accuracy of the paraxial
approximation is related to the distance $d$ between the point
where we want to know the field and the electron trajectory in the
space-frequency domain \cite{OURF}. When $d \gtrsim L_f$ this
accuracy can be estimated to be of order $\lambdabar/L_f$. It
quickly decreases as $L_f \gg d \gg \lambdabar$ remaining,
\textit{at least}, of order of $\lambdabar/d$.

The fact that we can reconstruct the near-zone field starting from
the knowledge of the far-zone field and from the propagation
equation looks paradoxical. In fact, in the far zone, all
information about the velocity field in the Lienard-Wiechert
expressions for the field is lost. It is interesting to note, for
example, what is reported in \cite{WILL} concerning the velocity
term: "In the case of infrared synchrotron radiation, and THz
radiation in particular, this term is not small and must be
included in all calculations". Here we apparently seem to have
lost track of every information about the velocity term. As shown
in \cite{OURF}, the paradox is solved by the fact that, although
in the far-zone limit of Eq. (\ref{generalfin2}) includes
information about the Fourier Transform of the acceleration term
of the Lienard-Wiechert fields only, information about the
velocity term is included in the field propagation equation
through the Green's function Eq. (\ref{green}), which solves
Maxwell's equations as the Lienard-Wiechert expressions do.

\section{\label{zero} Sharp-edge approximation}

Let us consider the system depicted in Fig. \ref{geome}(a). In
this case of study the trajectory and, therefore, the space
integration in Eq. (\ref{generalfin}) can be split in three parts:
the two bends, which will be indicated with $b_1$ and $b_2$, and
the straight section $AB$.  One may write

\begin{eqnarray}
\vec{\widetilde{{E}}}(z, \vec{r}) &=&
\vec{\widetilde{{E}}}_{b1}(z, \vec{r})+
\vec{\widetilde{{E}}}_{AB}(z, \vec{r})+
\vec{\widetilde{{E}}}_{b2}(z, \vec{r})~, \label{splite}
\end{eqnarray}
with obvious meaning of notation. We will denote the length of the
segment $AB$ with $L$. This means that points $A$ and $B$ are
located at longitudinal coordinates $z_A= -L/2$ and $z_B= L/2$.

Recalling the geometry in Fig. \ref{geome}(a) we have $\gamma_z(z)
= \gamma$ for $z_A<z'<z_B$. With the help of Eq.
(\ref{generalfin}) we write the contribution from the straight
line $AB$

\begin{equation}
\vec{\widetilde{E}}_{AB}={i \omega e\over{c^2 z}}
\int_{-L/2}^{L/2} dz' \vec{\theta} \exp{\left\{\frac{i \omega}{c}
\left[ {\theta^2 z\over{2}} +
{z'\over{2}}\left({1\over{\gamma^2}}+\theta^2\right)\right]\right\}}
~, \label{ABcontre}
\end{equation}
where we assumed that $\vec{r}_0(z')=0$, i.e. that the particle
has zero offset and deflection. From Eq. (\ref{ABcontre}) one
obtains:

\begin{eqnarray}
\vec{\widetilde{E}}_{AB}&&=\frac{ i \omega e L}{c^2 z}
\exp\left[\frac{i \omega \theta^2 z}{2 c}\right] \vec{\theta} ~
\mathrm{sinc}\left[\frac{\omega L}{4
c}\left(\theta^2+\frac{1}{\gamma^2}\right)\right]~.
\label{ABcontrint4e}
\end{eqnarray}
Note that Eq. (\ref{ABcontrint4e}) describes a spherical wave with
the center in the middle of the straight section. Moreover, it
explicitly depends on $L$.

In the previous Section we defined the formation length as the
length needed for the phase of the electric field seen by an
observer on the $z$ axis to overtake the phase of the sources of a
radiant. It follows from this definition, and from the phase in
Eq. (\ref{ABcontre}) that the formation length $L_{f}$ for the
straight section $AB$ can be written as $L_{f} \sim \min[\gamma^2
\lambdabar ,L]$. Then, either $L_{f} \sim \gamma^2 \lambdabar$ or
$L_{f} \sim L$. In both cases, with the help of the phase in the
integrand in Eq. (\ref{ABcontre}) we can formulate on a purely
mathematical basis an upper limit to the value of the observation
angle of interest for the straight line, $\theta_{x,y}^2
\lesssim{\lambdabar}/{L_f}$. Moreover, if $L_{f} \sim \gamma^2
\lambdabar$, the maximal angle of interest is independent of the
frequency and equal to $1/\gamma$, in agreement with what has been
said before.

According to the superposition principle, one should sum the
contribution due to the straight section to that from the bends.
However, as discussed in Section \ref{sec:param}, when $\delta \ll
1$ one can ignore the presence of the bending magnets with good
accuracy. Note that a direct confirmation of this fact can be
given by analyzing explicitly the field from the half bends, e.g.
$\vec{\widetilde{{E}}}_{b2}$. An expression for the quantity
$\vec{\widetilde{{E}}}_{b2}$ can be found from first principles,
applying Eq. (\ref{generalfin}) to the case of a half bend.

In the framework of the paraxial approximation we obtain for $z >
L/2$:

\begin{equation}
s(z)  \simeq z + \frac{(z-L/2)^3}{6R^2}~~,~~~~\vec{r}(s) =
-R\left[1-\cos\left(\frac{s-L/2}{R}\right)\right] \vec{e}_x~,
\label{sz}
\end{equation}
$\vec{e}_x$ (or $\vec{e}_y$) being a unit vector along the $x$ (or
$y$) direction. Substitution in Eq. (\ref{generalfin}) and use of
Eq. (\ref{moregen}) gives\footnote{Usually, in textbooks, the $z$
axis is chosen in such a way that $\theta_x = x/z = 0$, i.e. it is
not fixed, but depends on the observer position. This can always
be done, and simplifies calculations. However, since the wavefront
is not spherical, this way of proceeding can hardly help to obtain
the phase of the field distribution on a plane perpendicular to a
$\textit{fixed}$ $z$ axis. Calculations in our (fixed) coordinate
system is more complicated and can be found in e.g. in
\cite{JAPP}. After some algebraic manipulation and change of
variables we obtain Eq. (\ref{srtwob}).}:

\begin{eqnarray}
\vec{\widetilde{{E}}}_{b2}(z, \vec{\theta}) &=& \frac{i \omega
e}{c^2 z} \exp\left[i\Phi_s\right] \exp\left[i\Phi_0\right]
\int_{R \theta_x}^{\infty} dz' \left({z'\over{R}}
{\vec{e}_x}+\theta_y {\vec{e}_y}\right) \cr && \times
\exp\left\{{\frac{i
\omega}{c}\left[{z'\over{2\gamma^2}}\left(1+\gamma^2\theta_y^2\right)
+{z'^3\over{6R^2}}\right]}\right\}~,\label{srtwob}
\end{eqnarray}
where
\begin{eqnarray}
&&\Phi_s = \frac{\omega z ( \theta_x^2+\theta_y^2
)}{2c}~~~\mathrm{and}\cr && \Phi_0 = -\frac{\omega R
\theta_x}{2c}\left( {1\over{\gamma^2}} +{\theta_x^2\over{3}}
+\theta_y^2 \right)+\frac{\omega L}{4
c}\left(\frac{1}{\gamma^2}+\theta_x^2+\theta_y^2\right)~.\label{phio}
\end{eqnarray}
Analysis of the phase term in $z'^3$ Eq. (\ref{srtwob}) shows that
the integrand starts to exhibit oscillatory behavior within
distances of order of $L_{fb} = \sqrt[3]{R^2 \lambdabar}$, that is
the radiation formation length for the bending magnet at
$\lambdabar \gg R/\gamma^3$. Similarly, we have seen from Eq.
(\ref{ABcontre}), and has been also shown with qualitative
arguments in Section \ref{sec:param}, the formation length for the
straight section amounts to $L_{f} = \min[L,\gamma^2\lambdabar]$.
The ratio $L_{fb}/L_f$ is responsible for the relative weight of
ER compared to bending magnet field contribution. Note that
strictly speaking, when $\phi \gg 1$, $L_{fb}/L_f$ is equal to
$\delta \cdot \phi$ (and not to $\delta$). However, $\delta \cdot
\phi \ll 1$ always, to insure that $\lambdabar \gg \lambdabar_c$.
As a result, in all generality, it is possible to talk about ER if
and only if $\delta \ll 1$ and $\delta \cdot \phi \ll 1$ (or
$\sqrt[3]{R^2 \lambdabar}/L \ll 1$, $\lambdabar/\lambdabar_c \ll
1$ in terms of dimensional parameters).

When $\delta \gtrsim 1$, one cannot talk about pure ER. One must
account for bending magnet contributions as well. Then,
expressions presented here for the electric field from the
straight section can be seen as partial contributions, to be added
to bending magnet contributions calculated elsewhere.

Note that in this paper we first introduced a measure of "how
sharp" the edges are through the parameter $\delta$ and, with
this, we specified the region of applicability for ER theory.

It should also be noted that bending magnets at the straight
section ends act like switchers, i.e. they switch on and off
radiation seen by an observer. Observers see uniform intensity
from a bend along the horizontal direction. However, not all parts
of the trajectory contribute to the radiation seen by a given
observer, because radiation contributions from different parts of
the bend is highly collimated, hence the switching function. Since
we are not interested in electromagnetic sources responsible for
field contributions that are not seen by the observer, we may say
that bends switch on and off electromagnetic sources as well.

Finally, it should be remarked that the far-zone asymptotic in Eq.
(\ref{ABcontre}) is valid at observation positions $z \gg L$. This
is a necessary and sufficient condition for the vector $\vec{n}$
pointing from source to observer, to be considered constant. This
result is independent of the formation length. When $L \lesssim
\gamma^2\lambdabar$ we can say that an observer is the far zone if
and only if it is located many formation lengths away from the
origin. This is no more correct when $L \gg \gamma^2\lambdabar$.
In this case the observer can be located at a distance $z \gg
\gamma^2 \lambdabar$, i.e. many formation lengths away from the
origin of the reference system, but still at $z \sim L$, i.e. in
the near zone. As we see here, the formation length $L_f$ is
often, but not always related to the definition of the far (or
near) zone. In general, the far (or near) zone is related to the
characteristic size of the system, in our case $L$. In its turn
$L_f \lesssim L$, which includes, when $\gamma^2 \lambdabar \ll
L$, the situation $L_f \ll L$.

Since in the following we will only deal with a contribution of
the electric field, i.e. that from the straight section
$\vec{\widetilde{{E}}}_{AB}$, from now on, for simplicity, we will
omit the subscript $AB$.

\begin{figure}
\begin{center}
\includegraphics*[width=100mm]{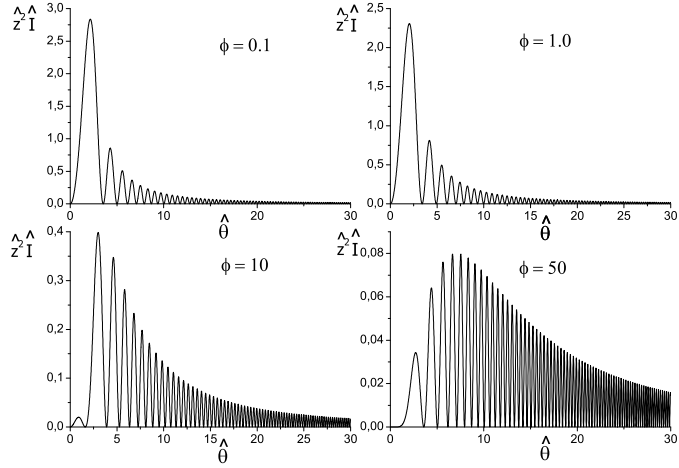}
\caption{\label{Dirfar2} Normalized angular spectral flux of the
radiation from the setup in Fig. \ref{geome}(a) for different
values of $\phi$.}
\end{center}
\end{figure}

The radiation energy density as a function of angles and
frequencies $\omega$, i.e. the angular spectral flux, can be
written as

\begin{eqnarray}
\frac{d W}{d\omega d \Omega} = \frac{c z^2}{4 \pi^2}
\left|\vec{\widetilde{E}}\right|^2~, \label{endene}
\end{eqnarray}
$d \Omega$ being the differential of the solid angle $\Omega$.
%
Substituting Eq. (\ref{ABcontrint4e}) in Eq. (\ref{endene}) it
follows that \cite{CHUS}

\begin{eqnarray}
&&\frac{d W}{d\omega d \Omega} = \frac{ e^2 \omega^2 L^2}{4 \pi^2
c^3} {\theta^2} ~ \mathrm{sinc}^2\left[\frac{L \omega}{4
c}\left(\theta^2+\frac{1}{\gamma^2}\right)\right]~.
\label{enden2e}
\end{eqnarray}
It is now straightforward to introduce the same normalized
quantities defined in Section \ref{sec:param}: $\vec{\hat{\theta}}
= \sqrt{L/\lambdabar} ~\vec{\theta}$, and ${\phi} = {L}/({\gamma^2
\lambdabar})$. We may write the angular spectral flux in
normalized units as

\begin{eqnarray}
\hat{I} \equiv \frac{c^3 L}{\omega e^2} \left
|\vec{\widetilde{E}}\right|^2 ~,\label{Ihat}
\end{eqnarray}
so that

\begin{eqnarray}
\hat{z}^2 \hat{I} =  {\hat{\theta}^2} ~
\mathrm{sinc}^2\left[\frac{1}{4
}\left(\hat{\theta}^2+{\phi}\right)\right]~, \label{normI1}
\end{eqnarray}
where $\hat{z} = z/L$. Eq. (\ref{normI1}) is plotted in Fig.
\ref{Dirfar2} for several values of ${\phi}$ as a function of the
normalized angle $\hat{\theta}$. The natural angular unit is
evidently $\sqrt{\lambdabar/L}$.

The angular spectral flux, once integrated in angles, is
divergent, as one can see from $\hat{z}^2 \hat{I} \propto
1/\hat{\theta}$. This feature is related with the limit of
applicability of the sharp-edge approximation. Note that within
the framework of the paraxial approximation alone, the integrated
angular spectral flux calculated with Eq. (\ref{generalfin}) does
not have any singularity, whatever the electron trajectory is. The
paraxial approximation, as discussed above, is related to the
large parameter $\gamma^2$. However, our ER theory is related to
another parameter, $\delta \ll 1$, which controls the accuracy of
the sharp-edge approximation. It is within the framework of the
sharp-edge approximation that the integrated flux is
logarithmically divergent. Accounting for the presence of the bend
would simply cancel this divergence.

We can now justify findings in the previous Section. From Eq.
(\ref{normI1}) we see that, in the limit for $\phi \ll 1$, the
radiation profile is a universal function, and peaks at
$\hat{\theta} \sim 2.2$. When, instead, $\phi \gtrsim 1$,
radiation is much better collimated peaking at $\hat{\theta} \sim
\sqrt{\phi}$ corresponding to $\theta \sim 1/\gamma$.

\begin{figure}
\begin{center}
\includegraphics*[width=100mm]{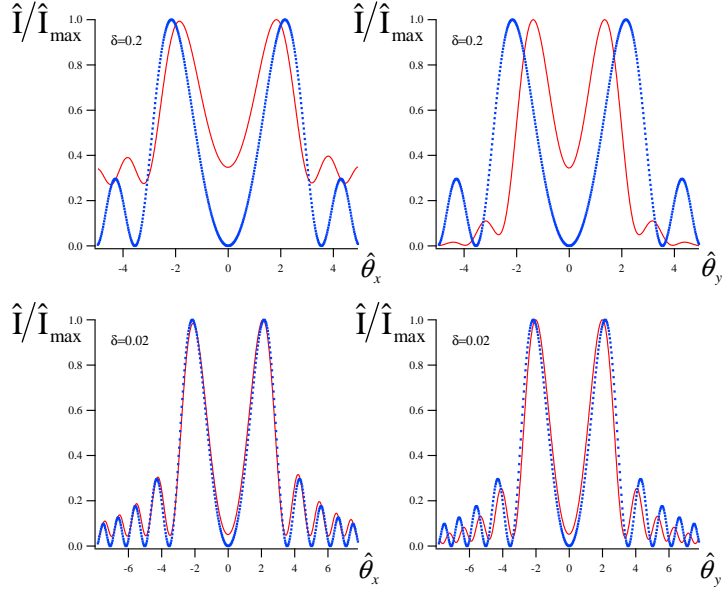}
\caption{Angular spectral flux as a function of the normalized
angle $\hat{\theta}$ for two different edge length parameters
$\delta = 0.2$ and $\delta = 0.02$. Here the straight section
length parameter $\phi \simeq 0.01$. Left and right plots are
obtained cutting the spectral flux profile at $\hat{\theta}_y = 0$
and $\hat \theta_x = 0$ respectively (i.e. electron motion is in
xz plane). The dotted curves are calculated with the analytical
formula Eq. (\ref{normI1}). Solid lines are the results of
numerical calculations with computer code SRW. \label{Diracc} }
\end{center}
\end{figure}
The behavior of the far-field emission described here is
well-known in literature. Nonetheless, the accuracy of the
asymptotic expression for $\delta \ll 1$, Eq. (\ref{normI1}), has
never been discussed: the parameter $\delta$ has been introduced
here for the first time. Numerical calculations were never used
before to scan the parameter space in $\delta$ and to provide an
universal algorithm for estimating the accuracy of the ER theory.
We can study such accuracy now by comparing asymptotical results
with SRW outcomes at different values of $\delta$. This comparison
is illustrated in Fig. \ref{Diracc}. It can be seen that edge
radiation approximation provides good accuracy for $\delta
\lesssim 0.01$.

\begin{figure}
\begin{center}
\includegraphics*[width=100mm]{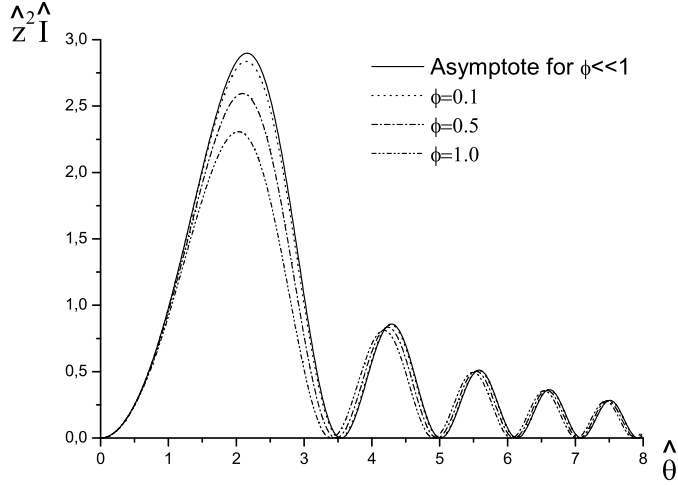}
\caption{Angular spectral flux as a function of the normalized
angle $\hat{\theta}$ for different straight-section length
parameters $\phi$ calculated after Eq. (\ref{normI1}) and
comparison with the asymptotic limit for $\phi \ll 1$ in Eq.
(\ref{normI2}). \label{phiaccu} }
\end{center}
\end{figure}
For completeness, and within the limiting case for $\delta \ll 1$,
it is interesting to study the accuracy of the asymptotic
expression for $\phi \ll 1$ of Eq. (\ref{normI1}). In this case
one does not need comparisons with SRW results, because the
asymptotic limit of Eq. (\ref{normI1}) is simply

\begin{eqnarray}
\hat{z}^2 \hat{I} =  {\hat{\theta}^2} ~
\mathrm{sinc}^2\left[\frac{\hat{\theta}^2}{4 }\right]
~.\label{normI2}
\end{eqnarray}
Results are shown in Fig. \ref{phiaccu}. It can be seen that the
asymptotic expression Eq. (\ref{normI2}) provides good accuracy
for $\phi \lesssim 0.1$.

From now on, we will only consider the asymptote for sharp-edges
$\delta \ll 1$ in the far-zone. This is the starting point for
further investigations of near-zone ER, based on the use of
virtual source techniques.

\section{\label{sub:vire} Near field Edge Radiation theory}

\subsection{\label{subsub:one} Edge Radiation as a field from a
single virtual source}

Eq. (\ref{virfiemody}) and Eq. (\ref{ABcontrint4e}) allow one to
characterize the virtual source through

\begin{eqnarray}
\vec{\widetilde{E}}(0,\vec{r}) =- \frac{\omega^2 e L}{2\pi c^3 }
\int d\vec{\theta} ~\vec{\theta}~ \mathrm{sinc}\left[\frac{\omega
L}{4 c} \left(\theta^2+\frac{1}{\gamma^2}\right)
\right]\exp\left[\frac{i \omega}{c}\vec{r}\cdot \vec{\theta}
\right] ~.\label{vir1e}
\end{eqnarray}
Eq. (\ref{vir1e}) is valid in any range of parameters, i.e. for
any choice of $\phi$. However, the Fourier transform in Eq.
(\ref{vir1e}) is difficult to calculate analytically in full
generality. Simple analytical results can be found in the
asymptotic case for $\phi \ll 1$, i.e. for $L/(\gamma^2
\lambdabar) \ll 1$. In this limit, the right hand side of Eq.
(\ref{vir1e}) can be calculated with the help of polar
coordinates. An analytic expression for the field amplitude at the
virtual source can then be found and reads:

\begin{eqnarray}
\vec{\widetilde{E}}(0,\vec{r}) = - i \frac{4  \omega e}{c^2 L}
~\vec{r} ~\mathrm{sinc}\left(\frac{\omega{r}^2}{c L}\right)
~,\label{vir3e}
\end{eqnarray}
where $r^2 = \left|\vec{r}\right|^2$ as usual. It is useful to
remark, for future use, that similarly to the far-field emission
Eq. (\ref{ABcontrint4e}), also the field in Eq. (\ref{vir3e})
explicitly depend on $L$.

It is interesting to comment on the meaning of the phase in Eq.
(\ref{vir3e}), i.e. on the factor $-i$ in front of the right hand
side. Such phase is linked with the (arbitrary) choice of phase of
the harmonic of the charge density at $z=0$. In particular, such
phase was chosen to be zero at $z=0$. Propagating Eq.
(\ref{vir3e}) to the far-zone, one obtains Eq.
(\ref{ABcontrint4e}). In other words, the plane wavefront
transforms into a spherical wavefront centered at $z=0$. Note that
there is an imaginary unit $i$ in front of Eq.
(\ref{ABcontrint4e}), meaning that an extra minus sign, i.e. a
phase shift of $\pi$, results from the propagation of Eq.
(\ref{vir3e}). This extra phase shift of $\pi$ represent the
analogous of the Guoy phase shift in laser physics, and is in
agreement with our interpretation of the virtual source in Eq.
(\ref{vir3e}) as the waist of a laser-like beam. Note that, while
for azimuthal-symmetric beams the Guoy phase shift is known to be
$\pi/2$, this result is not valid in our case where the cartesian
components of the field depend on the azimuthal angle.

We define the normalized transverse position $\vec{\hat{r}}=
\vec{r}/\sqrt{\lambdabar L}$. Moreover, since the source is
positioned at ${z}=0$, we indicate the normalized spectral flux at
the virtual source as $\hat{I}_s$, defined similarly as $\hat{I}$
in Eq. (\ref{Ihat}). It follows that

\begin{figure}
\begin{center}
\includegraphics*[width=100mm]{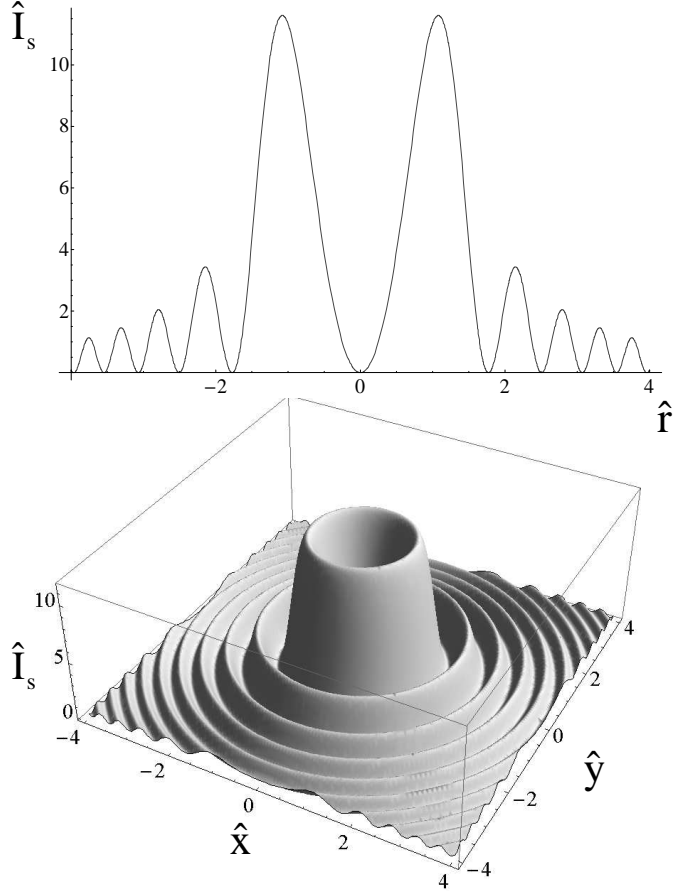}
\caption{\label{viredge}  Normalized spectral flux at the virtual
source, $\hat{I}_s$, as a function of $\hat{r}$ (upper plot) and
3D view as a function of $\hat{x}$ and $\hat{y}$.}
\end{center}
\end{figure}

\begin{eqnarray}
\hat{I}_s(\hat{r}) = 16 ~{\hat{r}}^2
\mathrm{sinc}^2\left({\hat{r}}^2\right) ~.\label{vir4ei}
\end{eqnarray}
The profile in Eq. (\ref{vir4ei}) can be detected (aside for
scaling factors) by imaging the virtual plane with an ideal lens,
and is plotted in Fig. \ref{viredge}.

Note that Eq. (\ref{vir3e}) describes a virtual source
characterized by a plane wavefront. Application of the propagation
formula, Eq. (\ref{fieldpropback}), to Eq. (\ref{vir3e}) allows
one to reconstruct the field both in the near and in the far
region. We obtain the following result:

\begin{eqnarray}
\vec{\widetilde{E}}({z},\vec{\theta}) &&= -\frac{2 e}{z c} \frac{
\vec{\theta}}{\theta^2}\exp\left[\frac{i \omega z \theta^2}{2 c
}\right] \cr&&\times\left[ \exp\left(-\frac{i \omega z {\theta}^2
}{2 c (1+2{z}/L)} \right)-\exp\left(\frac{i \omega z {\theta}^2
}{2 c (-1+2{z}/L)} \right) \right] ~,\label{vir4e}
\end{eqnarray}
where we defined $\vec{\theta}=\vec{r}/z$, \textit{independently}
of the value of $z$. This definition makes sense whenever $z \ne
0$, and yields usual angular distributions in the far zone, for $z
\gg L$. Eq. (\ref{vir4e}) solves the field propagation problem for
both the near and the far field in the limit for $\phi \ll 1$.

Eq. (\ref{vir4e}) is singular at $\vec{r} =0$ (i.e.
$\vec{\theta}=0$) and ${z} = L/2$. Within our sharp-edge
approximation, this singularity is mathematically related to the
divergence of the integrated spectral flux in the far zone, which
has been discussed above. If one goes beyond the applicability
region of the sharp-edge approximation by specifying the nature of
edges and calculating the field within the framework of the
paraxial approximation alone (i.e. with Eq. (\ref{generalfin})),
one sees that the integrated angular spectral flux is not
divergent anymore, and that the field reconstructed at the point
$\vec{r} =0$, ${z} = L/2$ using this far-field data has no
singularity at all.

Note that in the limit for ${z} \gg L$ Eq. (\ref{vir4e})
transforms to

\begin{eqnarray}
\vec{\widetilde{E}}&&=\frac{ i \omega e L}{c^2 z}
\exp\left[\frac{i \omega \theta^2 z}{2 c}\right] \vec{\theta} ~
\mathrm{sinc}\left[\frac{\omega L \theta^2}{4 c}\right]
\label{lllim}
\end{eqnarray}
corresponding to the limit of Eq. (\ref{ABcontrint4e}) for $L\ll
\gamma^2 \lambdabar$, i.e. $\phi \ll 1$.

\begin{figure}
\begin{center}
\includegraphics*[width=140mm]{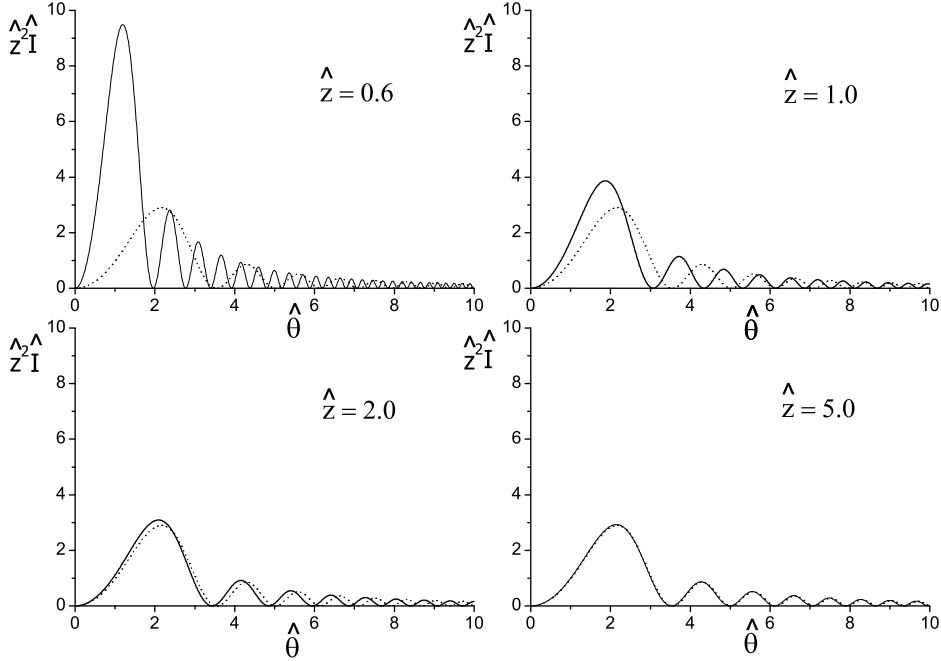}
\caption{\label{Anyd} Evolution of $\hat{z}^2 \hat{I}$ for edge
radiation in the limit for ${\phi} \ll 1$. These profiles,
calculated with Eq. (\ref{vir5e}), are shown as a function of
angles at different observation distances $\hat{z}=0.6$,
$\hat{z}=1.0$, $\hat{z}=2.0$ and $\hat{z}=5.0$ (solid lines). The
dashed line always refers to the far-zone asymptote, Eq.
(\ref{normI2}). }
\end{center}
\end{figure}
The normalized angular spectral flux associated with Eq.
(\ref{vir4e}) is given by

\begin{eqnarray}
\hat{z}^2 {\hat{I}}\left(\hat{z},\hat{\theta}\right) &=&
\frac{4}{\hat{\theta}^2} \left| \left[ \exp\left(-\frac{i
\hat{\theta}^2 \hat{z}}{2 (1+2\hat{z})} \right)-\exp\left(\frac{i
\hat{\theta}^2 \hat{z}}{2 (-1+2\hat{z})} \right) \right]\right|^2
~.\label{vir5e}
\end{eqnarray}
This notation is particularly suited to discuss near and far zone
regions. Eq. (\ref{vir5e}) reduces to Eq. (\ref{normI2}) when
$\hat{z} \gg 1$. To sum up, when $\phi \ll 1$ we have only two
regions of interest with respect to $\hat{z}$:

\begin{itemize}
\item \textbf{Far zone.} In the limit for $\hat{z} \gg 1$ one has the far field case, and Eq. (\ref{vir5e}) simplifies to
Eq. (\ref{normI2}).

\item \textbf{Near zone.}  When $\hat{z} \lesssim 1$
instead, one has the near field case, and Eq. (\ref{vir5e}) must
be used.
\end{itemize}
Of course, it should be stressed that in the case $\hat{z}
\lesssim 1$ we still hold the assumption that the sharp-edge
approximation of is satisfied. It is interesting to study the
evolution of the normalized angular spectral flux for edge
radiation along the longitudinal axis. This gives an idea of how
good the far field approximation ($\hat{z} \gg 1$) is. A
comparison between $\hat{z}^2 \hat{I}$ at different observation
points is plotted in Fig. \ref{Anyd}.

The case $\phi \ll 1$ studied until now corresponds to a short
straight section, in the sense that $L \ll \gamma^2 \lambdabar$.
When this condition is not satisfied, we find that the integral on
the right hand side of Eq. (\ref{vir1e}) is difficult to calculate
analytically. Thus, the single-source method used here is
advantageous in the case $\phi \ll 1$ only. However, one can use
numerical techniques to discuss the case for any value of $\phi$.
With the help of polar coordinates, the right hand side of Eq.
(\ref{vir1e}) can be transformed in a one-dimensional integral,
namely

\begin{figure}
\begin{center}
\includegraphics*[width=100mm]{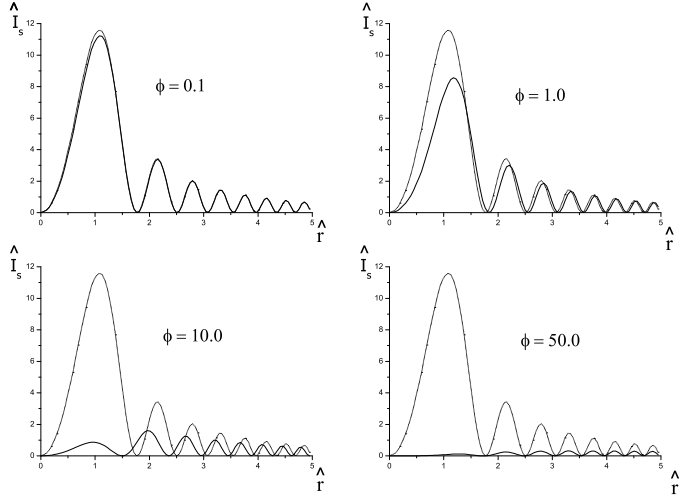}
\caption{\label{Intvirphi} Normalized spectral flux at the virtual
source for the setup in Fig. \ref{geome}(a). These profiles are
shown for $\phi=0.1$, $\phi=1.0$, $\phi=10.0$ and $\phi=50.0$
(solid lines). Solid curves are calculated with the help of Eq.
(\ref{vir2e1d}). The dotted lines show comparison with the
asymptotic limit for $\phi\ll 1$, shown in Fig. \ref{viredge} and
calculated using Eq. (\ref{vir4ei}).}
\end{center}
\end{figure}

\begin{figure}
\begin{center}
\includegraphics*[width=100mm]{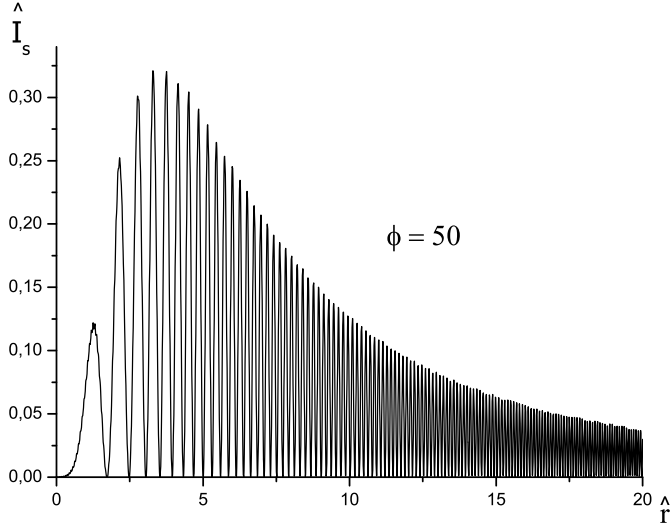}
\caption{\label{phil} Normalized spectral flux at the virtual
source for the setup in Fig. \ref{geome}(a) for $\phi=50$
(enlargement of the bottom right graph in Fig. \ref{Intvirphi}).}
\end{center}
\end{figure}

\begin{eqnarray}
\vec{\widetilde{E}}(0,\vec{r}) = - \frac{4 i \omega e} {c^2}
\frac{\vec{r}}{r} \int_0^\infty
\frac{{\theta}^2}{{\theta}^2+1/\gamma^2} \sin\left[\frac{\omega
L}{4 c} \left(\theta^2 + \frac{1}{\gamma^2}\right)\right]
J_1\left(\frac{\omega{\theta}{r}}{c}\right)
d{\theta}~,\label{vir2e1d}
\end{eqnarray}
yielding

\begin{eqnarray}
\hat{I}_s(\hat{r}) = \left|\int_0^\infty
\frac{4\hat{\theta}^2}{\hat{\theta}^2+\phi} \sin\left[
\frac{\hat{\theta}^2 + \phi}{4}\right]
J_1\left({\hat{\theta}\hat{r}}\right)
d\hat{\theta}\right|^2~.\label{virIfinal}
\end{eqnarray}
We calculated the spectral flux associated with the virtual source
for values $\phi=0.1$, $\phi=1$, $\phi=10$ and $\phi=50$, the same
values chosen for Fig. \ref{Dirfar2}. We plot these distributions
in Fig. \ref{Intvirphi}. It is also instructive to make a
separate, enlarged plot of the case $\phi = 50$, that is in the
asymptotic case for $\phi \gg 1$. This is given in Fig.
\ref{phil}. Fine structures are now evident, and are consistent
with the presence of fine structures in Fig. \ref{Dirfar2} for the
far zone. 

We managed to specify the field at the virtual source by means of
numerical techniques, even in the case $\phi \gg 1$ (see Fig.
\ref{phil}). Once the field at the virtual source is specified for
any value of $\phi$, Fourier Optics can be used to propagate it.
However, we prefer to proceed following another path. There is, in
fact, an alternative way to obtain the solution to the field
propagation problem valid for any value of $\phi$, and capable of
giving a better physical insight for large values of $\phi$.

\subsection{\label{sub:singles} Edge Radiation as a superposition of the field from
two virtual sources}

Consider the far field in Eq. (\ref{ABcontrint4e}). This can also
be written as

\begin{eqnarray}
\vec{\widetilde{E}}\left({z},\vec{\theta}\right) =
\vec{\widetilde{E}}_1\left({z},\vec{\theta}\right)
+\vec{\widetilde{E}}_2\left({z},\vec{\theta}\right) ~
\label{Efarsum2a}~,
\end{eqnarray}
where

\begin{eqnarray}
\vec{\widetilde{E}}_{1,2}\left({z},\vec{\theta}\right) = \mp
\frac{2 e\vec{{\theta}}}{c {z} (\theta^2 + 1/\gamma^2)}
\exp\left[\mp\frac{i \omega L}{4 c \gamma^2}\right]
\exp\left[\frac{i \omega L \theta^2}{2 c
}\left(\frac{z}{L}\mp\frac{1}{2}\right)\right] . \label{Efarsum2}
\end{eqnarray}
When $z \gg L$, the two terms $\vec{\widetilde{E}}_{1}$ and
$\vec{\widetilde{E}}_{2}$ represent two spherical waves
respectively centered at ${z}_{s1} = - L/2$ and ${z}_{s2} = L/2$,
that is at the edges between straight section and
bends\footnote{At first glance this statement looks
counterintuitive. In order to find where the spherical wave is
centered, one needs to know where the phase becomes zero. Now,
when $z=L/2$, the phase in $\theta^2$ for
$\vec{\widetilde{E}}_{2}$ in Eq. (\ref{Efarsum2}) is not zero,
hinting to the fact that the spherical wave
$\vec{\widetilde{E}}_{2}$  is not centered at $z=L/2$. This last
observation, however, is misleading. In fact, one should account
for the fact that the definition $\vec{\theta} = \vec{r}/z$ is not
natural for $\vec{\widetilde{E}}_{1}$ and
$\vec{\widetilde{E}}_{2}$. In fact, according to it,
$\vec{\theta}$ is measured from the center of the straight
section, while it should be measured from $z=-L/2$ for
$\vec{\widetilde{E}}_{1}$ and from $z=L/2$ for
$\vec{\widetilde{E}}_{2}$. Also note that Eq. (\ref{Efarsum2}) is
only valid in the limit $z \gg L$.}. Analysis of Eq.
(\ref{Efarsum2}) shows that both contributions to the total field
are peaked at an angle of order $1/\gamma$. While the amplitude of
the total field explicitly depends on $L$, the two expressions
$\vec{\widetilde{E}}_{1}$ and $\vec{\widetilde{E}}_{2}$ exhibit
dependence on $L$ through phase factors only. This fact will have
interesting consequences, as we will discuss later. The two
spherical waves represented by $\vec{\widetilde{E}}_{1}$ and
$\vec{\widetilde{E}}_{2}$ may be thought as originating from two
separate virtual sources located at the edges between straight
section and bends. One may then model the system with the help of
two separate virtual sources, and interpret the field at any
distance as the superposition of the contributions from two edges.
It should be clear that contributions from these edges are linked
with an integral of the trajectory followed by the electron. Thus
the word "edge" should be considered as a synonym for "virtual
source" here.

Let us specify analytically the two virtual sources at the edges
of the setup. To this purpose, we take advantage of Eq.
(\ref{virfiemody}) with $z_s = {z}_{s(1,2)}$, separately
substituting $\vec{\widetilde{E}}_{1}$ and
$\vec{\widetilde{E}}_{2}$ and using polar coordinates. We find the
following expressions for the field at the virtual source
positions ${z}_{s1} = - L/2$ and ${z}_{s2}=L/2$:

\begin{eqnarray}
\vec{\widetilde{E}}_{s1,s2}\left(\mp \frac{L}{2},\vec{r}\right) =
\pm \frac{2 \omega e }{c^2 \gamma} \exp\left[\mp \frac{i \omega L
}{4\gamma^2 c}\right] \frac{\vec{r}}{r} K_1\left(\frac{\omega r}{c
\gamma} \right)~,\label{virpm05}
\end{eqnarray}
where $K_1(\cdot)$ is the modified Bessel function of the first
order. Analysis of Eq. (\ref{virpm05}) shows a typical scale
related to the source dimensions of order $\lambdabar \gamma$ in
dimensional units, corresponding to $1/\sqrt{\phi}$ in normalized
units. Also, the fact that the field in the far zone, Eq.
(\ref{ABcontrint4e}), exhibits dependence on $L$ only through
phase factors implies that the field at the virtual sources, Eq.
(\ref{virpm05}), exhibits dependence on $L$ only through phase
factors (and viceversa).

Application of the propagation formula Eq. (\ref{fieldpropback})
allows to calculate the field at any distance ${z}$ in free-space.
Of course, Eq. (\ref{virpm05}) can also be used as input to any
Fourier code to calculate the field evolution in the presence of
whatever optical beamline. However, here we restrict ourselves to
the free-space case. In order to simplify the presentation of the
electric field we take advantage of polar coordinates and we use
the definition $\vec{\hat{E}} \equiv \vec{\widetilde{E}}
\sqrt{\lambdabar L} ~c/e$ (so that $\hat{I}$, introduced in Eq.
(\ref{Ihat}), is given by $\hat{I} = |\vec{\hat{E}}|^2$) for the
field in normalized units. We obtain:

\begin{eqnarray}
\hat{z} \vec{\hat{E}}\left(\hat{z},\vec{\hat{\theta}}\right) &=&
\left. \left\{\frac{\vec{\hat{\theta}}}{\hat{\theta}} \frac{2
\hat{z} \sqrt{\phi} \exp\left[i\phi/4\right] }{\hat{z}-1/2}
\exp\left[\frac{i\hat{\theta}^2 \hat{z}^2
}{2\left(\hat{z}-1/2\right)}\right]\right.\right.\cr &&
\left.\left. \times \int_0^{\infty} d\hat{r}' \hat{r}'
K_1\left(\sqrt{\phi}\hat{r}'\right) J_1
\left(\frac{\hat{\theta}\hat{r}' \hat{z}}{\hat{z}-1/2}\right)
\exp\left[\frac{i\hat{r}^{'2} }{2\left(\hat{z}-1/2\right)}\right]
\right\} \right.\cr &&\left. -
\left\{\frac{\vec{\hat{\theta}}}{\hat{\theta}} \frac{2 \hat{z}
\sqrt{\phi} \exp\left[-i\phi/4\right] }{\hat{z}+1/2}
\exp\left[\frac{i\hat{\theta}^2 \hat{z}^2
}{2\left(\hat{z}+1/2\right)}\right]\right.\right.\cr &&
\left.\left. \times \int_0^{\infty} d\hat{r}' \hat{r}'
K_1\left(\sqrt{\phi}\hat{r}'\right) J_1
\left(\frac{\hat{\theta}\hat{r}' \hat{z}}{\hat{z}+1/2}\right)
\exp\left[\frac{i\hat{r}^{'2} }{2\left(\hat{z}+1/2\right)}\right]
\right\} \right.~.\label{fieldtot}
\end{eqnarray}
In the limit for $\hat{z} \gg 1$, using Eq. (\ref{fieldtot}) and
recalling $\int_0^\infty d\hat{r}' ~ \hat{r}' K_1(\sqrt{\phi}
\hat{r}') J_1(\hat{\theta}\hat{r}') = \hat{\theta}/[\sqrt{\phi}
(\hat{\theta}^2 +\phi)]$ we obtain back Eq. (\ref{normI1}).
Similarly, in the limit for $\phi \ll 1$, and using the fact that
$K_1(\sqrt{\phi} \hat{r}) \simeq 1/(\hat{r} \sqrt{\phi})$ one
recovers Eq. (\ref{vir4e}). In general, the integrals in Eq.
(\ref{fieldtot}) cannot be calculated analytically, but they can
be integrated numerically.


\subsection{\label{sub:classi} Classification of regions of observation}

Qualitatively, we can deal with two limiting cases of the theory,
the first for $\phi \ll 1$ and the second for $\phi \gg 1$. As for
the case of a single virtual source, there are no constraints, in
principle, on the value of $\phi$. However, as we will see, the
two-source method gives peculiar advantages in the case $\phi \gg
1$, while, has we have seen before, the case $\phi \ll 1$ is
better treated in the framework of a single source.

\subsubsection{Case $\phi \ll 1$}

Let us briefly discuss the case ${\phi} \ll 1$ in the framework of
the two-source method. In this case, one obviously obtains back
Eq. (\ref{vir4e}).  An alternative derivation has been shown in
Section \ref{subsub:one}. As one can see Eq. (\ref{vir4e}) is
independent of ${\phi}$. In Fig. \ref{Anyd} we plotted results for
the propagation according to Eq. (\ref{vir5e}). Radiation profiles
are shown as a function of angles $\hat{\theta}$ at different
observation distances $\hat{z}=0.6$, $\hat{z}=1.0$, $\hat{z}=2.0$
and $\hat{z}=5.0$. As discussed before, one can recognize two
observation zones of interest: the near and the far zone. As it
can be seen from Eq. (\ref{vir4e}), the total field is given, both
in the near and in the far zone, by the interference of the
virtual source contributions. The virtual sources themselves are
located at the straight section edges. Eq. (\ref{fieldtot}) shows
that the transverse dimension of these virtual sources is given by
$\gamma \lambdabar$ in dimensional units. This is the typical
scale in $r'$ after which the integrands in $d \hat{r}'$ in Eq.
(\ref{fieldtot}) are suppressed by the function $K_1$. Thus, the
sources at the edges of the straight section have a dimension that
is independent of $L$. In the center of the setup instead, the
virtual source has a dimension of order $\sqrt{\lambdabar L}$ as
it can be seen Eq. (\ref{vir4ei}). When $\phi\ll 1$ the source in
the center of the setup is much smaller than those at the edge.
This looks paradoxical. The explanation is that the two
contributions due to edge sources interfere in the center of the
setup. In particular, when $\phi\ll 1$ they nearly compensate, as
they have opposite sign. As a result of this interference, the
single virtual source in the center of the setup (and its far-zone
counterpart) has a dimension dependent on $L$ (in non-normalized
units)  while for two virtual sources at the edges (and in their
far-zone counterpart) the dependence on $L$ is limited to phase
factors only. Due to the fact that edges contributions nearly
compensate for $\phi\ll 1$ one may say that the single-source
picture is particularly natural in the case $\phi \ll 1$.

\subsubsection{Case $\phi \gg 1$}

Let us now discuss the case ${\phi} \gg 1$. In this situation the
two-sources picture becomes more natural.  We indicate with
${d}_{1,2} = {z} \pm L/2$ the distances of the observer from the
edges. From Eq. (\ref{ABcontre}) we know that when ${\phi} \gg 1$
the formation length is $L_f = \gamma^2 \lambdabar$, much shorter
than the system dimension $L$. As a result, one can recognize four
regions of observation of interest.

\begin{figure}
\begin{center}
\includegraphics*[width=140mm]{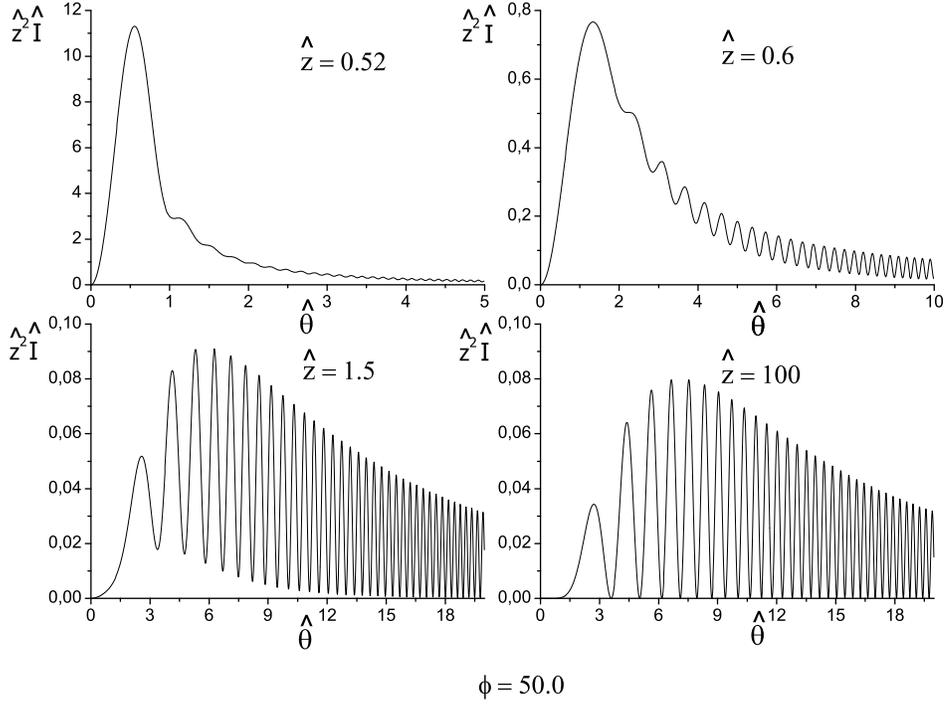}
\caption{\label{evolve} $\hat{z}^2 \hat{I}$ at $\phi=50$. These
profiles are shown as a function of angles at different
observation distances $\hat{z}=0.52$, $\hat{z}=0.6$, $\hat{z}=1.5$
and $\hat{z}=100.0$.}
\end{center}
\end{figure}

In Fig. \ref{evolve} we plotted, in particular, results for the
propagation in case $\phi=50$. In this case, for arbitrary
$\hat{z}$, integrals in Eq. (\ref{fieldtot}) cannot be calculated
analytically, but they can be integrated numerically.  Radiation
profiles are shown as a function of angles $\hat{\theta}$ at
different observation distances $\hat{z}=0.52$, $\hat{z}=0.6$,
$\hat{z}=1.5$ and $\hat{z}=100.0$.

\begin{itemize}
\item \textbf{Two-edge radiation, far zone: $d_{1,2} \gg L$ (i.e. $z \gg L$). Eq.
(\ref{ABcontrint4e}) and Eq. (\ref{enden2e}) should be used.} When
$d_{1,2} \gg L$ we are summing far field contributions from the
two edge sources. This case is well represented in Fig.
\ref{evolve} for $\hat{z} = 100$, where interference effects
between the two edges contribution are clearly visible.
\item \textbf{Two-edge radiation, near zone: $d_{1,2} \sim L$. Eq. (\ref{fieldtot}) should be used.}  When $d_{1,2} \sim L$ the observer is located far
away with respect to the formation length of the sources. Both
contributions from the sources are important, but that from the
nearest source begins to become the main one, as $d_1$ and $d_2$
become sensibly different. This case is well represented in Fig.
\ref{evolve} for $\hat{z} =1.5$.
\item \textbf{Single-edge radiation, far zone: $\gamma^2\lambdabar \ll d_2 \ll L$ and $r\ll L/\gamma$. Eq. (\ref{Efarsum3}) should be used.} When
$\gamma^2\lambdabar \ll d_2 \ll L$, the contribution due to the
near edge becomes more and more important. Such tendency is
clearly depicted in Fig. \ref{evolve} for $\hat{z} =0.6$.
Interference tends to disappear as the near edge becomes the
dominant one. In this case, one finds that the electric field in
Eq. (\ref{fieldtot}) reduces to

\begin{eqnarray}
\vec{\widetilde{E}}\left({z},\vec{\xi}\right) =  \frac{2 e
\gamma^2\vec{{\xi}}}{c {(z-L/2)} (\gamma^2 \xi^2 + 1)}
\exp\left[\frac{i \omega L}{4 c \gamma^2}\right] \exp\left[\frac{i
\omega L \xi^2}{2 c }\left(\frac{z}{L}-\frac{1}{2}\right)\right]
~, \label{Efarsum3}
\end{eqnarray}
where $\xi=r/(z-L/2)=r/d_2$. Note that $\xi$ is used here in place
of $\theta$, because by definition $\theta = r/z$, where $z$ is
calculated from the center of the straight section, whereas the
definition of $\xi$ is related to the edge position at $z=L/2$. It
should be remarked that Eq. (\ref{Efarsum3}) constitutes the field
contribution from the downstream edge of the straight section, and
that the contribution from the upstream edge (at $z=-L/2$) can be
found from Eq. (\ref{Efarsum3}) by performing everywhere in Eq.
(\ref{Efarsum3}), i.e. also in $\xi$, the substitution
$L/2\longrightarrow -L/2$, and by changing an overall sign. Eq.
(\ref{Efarsum3}) corresponds to an angular spectral flux
\cite{CHUS}

\begin{eqnarray}
\frac{d W}{d\omega d \Omega}=  \frac{e^2 }{c \pi^2} \frac{\gamma^4
{\xi}^2} {\left( \gamma^2 \xi^2+1\right)^2} ~.\label{Isefar0}
\end{eqnarray}
It is important to specify the region of applicability of Eq.
(\ref{Efarsum3}) in the transverse direction. For a single edge in
the far zone, the amplitude of the field decreases as $r^{-1}$, as
can be checked by substituting the definition of $\vec{\xi}$ in
Eq. (\ref{Efarsum3}), and does not depend on  $z$ nor $\gamma$ for
angles of observation larger than $1/\gamma$. Since $d_2 \ll L$,
such dependence holds for the upstream edge at $r \gtrsim
L/\gamma$, and for the downstream edge at $r \sim d_2/\gamma \ll
L/\gamma$. As a result, for $r \gtrsim L/\gamma$, the
contributions from the two edges are comparable, and a single-edge
asymptote cannot be used. It follows that Eq. (\ref{Efarsum3})
applies for $r \ll L/\gamma$.

\item \textbf{Single-edge radiation, near zone: $\sqrt[3]{\lambdabar R^2} \ll d_2 \lesssim \gamma^2
\lambdabar$, $r \ll \gamma \lambdabar$. Eq. (\ref{Isefar}) should
be used.} When $\sqrt[3]{\lambdabar R^2} \ll d_2 \lesssim \gamma^2
\lambdabar$ we have the contribution from a single edge in the
near zone.

As ${d}_2$ becomes smaller and smaller the maximum in the
radiation profile increases (see Fig. \ref{evolve}). This behavior
is to be expected. In fact, on the one hand the virtual source
exhibits a singular behavior at $r = 0$, while on the other hand
the integral in Eq. (\ref{fieldpropback}) must reproduce the
virtual source for $z \longrightarrow z_s$. In other words, for $z
\longrightarrow z_s$, the propagator must behave like a Dirac
$\delta$-distribution. However, the way such asymptote is realized
is not trivial. At any finite distance $d_2$ from the source, Eq.
(\ref{fieldpropback}) eliminates the singularity of the Bessel
$K_1$ function. This means that the maximum value in
$|\vec{\widetilde{E}}|^2$ increases as $d_2$ decreases, but it
always remains finite. In particular, at ${\theta}=0$,
$|\vec{\widetilde{E}}|^2=0$. However, by conservation of energy
the integral of $|\vec{\widetilde{E}}|^2=0$ over transverse
coordinate must diverge at any finite distance from the source,
because the field diverges at the source
position. 

Note that when $r \ll \gamma \lambdabar$ and $d_2 \ll \gamma^2
\lambdabar$, the integral pertaining the near (downstream) edge
(at $z_{s2}=L/2$) in Eq. (\ref{fieldtot}) can be calculated
analytically. In fact, the Bessel $K_1$ function in the integrand
can be expanded for small values of the argument when $\hat{r}'
\ll 1/\sqrt{\phi}$. When this is not the case ($\hat{r}' \gtrsim
1/\sqrt{\phi}$) the phase factor under the integral sign makes the
integrand exhibiting oscillatory behavior (because $\hat{z}-1/2
\ll 1/\phi$, since $d_2 \theta \ll \gamma^2 \lambdabar$).
Contributions to the integrals are therefore negligible. As a
result, in this case one can use the expansion $K_1(\sqrt{\phi}
\hat{r}') \sim 1/(\sqrt{\phi} \hat{r}')$. Then, using the fact
that $\int_0^\infty d x J_1(A x) \exp[i B r^2] = 1/A \{1-\exp[-i
A^2/(4B)]\}$ (for $A$ and $B$ positive), one obtains \cite{CHUN}

\begin{eqnarray}
\vec{\widetilde{E}}\left(z,\vec{\xi}\right) &=& \frac{2 i e
\vec{\xi}}{c \xi^2 (z-L/2)} \exp\left[\frac{i \omega L}{4 c
\gamma^2 }\right] \exp\left[\frac{i \omega \xi^2 (z-L/2)}{4
c}\right] \cr && \times \sin\left[\frac{\omega \xi^2(z-L/2)}{4
c}\right]~.\label{Isefar}
\end{eqnarray}
Note that while the modulus of Eq. (\ref{Isefar}) is independent
of ${\phi}$, its region of applicability is related to ${\phi}$
and the asymptotic expression deviates from Eq. (\ref{fieldtot})
for smaller value of $\hat{\theta}$ when ${\phi}$ is larger. In
fact, Eq. (\ref{Isefar}) is valid only when $\hat{z}-1/2 \ll
1/\phi$ and $\hat{z} \hat{\theta} \ll 1/\sqrt{\phi}$ (i.e. $r \ll
\gamma\lambdabar$ and $d_2 \ll \gamma^2 \lambdabar$).

It is interesting to remark here that Eq. (\ref{vir4e}), which was
derived for $\phi \ll 1$, reduces to Eq. (\ref{Isefar}) when $d_2
\ll L$ and $r \ll \sqrt{\lambdabar L}$. It follows that the
validity of Eq. (\ref{Isefar}) has a wider region of applicability
than that considered here. In fact, it may be applied whenever
$\sqrt[3]{\lambdabar R^2} \ll d_2 \ll \min(L,\gamma^2 \lambdabar)$
and $r \ll \sqrt{\lambdabar \min(L,\gamma^2 \lambdabar)}$.

If we propagate  Eq. (\ref{Isefar}) to the far zone we obtain an
asymptote which is valid only for angles much larger than
$1/\gamma$ (i.e. Eq. (\ref{Isefar}) is an asymptote for high
values of spatial frequencies). The modulus of Eq. (\ref{Isefar})
does not depend on $\gamma$ (while in the non-asymptotic case
radiation for any value of $z$ must depend on $\gamma$, because
the far-field radiation from a single edge depends on $\gamma$
too) nor it includes information about distribution in the far
zone within angles comparable with $1/\gamma$. Thus, the
applicability of this high-spatial frequency asymptote depends on
what practical (or theoretical) problem we try to solve. It is
useful, for example, if we discuss about a sample in the very near
zone. However, if we discuss about design of beam line with an
acceptance angle comparable with $1/\gamma$ (which is equivalent
to some spatial-frequency filter) the asymptotic expression in Eq.
(\ref{Isefar}) cannot be applied anymore, and one should use exact
results from the propagation integral, i.e. the near-field
expression Eq. (\ref{fieldtot}).

\end{itemize}

The above-given classification in zones of interest with
asymptotical expressions for the electric field constitutes an
important result of our paper. In fact, expressions for the
electric field without explicit specification of their region of
applicability are incomplete, and have no practical nor
theoretical utility. From this viewpoint, it is interesting to
compare our results with literature. We will limit our discussion
to a comparison with recent review \cite{WILL}, which summarizes
up-to-date understanding of ER within the SR community.

One result in \cite{WILL} (Eq. (26)) corresponds to the square
modulus of our Eq. (\ref{Isefar}), the single edge near-zone case.
The region of applicability specified in \cite{WILL} for such
result\footnote{Converted to our notation.} is $\lambdabar \ll d_2
\lesssim \gamma^2 \lambdabar$ and $L \longrightarrow \infty$. A
first problem in applying this prescription is intrinsic in the
condition $L \longrightarrow \infty$, as there is no comparison of
$L$ to any other characteristic length. Secondly, this result is
independent of $\gamma$ and $L$. As  a result, it cannot be valid
for arbitrary transverse distance $r$. In fact, since the far-zone
field depends on both $\gamma$ and $L$, should we propagate Eq.
(\ref{Isefar}) in the far zone, we could never obtain an outcome
dependent on $\gamma$ and $L$. It follows that, as we discussed
above, Eq. (\ref{Isefar}) is valid for arbitrary $L$ (under the
sharp-edge limit $\sqrt[3]{\lambdabar R^2} \ll L$), but its region
of applicability depends on $L$ or $\gamma$: $\sqrt[3]{\lambdabar
R^2} \ll d_2 \ll \min(L,\gamma^2 \lambdabar)$ and $r \ll
\sqrt{\lambdabar \min(L,\gamma^2 \lambdabar)}$. Note that the
requirement $r \ll \lambdabar \gamma$ for the applicability of Eq.
(\ref{Isefar}) is present in the original paper \cite{CHUN}, but
has been omitted in some later publications, e.g.
\cite{BOS4,BOS7,BOSS}. Because of this the dependence of near-zone
single-edge radiation upon $\gamma$ was unclear, and the
asymptotic expression Eq. (\ref{Isefar}) started to be considered
by other authors, as in \cite{WILL}, without proper requirements
on $r$.

A second result in \cite{WILL} for the case of a finite straight
section length $L$ (Eq. (27)) corresponds to our Eq.
(\ref{vir5e}). This fact can be proved with the help of
straightforward mathematical steps. The region of applicability of
Eq. (27) in \cite{WILL} is specified by the words "under
conditions of validity of equation (26)", i.e. $\lambdabar \ll d_2
\lesssim \gamma^2 \lambdabar$. In contrast to this, we have seen
that Eq. (\ref{vir5e}) is valid for arbitrary $d_2 \gg
\sqrt[3]{\lambdabar R^2}$, but includes limitations on $L$ in the
form: $\sqrt[3]{\lambdabar R^2} \ll L \ll \gamma^2 \lambdabar$.

The following result in \cite{WILL}, Eq. (28), is the far-zone
single-edge result, i.e. our Eq. (\ref{Isefar0}). Eq. (28) in
\cite{WILL} is presented without region of applicability while, as
we have seen, it is valid for $\gamma^2\lambdabar \ll d_2 \ll L$
and $r\ll L/\gamma$. Note that in this case the denomination "far
zone" is not related with the usual understanding $d_2
\longrightarrow \infty$, as is the case for the usual far-zone
expression for two-edges, but it includes a limit on $d_2$, i.e.
$d_2 \ll L$.

The final result in \cite{WILL} is Eq. (29), that is our far-zone
two-edge case, Eq. (\ref{enden2e}). Also Eq. (29) is presented
without region of applicability while, as we have seen,  $d_2 \gg
L$.

It should be appreciated how our analysis of ER through the
parameter $\delta$ allowed us to define "how sharp" the edges are,
and to specify the region of applicability for ER theory.

As a final remark, note that in literature the single edge
far-zone case is usually presented as the simplest and fundamental
case, while in our view this is the most complicated and
misleading case to discuss. For the sake of exemplification,
consider the usual assumption made in this case, i.e. $L
\longrightarrow \infty$. In order to discuss the far-zone
expression, one needs $z \longrightarrow \infty$, and comparing
$z$ to $L$ becomes impossible. In other words, $L$ drops out of
from the problem parameters and, as in \cite{WILL}, it never
appears in the condition for the region of applicability anymore.
In contrast to this, our simplest model is the two-edge far-zone
model, whose region of applicability is: $z \gg L$ (or $d_2 \gg
L$). It is independent of $\gamma$ and it is much easier (although
an extra-limitation on angles $\theta \ll \sqrt[3]{\lambdabar/R}$
should be included, due to the sharp-edge approximation). After
introduction of this model, the following natural step was to
generalize it to the near zone, introducing more complicated
regions of applicability discussed before.

\section{\label{sec:five}  Transition undulator radiation}

\begin{figure}
\begin{center}
\includegraphics*[width=100mm]{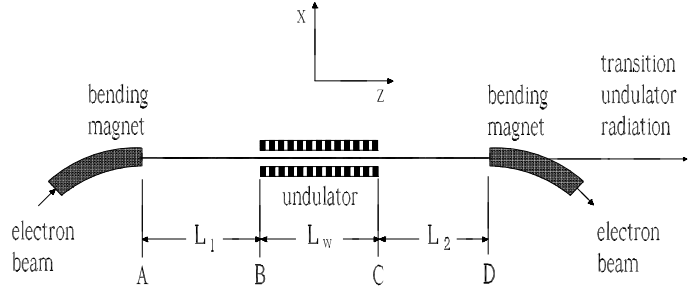}
\caption{\label{geomu} Transition undulator radiation geometry.}
\end{center}
\end{figure}
In this Section we apply the method of virtual sources to the more
complicated case of an undulator setup.

Instead of the setup in Fig. \ref{geome}(a), we now consider the
system depicted in Fig. \ref{geomu} and we consider a single
particle moving along the system. The electron enters the setup
via a bending magnet, passes through a straight section (segment
$AB$), an undulator (segment $BC$), and another straight section
(segment $CD$). Finally, it leaves the setup via another bend.
Radiation is collected at a distance $z$ from the center of the
reference system, located in the middle of the undulator. The
passage of the electron through the setup results in collimated
emission of radiation in the range $\lambdabar \gg \lambdabar_r$
and $\lambdabar \gg \lambdabar_c$, where $\lambdabar_r$ is the
resonance wavelength of the fundamental harmonic of the undulator,
i.e. the extra characteristic length introduced in the setup. This
kind of radiation is known in literature as Transition Undulator
Radiation (TUR) \cite{KIM1,KINC,BOS4,CAST,ROY2}. We will retain
this name although, as we will see, what we are really discussing
about is edge radiation from an undulator setup.

In our case of study the trajectory and, therefore, the space
integration in Eq. (\ref{generalfin}) can be split in five parts:
the two bends, which will be indicated with $b_1$ and $b_2$, the
two straight sections $AB$ and $CD$ and the undulator $BC$. One
may write

\begin{eqnarray}
\vec{\widetilde{{E}}}(z, \vec{r}) &=&
\vec{\widetilde{{E}}}_{b1}(z, \vec{r})+
\vec{\widetilde{{E}}}_{AB}(z, \vec{r})+
\vec{\widetilde{{E}}}_{BC}(z, \vec{r})+
\vec{\widetilde{{E}}}_{CD}(z, \vec{r})+
\vec{\widetilde{{E}}}_{b2}(z, \vec{r})~,\cr && \label{split}
\end{eqnarray}
with obvious meaning of notation.

We will denote the length of the segment $AD$ with
$L_\mathrm{tot}$, while we will indicate the length of the
straight section $AB$ with $L_1$, the length of the straight
section $CD$ with $L_2$ and the length of the undulator with
$L_w$. It follows that $L_\mathrm{tot} = L_1+L_w+L_2$. This means
that point $A$ is located at longitudinal coordinate $z_A= -L_1 -
L_w/2$, while $B$, $C$ and $D$ are located respectively at $z_B=
-L_w/2$, $z_C = L_w/2$ and $z_D =  L_w/2 +L_2$.

We will describe the field from our TUR setup as a superposition
of three laser-like beam from straight sections and undulator. As
before, with the help of Eq. (\ref{generalfin}) we will first
derive an expression for the field in the far zone. Then we will
calculate the field distribution at the virtual source with the
help of Eq. (\ref{virfiemody}). Finally, Eq.
(\ref{fieldpropbackx}) will allow us to find an expression for the
field both in the near
and in the far zone.  
%

\subsection{\label{sub:far} Far field from the undulator setup}

Let us describe the far field from the undulator setup in Fig.
\ref{geomu} by separately characterizing different field
contributions and finally adding them together.

\subsubsection{\label{sub:undu} Field contribution calculated along
the undulator}

We first consider the contribution $\vec{\widetilde{{E}}}_{BC}$
from the undulator. Assuming a planar undulator with $N_w$ periods
we write the following expression for the transverse velocity of
an electron:

\begin{equation}
\vec{v}_\bot(z) = - {c K\over{\gamma}} \sin{\left(k_w z\right)}
\vec{x}~. \label{vuz}
\end{equation}
Here $K=(\lambda_w e H_w) / (2 \pi m_\mathrm{e} c^2)$ is the
undulator parameter, $m_\mathrm{e}$ being the electron mass and
$H_w$ being the maximum of the magnetic field produced by the
undulator on the $z$ axis. Moreover, $k_w=2\pi/\lambda_w$, where
$\lambda_w$ is the undulator period, so that the undulator length
is $L_w = N_w \lambda_w$. The transverse position of the electron
is

\begin{equation}
\vec{r}_0(z) = \frac{K}{\gamma k_w} \cos{\left(k_w z\right)}
\vec{x}~. \label{erz}
\end{equation}
We can now substitute Eq. (\ref{erz}) and Eq. (\ref{vuz}) in Eq.
(\ref{generalfin}). Such substitution leads to an expression valid
in the far zone. We obtain

\begin{eqnarray}
\vec{\widetilde{{E}}}_{BC}(z, \vec{r},\omega) &=& \frac{i \omega
e}{c^2 z} \int_{z_B}^{z_C} dz' {\exp{\left[i \Phi_{BC}\right]}}
\left\{\left[{K\over{\gamma}} \sin\left(k_w
z'\right)+\theta_x\right]{\vec{x}} +\theta_y{\vec{y}}\right\}~.
\label{unduraddd}
\end{eqnarray}
Here

\begin{eqnarray}
\Phi_{BC} &=& \frac{\omega}{c} \left\{{\theta^2\over{2}} z+
\frac{z'}{2}\left(\frac{1}{\bar{\gamma}_z^2}+\theta^2\right) -
\frac{K\theta_x}{\gamma k_w }\cos(k_w z') - \frac{K^2}{8\gamma^2
k_w } \sin(2 k_w z') \right\} ~,\label{phitundu}
\end{eqnarray}
where the average longitudinal Lorentz factor $\bar{\gamma}_z$ in
Eq. (\ref{phitundu}) turns out to be $\bar{\gamma}_z =
{\gamma}/{\sqrt{1+{K^2}/{2}}}$, and is always smaller than
$\gamma$ because the average longitudinal velocity of the electron
inside the undulator is smaller than that along the straight
sections.

In this paper we will be interested up to frequencies much lower
than the resonance frequency, i.e. $\lambdabar \gg \lambdabar_r$,
with $\lambdabar_r = 1/(2 \bar{\gamma}_z^2 k_w)$.

We can show that this condition is analogous, for TUR radiation,
to condition $\phi \cdot \delta \ll 1$ for the simple edge
radiation setup in Fig. \ref{geome}(a). In order to do so, we
first need to discuss the formation length associated with the
undulator, i.e. with Eq. (\ref{unduraddd}). The definition of
formation length was introduced before as the value of $z_2-z_1$
for which the right hand side of Eq. (\ref{moregen}) is of order
unity. However, the physical meaning of formation length is
related with the integration range in $z'$ such that the integrand
in Eq. (\ref{unduraddd}) exhibits an oscillatory behavior. In our
case, not only the phase in Eq. (\ref{phitundu}), but also the
$\sin(\cdot)$ term in Eq. (\ref{unduraddd}) have an oscillatory
character, and must be taken into account when calculating the
formation length. As a result, in the long-wavelength asymptotic
($\lambdabar \gg \lambdabar_r$), we deal with a situation where
the $\sin(\cdot)$ and the $\vec{\theta}$ terms in Eq.
(\ref{unduraddd}) have different formation lengths. From the
$\sin(\cdot)$ term in Eq. (\ref{unduraddd}) and from Eq.
(\ref{phitundu}) follows a formation length $L_f = \lambdabar_w$,
and a characteristic angle $\theta \sim
\sqrt{\lambdabar/\lambdabar_w}$. However, the TUR contribution is
given by the terms in $\theta_x$ and $\theta_y$ in Eq.
(\ref{unduraddd}). For these terms, from Eq. (\ref{phitundu})
follows a formation length $L_f \sim \min(\bar{\gamma}_z^2
\lambdabar, N_w \lambdabar_w)$, in the limit for $\lambdabar \gg
\lambdabar_r$.  In this limit, the TUR contribution is collimated
to angles $\theta^2 \ll \lambdabar/\lambdabar_w$. Only within
these conditions one can properly talk about TUR. Note that $L_f
\gg \lambdabar_w$ always, because $\bar{\gamma}_z^2 \lambdabar
\sim \lambdabar_w \lambdabar/\lambdabar_r \gg \lambdabar_w$ and we
assume $N_w \gg 1$. The expression for the formation length $L_f$
above is analogous to that for edge radiation, which is given by
$\min(\gamma^2 \lambdabar, L)$. The analogous of the $\phi$
parameter is now given by $\phi_w=\lambdabar_w
N_w/(\bar{\gamma}_z^2 \lambdabar)$, while the analogous of the
$\delta$ parameter is $\delta_w = \lambdabar_w/(\lambdabar_w N_w)
= 1/N_w \ll 1$, as $N_w \gg 1$. It follows that $\phi_w \cdot
\delta_w = \lambdabar_w/(\bar{\gamma}_z^2 \lambdabar) \sim
\lambdabar_r/\lambdabar \ll 1$.

For $\phi_w \cdot \delta_w \ll 1$ and $\delta_w \ll 1$ the
contribution due to the term in $\sin(k_w z')$ in Eq.
(\ref{unduraddd}) can always be neglected when compared with the
maximal field magnitude of the terms in $\theta_{x,y}$. Similarly,
in Eq. (\ref{phitundu}), phase terms in $\cos(k_w z')$ and
$\sin(2k_w z')$ can also be neglected. As a result, Eq.
(\ref{unduraddd}) can be simplified as

\begin{eqnarray}
\vec{\widetilde{{E}}}_{BC}(z, \vec{r},\omega) &=& \frac{i \omega
e}{c^2 z} \int_{z_B}^{z_C} dz' {\exp{\left[i \Phi_{BC}\right]}}
\left(\theta_x {\vec{x}} +\theta_y{\vec{y}}\right) \label{undurad}
\end{eqnarray}
where

\begin{eqnarray}
\Phi_{BC} &=& \frac{\omega}{c} \left[{\theta^2\over{2}} z+
\frac{z'}{2 }\left(\frac{1}{\bar{\gamma}_z^2}+\theta^2\right)
\right]~. ~\label{phitundu4}
\end{eqnarray}

\subsubsection{\label{sub:str} Field contribution calculated along
the straight sections}

With the help of Eq. (\ref{generalfin}) we write the contribution
from the straight line $AB$ as

\begin{equation}
\vec{\widetilde{E}}_{AB}={i \omega  e\over{c^2 z}}
\int_{z_A}^{z_B} dz' \exp{\left[i\Phi_{AB}\right]} \left(\theta_x
{\vec{x}}+\theta_y {\vec{y}}\right) \label{ABcontr}
\end{equation}
where $\Phi_{AB}$ in Eq. (\ref{ABcontr}) is given by

\begin{equation}
\Phi_{AB} = \frac{\omega}{c} \left[ {\theta^2\over{2}} z +
{z'\over{2}}\left({1\over{\gamma^2}}+\theta^2\right) -
\frac{L_w}{4\bar{\gamma}_z^2}+ \frac{L_w}{4\gamma^2}\right]~,
\label{phiab}
\end{equation}
The contribution from the straight section $CD$ is similar to that
from the straight section $AB$ and reads

\begin{equation}
\vec{\widetilde{E}}_{CD}={i \omega e\over{c^2 z}} \int_{z_C}^{z_D}
dz' \exp{\left[i\Phi_{CD}\right]} \left(\theta_x
{\vec{x}}+\theta_y {\vec{y}}\right) \label{CDcontr}
\end{equation}
where $\Phi_{CD}$ in Eq. (\ref{CDcontr}) is given by

\begin{equation}
\Phi_{CD} = \frac{\omega}{c} \left[ {\theta^2\over{2}} z +
{z'\over{2}}\left(\frac{1}{\gamma^2}+\theta^2\right)
+\frac{L_w}{4\bar{\gamma}_z^2}-\frac{L_w}{4\gamma^2}\right]~.
\label{phicd}
\end{equation}
In general, the phases $\Phi_{CD}$ and $\Phi_{AB}$ start
exhibiting oscillatory behavior when $z'/(2 \gamma^2 \lambdabar)
\sim 1$, which gives a maximal integration range in the
longitudinal direction. Similarly as before, in general one has
that the formation lengths $L_\mathrm{fs1}$ and $L_\mathrm{fs2}$
for the straight sections $AB$ and $CD$ can be written as
$L_\mathrm{fs(1,2)} \sim \min\left[ \lambdabar
\gamma^2,L_{(1,2)}\right]$.

\subsubsection{\label{sub:long} Total field and energy spectrum of radiation}

The contributions for segment $AB$ and segment $CD$ are given by
Eq. (\ref{ABcontr}) and Eq. (\ref{CDcontr}). One obtains

\begin{eqnarray}
\vec{\widetilde{E}}_{AB}&&= \frac{i \omega e L_1}{c^2 z}
\exp\left[\frac {i \omega \theta^2 z}{2 c} \right] \vec{\theta}~
\mathrm{sinc}\left[\frac{\omega L_1}{4
c}\left(\frac{1}{\gamma^2}+\theta^2\right)\right] \cr &&\times
\exp\left[-\frac{i \omega L_w }{4
c}\left(\frac{1}{\bar{\gamma}_z^2}+\theta^2\right)\right]\exp\left[-\frac{i
\omega L_1}{4
c}\left(\frac{1}{\gamma^2}+\theta^2\right)\right]\label{ABfar}
\end{eqnarray}

Similarly,

\begin{eqnarray}
\vec{\widetilde{E}}_{CD}&&= \frac{i \omega e L_2}{c^2 z}
\exp\left[\frac {i \omega \theta^2 z}{2 c} \right] \vec{\theta}~
\mathrm{sinc}\left[\frac{\omega L_2}{4
c}\left(\frac{1}{\gamma^2}+\theta^2\right)\right] \cr &&\times
\exp\left[\frac{i \omega L_w }{4
c}\left(\frac{1}{\bar{\gamma}_z^2}+\theta^2\right)\right]\exp\left[\frac{i
\omega L_2 }{4 c}\left(\frac{1}{\gamma^2}
+\theta^2\right)\right]\label{CDfar}
\end{eqnarray}

Finally, the contribution for the segment $BC$ is obtained from
Eq. (\ref{undurad}). Calculations yield:

\begin{eqnarray}
\vec{\widetilde{E}}_{BC}&&= \frac{i \omega e L_w}{c^2 z}
\exp\left[\frac {i \omega \theta^2 z}{2 c} \right]\vec{\theta}~
\mathrm{sinc} \left[\frac{\omega L_w}{4
c}\left(\frac{1}{\bar{\gamma}_z^2}+\theta^2\right)\right]
\label{BCfar}
\end{eqnarray}
The total field produced by the setup is obtained by summing up
Eq. (\ref{ABfar}), Eq. (\ref{CDfar}) and Eq. (\ref{BCfar}). By
this, we are neglecting bending magnet contributions. A sufficient
condition (in addition to the already accepted ones, $N_w \gg 1$
and $\lambdabar/\lambdabar_r \gg 1$) is $\lambdabar \gg
R/\bar{\gamma}_z^3$. In fact, we may neglect bending magnet
contributions for $\delta_{1,2} \ll 1$, but in this setup
$L_{1,2}$ may be set to zero, in which case we should also impose
that the formation length of the bend be much shorter than
$\bar{\gamma}_z^2 \lambdabar$, which reduces to $\lambdabar \gg
R/\bar{\gamma}_z^3$.

As before, the observation angle is measured starting from the
center of the undulator, located at $z=0$, i.e. $\vec{\theta} =
\vec{r}/z$. The angular spectral flux can be written substituting
the resultant total field in Eq. (\ref{endene}). We obtain

\begin{eqnarray}
&&\frac{d W}{d\omega d \Omega} = \frac{ e^2}{ \pi^2 c} \frac{
\gamma^4 \theta^2}{\left(1+\gamma^2\theta^2\right)^2 } \Bigg|
-\exp\left[- i \frac{\omega L_w }{4
c\gamma^2}\left(1+\frac{K^2}{2}+\gamma^2\theta^2\right) \right]\cr
&&+\exp\left[- i \frac{\omega L_1}{2 c \gamma^2}
\left(1+\gamma^2\theta^2\right) - i \frac{\omega  L_w}{4 c
\gamma^2} \left(1+\frac{K^2}{2}+\gamma^2\theta^2\right) \right]
\cr &&
+\frac{1/\gamma^2+\theta^2}{1/\bar{\gamma}_z^2+\theta^2}\left\{
{-\exp\left[\frac{i\omega L_w}{4c\gamma^2}
\left({1}+\frac{K^2}{2}+\gamma^2\theta^2\right)\right]} \right.\cr
&&\left.+{\exp\left[-\frac{i\omega L_w}{4c\gamma^2}
\left(1+\frac{K^2}{2}+\gamma^2\theta^2\right)\right]}\right\}
+\exp\left[ i \frac{\omega L_w }{4
c\gamma^2}\left(1+\frac{K^2}{2}+\gamma^2\theta^2\right) \right]\cr
&& -\exp\left[ i \frac{\omega L_2 }{2 c\gamma^2}
\left(1+\gamma^2\theta^2\right)+\frac{i \omega L_w}{4 c
\gamma^2}\left(1+\frac{K^2}{2}+\gamma^2\theta^2\right)
\right]\Bigg|^2~, \label{enden2}
\end{eqnarray}
that is equivalent to the analogous expression in \cite{BOS4}.

Note that $L_1$, $L_2$ and $L_w$ can assume different values.
$\gamma$ and $\bar{\gamma}_z$ are also different. It may therefore
seem convenient to introduce different normalized quantities,
referring to the undulator and the straight lines. However, in the
end we are interested in summing up all contributions from
different sources, so that it is important to keep a common
definition of vertical displacement (or observation angle).
Therefore we prescribe the same normalization for all quantities:

\begin{equation}
\vec{\hat{\theta}} = \sqrt{\frac{L_\mathrm{tot}}{\lambdabar}}
\vec{\theta}~,\phi_t=\frac{L_\mathrm{tot}}{\gamma^2
\lambdabar}~~~\mathrm{and}~~~ \vec{\hat{r}} = \frac{\vec{r}
}{\sqrt{L_\mathrm{tot} \lambdabar}} ~. \label{rsnoe}
\end{equation}
Then, we introduce parameters $\hat{L}_1 = L_1/L_\mathrm{tot}$,
$\hat{L}_2 = L_2/L_\mathrm{tot}$, $\hat{L}_w =  L_w/
L_\mathrm{tot}$, $\phi_{1,2} = L_{1,2}/(\gamma^2 \lambdabar) =
\hat{L}_{1,2} \phi_t$ and $\phi_w = L_w/(\bar{\gamma}_z^2
\lambdabar)$, as seen above. Here it should be clear that $\phi_t$
has been introduced only for notational convenience, while real
parameters related to the physics of the problem are $\phi_{1,2}$.
Finally, we define $\hat{z}_{s}=z_s/L_\mathrm{tot}$. From Eq.
(\ref{enden2}) follows

\begin{eqnarray}
\hat{z}^2 \hat{I} &=&  \frac{4
\hat{\theta}^2}{(\phi_t+\hat{\theta}^2)^2} \Bigg| \exp\left[-
\frac{i\hat{L}_1 }{2}\left(\phi_t+\hat{\theta}^2\right) - \frac{i
}{4}\left(\phi_w+\hat{L}_w \hat{\theta}^2\right)
\right]-\exp\left[- \frac{i }{4}\left(\phi_w+\hat{L}_w
\hat{\theta}^2\right) \right] \cr &&
+\frac{\phi_t+\hat{\theta}^2}{\phi_w/\hat{L}_w+\hat{\theta}^2}\left\{
{-\exp\left[\frac{i
}{4}\left(\phi_w+\hat{L}_w\hat{\theta}^2\right)\right]}
+{\exp\left[- \frac{i
}{4}\left(\phi_w+\hat{L}_w\hat{\theta}^2\right)\right]}\right\}
\cr && -\exp\left[ \frac{i\hat{L}_2
}{2}\left(\phi_t+\hat{\theta}^2\right)+ \frac{i
}{4}\left(\phi_w+\hat{L}_w\hat{\theta}^2\right) \right]+\exp\left[
\frac{i }{4}\left(\phi_w+\hat{L}_w\hat{\theta}^2\right)
\right]\Bigg|^2~, \label{endenorm1}
\end{eqnarray}
where $\hat{I} = \left|\vec{\hat{E}}\right|^2$ and $\vec{\hat{E}}
\equiv \vec{\widetilde{E}} \sqrt{\lambdabar L_\mathrm{tot}}
~c/e$. Note that outside the undulator the longitudinal velocity
is nearer to $c$ than inside ($\bar{\gamma}_z^2 < \gamma^2$). It
follows that the contribution of the undulator is suppressed
compared with that of the straight sections, and in the case of
comparable lengths and $K^2 \gg 1$, the straight section
contribution becomes dominant.

\begin{figure}
\begin{center}
\includegraphics*[width=100mm]{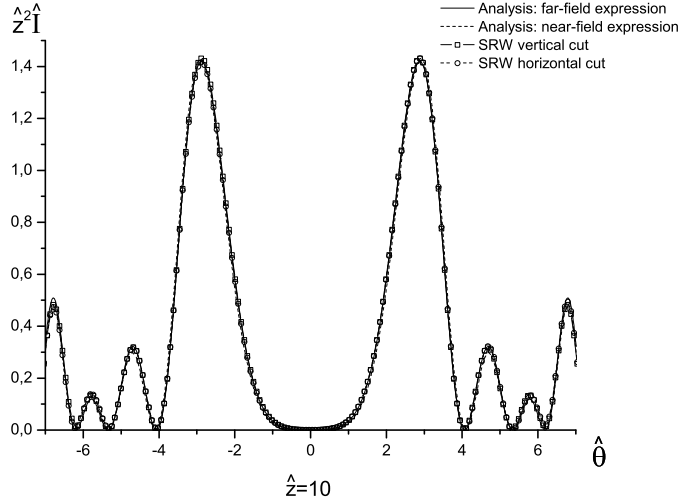}
\caption{\label{Fardtur} Cross-check of Eq. (\ref{endenorm2}) with
the help of SRW. Here $L_1=L_w=L_2 = 200$ m, $E = 17.5$ GeV,
$\lambda = 400$ nm, $R= 400$ m, and $K = 3.3$. Here $\lambda_w =
3.56$ cm. The observer is located at $z=6000$ m. Circles represent
horizontal and vertical cuts of the intensity profiles calculated
numerically with SRW. The solid curve is calculated with  Eq.
(\ref{endenorm2}). The dashed curve is obtained with the near-zone
expressions Eqs. (\ref{fieldtot1})-(\ref{fieldtot4}).}
\end{center}
\end{figure}

In the special case for $\hat{L}_1 = \hat{L}_2 = \hat{L}_w \equiv
\hat{L}/3$, $\phi_{1,2} = \phi_t/3$, and Eq. (\ref{endenorm1})
simplifies to

\begin{eqnarray}
\hat{z}^2 \hat{I} &=& \frac{4
\hat{\theta}^2}{(\phi_t+\hat{\theta}^2)^2} \Bigg| -4 i
\cos\left[\frac{\hat{L}}{12}\left(2\hat{\theta}^2+\phi_t\right)+\frac{\phi_w}{4}\right]
\sin\left[\frac{\hat{L}}{12}\left(\hat{\theta}^2+\phi_t\right)\right]\cr
&&- 2 i\frac{\phi_t+\hat{\theta}^2}{3
\phi_w/\hat{L}+\hat{\theta}^2} \sin\left[\frac{1}{12}\left(3
\phi_w+\hat{L}\hat{\theta}^2\right)\right] \Bigg|^2~.
\label{endenorm2}
\end{eqnarray}
Eq. (\ref{endenorm2}) can be readily evaluated. As an example, we
can calculate the intensity distribution of TUR emitted by the
SASE 1 European XFEL setup at a wavelength $\lambda = 400$ nm. We
assume that the XFEL operates at $17.5$ GeV. Setup parameters are
$L_1=L_w=L_2 = 200$ m, $R= 400$ m, $K = 3.3$ and $\lambda_w =
3.56$ cm \cite{XFEL}. In this case $\delta \sim 10^{-3}$ and
$\delta_w \sim 1/N_w \sim 10^{-4}$, while $\phi_t \simeq 8.0$ and
$\phi_w \simeq 52$. Results are plotted in Fig. \ref{Fardtur}. In
that figure, we also propose a comparison with outcomes from SRW
at
$z=6000$ m (vertical and horizontal cuts).  

In the far zone, well-accepted expressions for the TUR emission
are reported in literature \cite{KIM1,KINC,CAST, ROY2}, that are
equivalent to the following equation for the radiation energy
density as a function of angle and frequency:

\begin{eqnarray}
\frac{d W}{d \omega d\Omega} &=& \frac{e^2}{\pi^2 c}
\left[\frac{2\gamma^2 \theta K^2
}{(1+K^2/2+\gamma^2\theta^2)(1+\gamma^2\theta^2)}\right]^2
 \sin^2\left[\frac{\pi L_w}{2\gamma^2 \lambda}
\left(1+\frac{K^2}{2}+\gamma^2\theta^2\right)\right]. \cr &&
\label{wron3}
\end{eqnarray}
We will show that Eq. (\ref{wron3}) cannot be applied for TUR
calculations. In our understanding, there cannot be any range of
parameters in the setup in Fig. \ref{geomu} where Eq.
(\ref{wron3}) is valid.

In order to prove this it is sufficient to compare Eq.
(\ref{wron3}) with Eq. (\ref{enden2}). Eq. (\ref{wron3}) does not
depend on the straight section lengths $L_1$ or $L_2$, and can be
applied when $L_1,~ L_2 \longrightarrow 0$.

However, in that limit, Eq. (\ref{enden2}) reduces to:

\begin{eqnarray}
\frac{d W}{d\omega d \Omega} = \frac{ e^2}{ \pi^2 c}\left[ \frac{2
\gamma^2\theta}{1+K^2/2+\gamma^2\theta^2 }\right]^2
\sin^2\left[\frac{\pi L_w}{2 \gamma^2 \lambda }
\left({1}+\frac{K^2}{2}+\gamma^2\theta^2\right)\right] ~,
\label{enden2zero}
\end{eqnarray}
that is obviously different from Eq. (\ref{wron3}). Also note that
in the limit for $K \longrightarrow 0$ Eq. (\ref{wron3}) tends to
zero, whereas Eq. (\ref{enden2zero}) gives back Eq.
(\ref{enden2e}) as it must be.

\subsection{\label{sub:virt} Virtual source characterization and
field propagation}

There is a general need, in the FEL community, to extend the
current theory of TUR to cover the near zone. For instance, a
possible use of coherent TUR to produce visible light synchronized
with X-rays from an X-ray free-electron laser is discussed in
\cite{OURM}. As we have seen, TUR can be discussed as a more
complicated edge-radiation setup. Within the sharp-edge
approximation we have contributions from three parts, two straight
lines and the undulator. The undulator contribution is similar to
a straight line contribution, the only difference being a
different average longitudinal velocity of the electron. Then, the
far-zone region can be identified by distances $z \gg L_{tot}$.

\begin{figure}
\begin{center}
\includegraphics*[width=100mm]{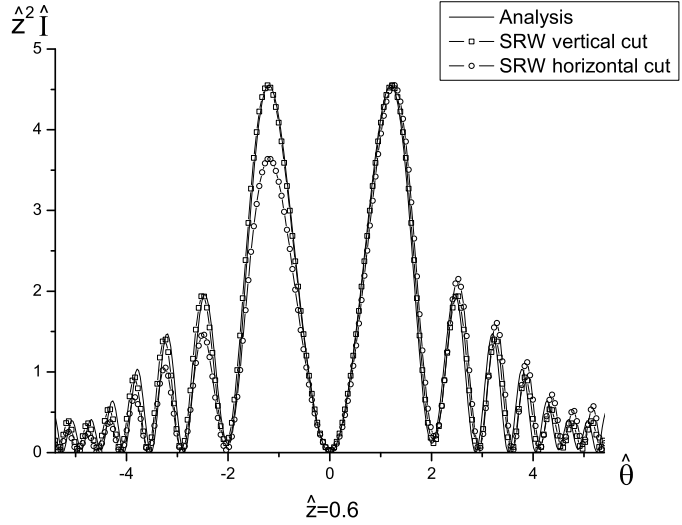}
\caption{\label{z0p6dtur} Cross-check of analytical results with
SRW. Here $L_1=L_w=L_2 = 200$ m, $E = 17.5$ GeV, $\lambda = 400$
nm, $R= 400$ m, and $K = 3.3$. The observer is located at $z=360$
m. Here $\lambda_w = 3.56$ cm. Horizontal and vertical cuts of the
intensity profiles are compared with results obtained with the
help of Eqs. (\ref{fieldtot1}) to (\ref{fieldtot4}).}
\end{center}
\end{figure}

\begin{figure}
\begin{center}
\includegraphics*[width=100mm]{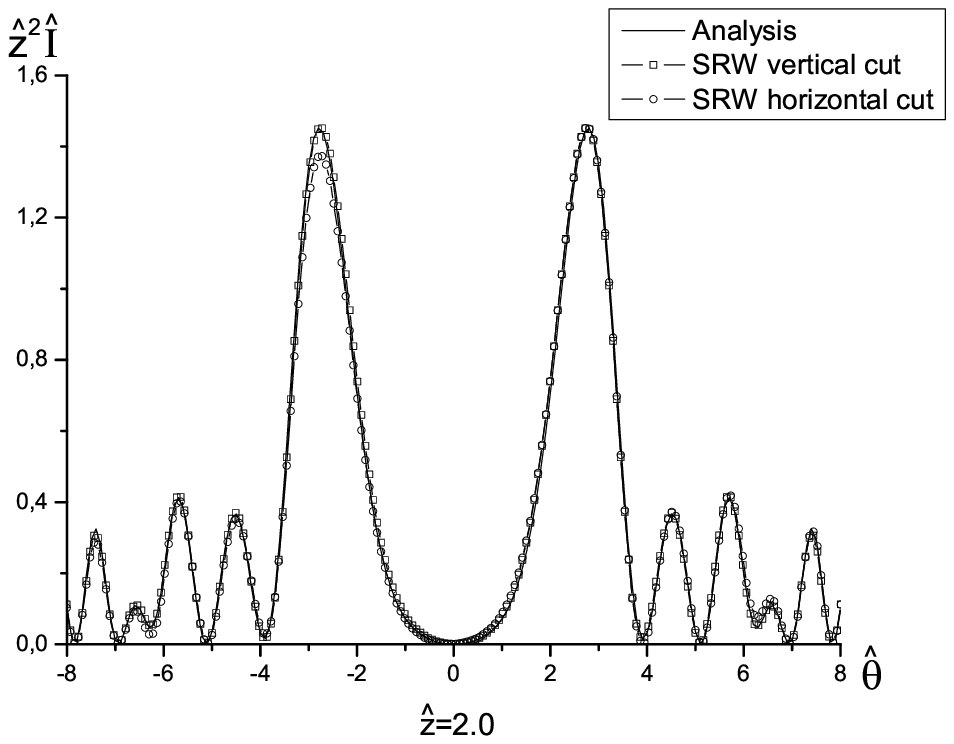}
\caption{\label{z2p0tur} Cross-check of analytical results with
SRW. Here $L_1=L_w=L_2 = 200$ m, $E = 17.5$ GeV, $\lambda = 400$
nm, $R= 400$ m, and $K = 3.3$. Here $\lambda_w = 3.56$ cm. The
observer is located at $z=1200$ m. Horizontal and vertical cuts of
the intensity profiles are compared with results obtained with the
help of  Eqs. (\ref{fieldtot1}) to (\ref{fieldtot4}).}
\end{center}
\end{figure}

Expressions in Eq. (\ref{ABfar}), Eq. (\ref{CDfar}) and Eq.
(\ref{BCfar}) can be interpreted as far field radiation from
separate virtual sources. For each far field contribution we use a
picture with two virtual sources, located at the ends of the
straight sections and of the undulator. This makes a total of six
sources. However, since the virtual source at the downstream
[upstream] edge of the first [second] straight section has the
same longitudinal position of the virtual source at the upstream
[downstream] edge of the undulator, i.e. $z=-L_w/2$ [$z=L_w/2$],
we combine them together, summing them up by superposition
principle. As a result, we are left with only four sources,
located at

\begin{equation}
{z}_{s1} = -\frac{{L}_w }{2  }-{L}_1 ~,~~~ {z}_{s2} = -\frac{{L}_w
}{2  }~,~~~ {z}_{s3} = \frac{{L}_w }{2  }~,~~~\mathrm{and}~~~
{z}_{s4} = \frac{{L}_w }{2}+{{L}_2} ~.\label{co3}
\end{equation}
We obtain an explicit expression for these sources with the help
of Eq. (\ref{ABfar}), Eq. (\ref{CDfar}) and Eq. (\ref{BCfar}),
proceeding analogously as in Section \ref{sub:singles}:

\begin{eqnarray}
\vec{\widetilde{E}}_{s1}\left(-\frac{L_w}{2}-L_1,\vec{r}\right) =
\frac{2 \omega e }{c^2 \gamma} \exp\left[- \frac{i \omega  L_1 }{2
c \gamma^2 }\right] \exp\left[-\frac{i \omega L_w }{4 c \gamma_z^2
}\right] \frac{\vec{r}}{r} K_1\left(\frac{\omega  r}{c \gamma}
\right)~,\label{virs1}
\end{eqnarray}
\begin{eqnarray}
\vec{\widetilde{E}}_{s2}\left(-\frac{L_w}{2},\vec{r}\right) &=& -
\frac{2 \omega  e }{c^2 \gamma} \exp\left[-\frac{i \omega  L_w }{4
c\gamma_z^2}\right] \frac{\vec{r}}{r} K_1\left(\frac{\omega r}{c
\gamma} \right) \cr && + \frac{2 \omega  e }{c^2 \gamma_z}
\exp\left[- \frac{i \omega L_w }{4 c \gamma_z^2}\right]
\frac{\vec{r}}{r} K_1\left(\frac{\omega  r}{c \gamma_z} \right)
~,\label{virs2}
\end{eqnarray}
\begin{eqnarray}
\vec{\widetilde{E}}_{s3}\left(\frac{L_w}{2},\vec{r}\right) &=& -
\frac{2 \omega  e }{c^2 \gamma_z} \exp\left[\frac{i \omega  L_w
}{4 c\gamma_z^2}\right] \frac{\vec{r}}{r} K_1\left(\frac{\omega
r}{c \gamma_z} \right)\cr && + \frac{2 \omega e }{c^2 \gamma}
\exp\left[\frac{i \omega L_w }{4
c\gamma_z^2}\right]\frac{\vec{r}}{r} K_1\left(\frac{\omega r}{c
\gamma} \right)~,\label{virs3}
\end{eqnarray}
\begin{eqnarray}
\vec{\widetilde{E}}_{s4}\left(\frac{{L}_w
}{2}+{{L}_2},\vec{r}\right) = - \frac{2 \omega  e }{c^2 \gamma}
\exp\left[\frac{i \omega L_2}{2 c\gamma^2}\right]\exp\left[\frac{i
\omega L_w }{4 c \gamma_z^2}\right] \frac{\vec{r}}{r}
K_1\left(\frac{\omega  r}{c \gamma} \right)~.\label{virs4}
\end{eqnarray}
In order to calculate the field at any distance $z$ we proceed in
analogy with Eq. (\ref{fieldtot}), applying the propagation
formula Eq. (\ref{fieldpropback}). As before, the above given
equations for the sources can also be used as input to any Fourier
code to calculate the field evolution in the presence of whatever
optical beamline. However, here we restrict ourselves to the
free-space case. In order to simplify the presentation of the
electric field we take advantage of polar coordinates and we use
the definition $\vec{\hat{E}} \equiv \vec{\widetilde{E}}
\sqrt{\lambdabar L_\mathrm{tot}} ~c/e$ (so that $\hat{I}$,
introduced in Eq. (\ref{Ihat}), is given by $\hat{I} =
|\vec{\hat{E}}|^2$) for the field in normalized units. Note that
here $\hat{z}  = z/L_\mathrm{tot}$. We obtain four field
contributions, one for each source:

\begin{eqnarray}
\hat{z} \vec{\hat{E}}_{1}\left(\hat{z},\vec{\hat{\theta}}\right)
&=& \left. -  \left\{\frac{\vec{\hat{\theta}}}{\hat{\theta}}
\frac{2 \hat{z} \sqrt{\phi_t} \exp[- i \hat{L}_1
\phi_t/2]\exp\left[-i \phi_w/4\right]
}{\hat{z}+\hat{L}_w/2+\hat{L}_1} \right.\right.\cr && \left.\left.
\times \int_0^{\infty} d\hat{r}' \hat{r}'
K_1\left(\sqrt{\phi_t}\hat{r}'\right) J_1
\left(\frac{\hat{\theta}\hat{r}'
\hat{z}}{\hat{z}+\hat{L}_w/2+\hat{L}_1}\right) \right. \right.\cr
&&\left. \left. \times \exp\left[\frac{i\hat{r}^{'2}
}{2\left(\hat{z}+\hat{L}_w/2+\hat{L}_1\right)}\right]
\exp\left[\frac{i\hat{\theta}^2 \hat{z}^2
}{2\left(\hat{z}+\hat{L}_w/2+\hat{L}_1\right)}\right]\right\}
\right.~.\label{fieldtot1}
\end{eqnarray}
\begin{eqnarray}
\hat{z} \vec{\hat{E}}_{2}\left(\hat{z},\vec{\hat{\theta}}\right)
&=& \left. \left\{\frac{\vec{\hat{\theta}}}{\hat{\theta}} \frac{2
\hat{z} \sqrt{\phi_t} \exp\left[-i \phi_w/4\right]
}{\hat{z}+\hat{L}_w/2} \exp\left[\frac{i\hat{\theta}^2 \hat{z}^2
}{2\left(\hat{z}+\hat{L}_w /2\right)}\right]\right.\right.\cr &&
\left.\left. \times \int_0^{\infty} d\hat{r}' \hat{r}'
K_1\left(\sqrt{\phi_t}\hat{r}'\right) J_1
\left(\frac{\hat{\theta}\hat{r}' \hat{z}}{\hat{z}+\hat{L}_w
/2}\right) \exp\left[\frac{i\hat{r}^{'2}
}{2\left(\hat{z}+\hat{L}_w /2\right)}\right] \right\} \right.\cr
&&\left. - \left\{\frac{\vec{\hat{\theta}}}{\hat{\theta}} \frac{2
\hat{z} \sqrt{\phi_w/\hat{L}_w} \exp\left[-i \phi_w/4\right]
}{\hat{z}+\hat{L}_w /2} \exp\left[\frac{i\hat{\theta}^2 \hat{z}^2
}{2\left(\hat{z}+\hat{L}_w /2\right)}\right]\right.\right.\cr &&
\left.\left. \times \int_0^{\infty} d\hat{r}' \hat{r}'
K_1\left(\sqrt{\phi_w/\hat{L}_w}\hat{r}'\right) J_1
\left(\frac{\hat{\theta}\hat{r}' \hat{z}}{\hat{z}+\hat{L}_w
/2}\right) \exp\left[\frac{i\hat{r}^{'2}
}{2\left(\hat{z}+\hat{L}_w /2\right)}\right] \right\}
\right.~.\label{fieldtot2b}
\end{eqnarray}
\begin{eqnarray}
\hat{z} \vec{\hat{E}}_{3}\left(\hat{z},\vec{\hat{\theta}}\right)
&=& - \left. \left\{\frac{\vec{\hat{\theta}}}{\hat{\theta}}
\frac{2 \hat{z} \sqrt{\phi_t} \exp\left[i \phi_w/4\right]
}{\hat{z}-\hat{L}_w/2} \exp\left[\frac{i\hat{\theta}^2 \hat{z}^2
}{2\left(\hat{z}-\hat{L}_w /2\right)}\right]\right.\right.\cr &&
\left.\left. \times \int_0^{\infty} d\hat{r}' \hat{r}'
K_1\left(\sqrt{ \phi_t}\hat{r}'\right) J_1
\left(\frac{\hat{\theta}\hat{r}' \hat{z}}{\hat{z}-\hat{L}_w
/2}\right) \exp\left[\frac{i\hat{r}^{'2}
}{2\left(\hat{z}-\hat{L}_w /2\right)}\right] \right\} \right.\cr
&&\left. +  \left\{\frac{\vec{\hat{\theta}}}{\hat{\theta}} \frac{2
\hat{z} \sqrt{\phi_w/\hat{L}_w} \exp\left[i \phi_w/4\right]
}{\hat{z}-\hat{L}_w /2} \exp\left[\frac{i\hat{\theta}^2 \hat{z}^2
}{2\left(\hat{z}-\hat{L}_w /2\right)}\right]\right.\right.\cr &&
\left.\left. \times \int_0^{\infty} d\hat{r}' \hat{r}'
K_1\left(\sqrt{\phi_w/\hat{L}_w}\hat{r}'\right) J_1
\left(\frac{\hat{\theta}\hat{r}' \hat{z}}{\hat{z}-\hat{L}_w
/2}\right) \exp\left[\frac{i\hat{r}^{'2}
}{2\left(\hat{z}-\hat{L}_w /2\right)}\right] \right\}
\right.~.\label{fieldtot3b}
\end{eqnarray}
\begin{eqnarray}
\hat{z} \vec{\hat{E}}_{4}\left(\hat{z},\vec{\hat{\theta}}\right)
&=& \left\{\frac{\vec{\hat{\theta}}}{\hat{\theta}} \frac{2 \hat{z}
\sqrt{ \phi_t} \exp\left[i\hat{L}_2 \phi_t/2\right]\exp\left[i
\phi_w/4\right] }{\hat{z}-\hat{L}_w/2 -  \hat{L}_2} \right.\cr &&
\left. \times \int_0^{\infty} d\hat{r}' \hat{r}'
K_1\left(\sqrt{\phi_t}\hat{r}'\right) J_1
\left(\frac{\hat{\theta}\hat{r}' \hat{z}}{\hat{z}-\hat{L}_w/2 -
\hat{L}_2 }\right) \right. \cr && \left. \times
\exp\left[\frac{i\hat{r}^{'2} }{2\left(\hat{z}-\hat{L}_w/2 -
\hat{L}_2\right)}\right] \exp\left[\frac{i\hat{\theta}^2 \hat{z}^2
}{2\left(\hat{z}-\hat{L}_w/2 -  \hat{L}_2
\right)}\right]\right\}~.\label{fieldtot4}
\end{eqnarray}
Eqs. (\ref{fieldtot1}) to (\ref{fieldtot4}) can be used to
calculate the field, and hence the intensity, at any position of
interest in the far and in the near zone. Obviously, in the far
zone, for $\hat{z} \gg 1$, the square modulus of their sum reduces
to Eq. (\ref{endenorm2}). As before, It is interesting to
cross-check Eqs. (\ref{fieldtot1}) to (\ref{fieldtot4}) with the
computer code SRW. We used the same numerical parameters as
before: $L_1=L_w=L_2 = 200$ m, $E = 17.5$ GeV, $\lambda = 400$ nm
and $R= 400$ m, with an undulator parameter $K = 3.3$.
Additionally, we chose different values ${z} = 360$m, ${z} =
600$m, ${z} = 1200$m (see Figs.  \ref{z0p6dtur} $\div$
\ref{z2p0tur}), and ${z} = 6000$m (see Fig. \ref{Fardtur}),
corresponding to $\hat{z}= 0.6$, $\hat{z}= 1.0$, $\hat{z}= 1.5$,
$\hat{z}= 2.0$ and $\hat{z}= 10$ (which is the far-zone case
treated before). Here the bending plane is the horizontal plane.
As the observation point becomes nearer to the edge of the magnet,
the influence of the bending magnet becomes more and more
important, an effect which is evident in the figures from the
horizontal cuts of SRW two-dimensional intensity profiles. From
Figs. \ref{Fardtur}$\div$\ref{z2p0tur} we can see that the
analytical result for the vertical cut is valid with good accuracy
up to $\hat{z}=0.6$.

\begin{figure}
\begin{center}
\includegraphics*[width=100mm]{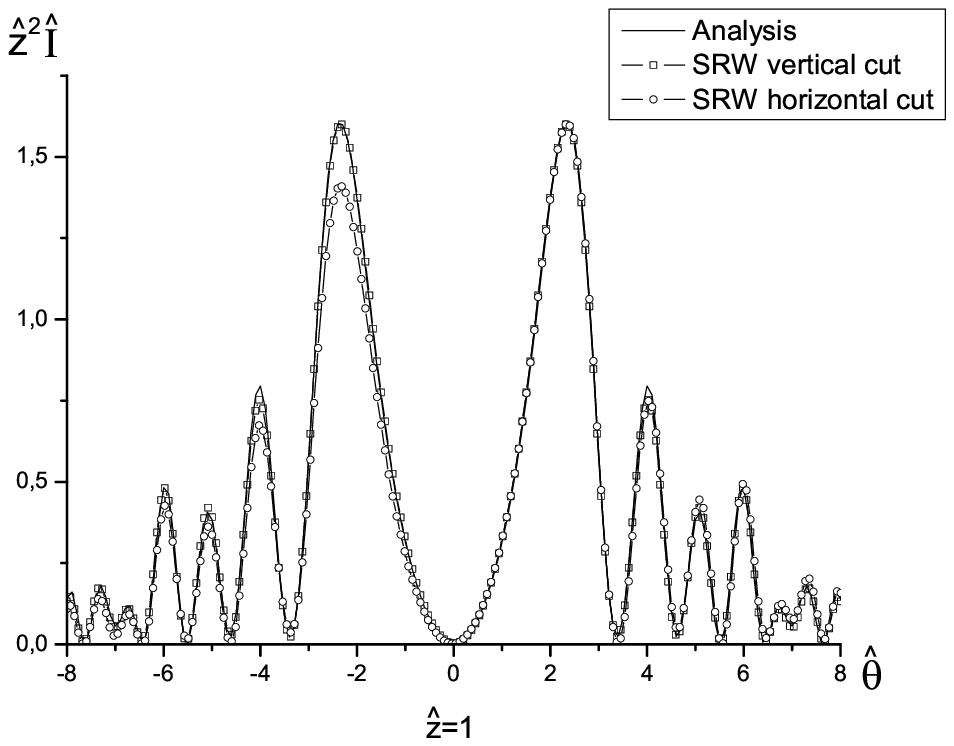}
\caption{\label{z1p0tur} Cross-check of analytical results with
SRW. Here $L_1=L_w=L_2 = 200$ m, $E = 17.5$ GeV, $\lambda = 400$
nm, $R= 400$ m, and $K = 3.3$. Here $\lambda_w = 3.56$ cm. The
observer is located at $z=600$ m. Horizontal and vertical cuts of
the intensity profiles are compared with results obtained with the
help of  Eqs. (\ref{fieldtot1}) to (\ref{fieldtot4}).}
\end{center}
\end{figure}

\section{\label{sec:waveguide} Edge radiation in a waveguide}

In the previous Sections we considered edge radiation propagating
in unbounded space. In this Section we extend previous
considerations to account for the presence of metallic
surroundings (like e.g. the electron vacuum chamber), which
effectively act like a waveguide.

The need to account for guiding structures in edge radiation
setups arises when one wants to operate in the sub-millimeter
wavelength range. Emission of edge radiation in the presence of
metallic boundaries has been a much less-treated subject in
literature, compared to the unbounded space case. To the best of
our knowledge, there is only one article reporting on edge
radiation from electrons in a planar overmoded metallic waveguide
\cite{BOS2}. The analysis in \cite{BOS2} is based on
Lienard-Wiechert fields, modified by the presence of finite
metallic boundary. To account for perfectly conducting plates, the
author of \cite{BOS2} uses a generalization of the method of
images, well-known from electrostatics \cite{JACK}. Instead of
this approach, here we prefer to use a mode-expansion technique.
We characterize radiation from particles in guiding structures
with the help of the proper tensor Green's function, automatically
accounting for boundary conditions. The Green's function itself
and the field are then conveniently expressed in terms of the
natural modes of the guiding structure. This approach is shown to
provide a convenient and methodical way to deal with the vectorial
character of our problem. We apply our method to the case of a
homogeneous waveguide with circular cross-section. This
configuration approximates the vacuum chamber to be used at XFEL
and Energy Recovery Linac (ERL) facilities. Our approach, however,
is very general and may be applied to other geometries, e.g. to
rectangular waveguides, which approximate vacuum chambers used at
SR facilities.

\begin{figure}
\begin{center}
\includegraphics*[width=140mm]{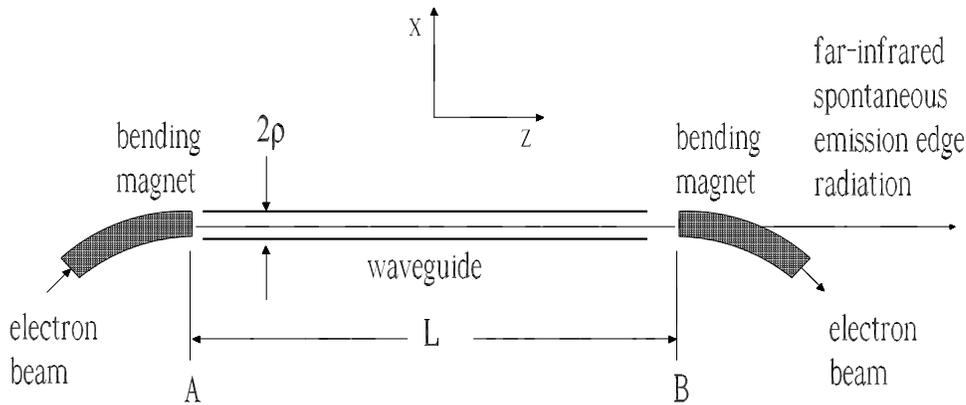}
\caption{\label{guide} Setup for edge radiation emission includes
the presence of a waveguide with circular cross-section of radius
$\rho$ along the straight section.}
\end{center}
\end{figure}

\subsection{Definition of setup and position of the problem}

To fix ideas we focus our attention on the setup in Fig.
\ref{guide}. Electrons travel through the usual edge radiation
setup, similarly as in Fig. \ref{geome}(a). The difference is that
now we account for the presence of a waveguide with typical
cross-section dimension $\rho$ along the straight section. Since
electrons are bent away, one may assume that the vacuum chamber
ends at the downstream bend position $z=L/2$. Our goal in this
Section is to calculate the field distribution at that position,
which should be subsequently propagated up to the experimental
station.

%
Under the assumption $\delta \ll 1$, which is always considered
valid, one can then extend the ER theory from unbounded space to
the case when a waveguide is present. In unbounded space, the
assumption $\delta \ll 1$ allowed us to apply the ER theory with
given constraints on the angular region of observation: in fact,
at large angles the SR contributions from bends is dominant
compared to the ER contribution to the field. In other words, as
we have seen, ER is a more collimated form of radiation, compared
to bending magnet radiation. The presence of a waveguide seems at
first glance to radically modify this viewpoint, because both SR
from bends and ER is collected and guided into the vacuum chamber.
However, we will see that this situation still carries a strong
analogy with the unbounded space case: while in unbounded space ER
theory can be applied within a given angular region dependent on
$\delta$, in the waveguide case ER theory can be applied within a
given wave-number range dependent on $\delta$. As said above,
after the field distribution is calculated at $z=L/2$, it has to
be propagated up to the experimental station. Radiation will be
collected within a finite acceptance angle. This finite acceptance
can be seen as a low-pass filter for spatial frequencies. If the
acceptance angle is small enough (and this depends on the value of
$\delta$) only spatial frequencies where ER is dominant will be
accepted. Thus, one can conclude that both in the case of
unbounded space and vacuum chamber, ER theory can be applied with
given constraints on the acceptance angle of the setup.

Despite the fact that equations for the edge radiation source are
significantly complicated by the presence of a waveguide, we find
that these complications result in the appearance of a single
dimensionless extra-parameter, namely the waveguide diffraction
parameter $\Omega = \rho^2/(\lambdabar L)$. When $\phi \ll 1$ and
$\delta \ll 1$, $\Omega$ is the only parameter of the problem and
has a simple interpretation as the ratio between the waveguide
cross-section area and the square of the radiation diffraction
size of the ER virtual source located in the middle of the
straight section in unbounded space.

The general theoretical framework of our extended ER theory relies
on the solution of Maxwell's equations with modified boundary
conditions, and follows closely the approach proposed in
\cite{OURW}. We wish to describe, in the space-frequency domain,
radiation from an ultrarelativistic electron moving on a given
trajectory inside a metallic vacuum chamber of arbitrary
cross-section, homogeneous along the $z$-axis. An example is
depicted in Fig. \ref{geobou}, where $\vec{n}$ is a vector field
defined on the boundary surface $S$, such that $|\vec{n}|=1$. At
any point, $\vec{n}$ is orthogonal to $S$ and points outwards and
at this stage the waveguide cross-section is still considered
arbitrary.

Since the electron is ultrarelativistic, the Lorentz factor
$\gamma$ obeys $1/\gamma^2 \ll 1$, and paraxial Maxwell's
equations can be used in unbounded space. The presence of a
waveguide introduces the extra-limitation $\lambdabar \ll \rho$.
Paraxial approximation can be used under these two constraints.
Then, the field equations read

\begin{equation}
\left\{
\begin{array}{l}
\left({\nabla_\bot}^2 \vec{\widetilde{E}}+ {2 i \omega \over{c}}
\partial_z \vec{\widetilde{E}}\right) = \frac{4
\pi e}{c}   \exp\left[{i \int_{0}^{z} d \bar{z} \frac{\omega}{2
\gamma_z^2(\bar{z}) c}}\right]
\left[\frac{i\omega}{c^2}\vec{v}_\bot(z) -\vec{\nabla}_\bot
\right]\delta\left(\vec{r}-\vec{r}_0(z)\right)
\\
\left(\vec{n} \times \vec{\widetilde{E}}\right)_{\Big|_S} = 0
\\
\left(\vec{\nabla}_\bot \cdot \vec{\widetilde{E}}\right)_{\Big|_S}
= 0~,
\end{array}\right.\label{pb1}
\end{equation}

where the second and the third equation define boundary conditions
at the waveguide surface S.

As has been demonstrated in \cite{FELB,EVN2,OURW} the field
$\vec{\widetilde{E}}$ can be written in terms of cartesian
components as (compare with Eq. (\ref{efielGfree}))

\begin{eqnarray}
\widetilde{E}^\alpha &=& \frac{4\pi e}{c} \int_{-\infty}^{z} dz'
\left\{ \frac{i\omega}{c^2} v_\bot^\beta(z')
G^\alpha_\beta\left(\vec{r},\vec{r}_0(z'), z-z'
\right)+\left[\partial'_\beta
G^\alpha_\beta\left(\vec{r},\vec{r'},
z-z'\right)\right]_{\vec{r'}=\vec{r}_0(z')} \right\}\cr &&\times
\exp\left[\frac{i\omega}{2c} \int_0^{z'}
\frac{d\bar{z}}{\gamma^2_z(\bar{z})} \right]~,\label{efielG}
\end{eqnarray}
where derivatives $\partial'_\beta$ are taken with respect to
$\vec{r'}$, and $-\infty$ has to be taken as the point where the
source harmonics begin to exist (in our case, at $z=-L/2$). Eq.
(\ref{efielG}) is the solution of the inhomogeneous field
equation, describing the field from a single electron. As before,
$\vec{r}_0(z')$ and $\vec{v}_\bot(z')$ fix transverse position and
velocity of the electron as a function of $z'$, and
$\gamma_z(\bar{z}) = [1-v_z(\bar{z})^2/c^2]^{-1/2}$, $v_z$
indicating the longitudinal velocity of the electron. Finally,
$G^\alpha_\beta$ are the cartesian components of a suitable
tensorial Green's function for a homogeneous waveguide (along the
$z$-axis):

\begin{eqnarray}
G\left(\vec{r},\vec{r'},z,z'\right) &&= \frac{c}{2 i \omega}
\sum_j \Bigg\{\exp\left[-\frac{i c \lambda^\mathrm{TE}_j
}{2\omega}(z-z')\right]\left[\vec{e}_x~
\partial_{y~}
\psi^\mathrm{TE}_j\left(\vec{r}\right)-\vec{e}_y~
\partial_x \psi^\mathrm{TE}_j\left(\vec{r}\right)
\right] \cr&& \otimes\left[\vec{e}_x~
\partial_{y'}
\psi^\mathrm{TE}_j\left(\vec{r'}\right)-\vec{e}_y~
\partial_{x'} \psi^\mathrm{TE}_j\left(\vec{r'}\right)\right]\Bigg\}\cr && +
\frac{c}{2 i \omega} \sum_j \Bigg\{\exp\left[-\frac{i c
\lambda^\mathrm{TM}_j }{2\omega}(z-z')\right]\left[\vec{e}_x~
\partial_{x}
\psi^\mathrm{TM}_j\left(\vec{r}\right)+\vec{e}_y~
\partial_{y~} \psi^\mathrm{TM}_j\left(\vec{r}\right)
\right]\cr&&\otimes\left[\vec{e}_x~
\partial_{x'}
\psi^\mathrm{TM}_j\left(\vec{r'}\right)+\vec{e}_y~
\partial_{y'} \psi^\mathrm{TM}_j\left(\vec{r'}\right)\right]\Bigg\}~,\label{Gfex}
\end{eqnarray}
where notation $\otimes$ indicates the tensor product. Here,
$\vec{e}_x$ and $\vec{e}_y$ are unit vectors along cartesian axes
$x$ and $y$, while $\psi^{\mathrm{TE},\mathrm{TM}}_j$ and
$\lambda^\mathrm{TE,TM}_j$ are eigenfunctions and eigenvalues
referring to the scalar problem:

\begin{figure}
\begin{center}
\includegraphics*[width=100mm]{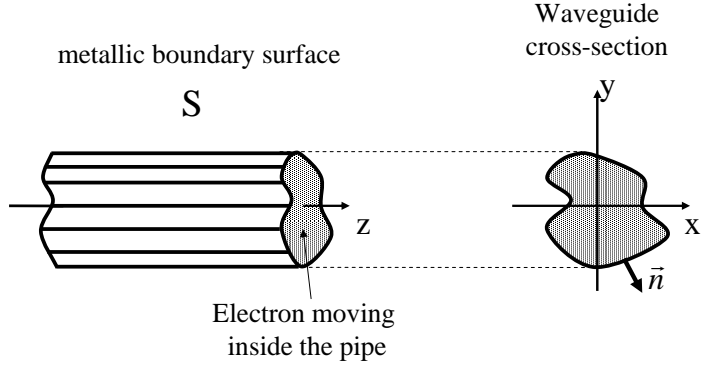}
\caption{Geometry of the problem. The electron is moving inside
the vacuum chamber. \label{geobou}}
\end{center}
\end{figure}

\begin{equation}
\left\{
\begin{array}{l}
\nabla_\bot^2 \psi_j^\mathrm{TE,TM}(\vec{r}) +
\lambda_j^\mathrm{TE,TM} ~\psi_j^\mathrm{TE,TM}(\vec{r}) = 0
\\
\vec{n} \cdot \left(\vec{\nabla}_\bot
\psi_j^\mathrm{TE}\right)_{\Big|_S} = 0
\\
\left(\psi_j^\mathrm{TM}\right)_{\Big|_S} = 0~,
\end{array}\right.\label{pbeigpsiH}
\end{equation}
TE and TM respectively standing for Transverse Electric and
Transverse Magnetic, with normalization condition

\begin{eqnarray}
\int_s d\vec{r} \left|\vec{\nabla} \psi_j^\mathrm{TE,TM}\right|^2
=1 ~.\label{normcb}
\end{eqnarray}
Note that here we derived the Green's function Eq. (\ref{Gfex})
directly in paraxial approximation starting from Eq. (\ref{pb1})
for the transverse components of the field.

Simplifications to Eq. (\ref{efielG}) apply remembering that in
our case $\vec{v}_\bot(z')=0$ and $\vec{r}_0(z')=0$. Setting
$z=L/2$ one obtains:

\begin{eqnarray}
\widetilde{E}^\alpha &=& \frac{4\pi e}{c} \int_{-L/2}^{L/2} dz'
\left[\partial'_\beta G^\alpha_\beta\left(\vec{r},\vec{r'},
z-z'\right)\right]_{\vec{r'}=0} \exp\left[\frac{i \omega z'}{2 c
\gamma^2}\right]~.\label{efielG2}
\end{eqnarray}
In a waveguide that is homogenous along the $z$-axis, only TM
modes turn out to be driven by the uniform motion of the
space-charge distribution. As a result, we can focus on TM modes
only. This may be directly seen by calculating $\partial'_\beta
G^\alpha_\beta$. A compact mathematical way proving this fact
without cumbersome calculations is to note that TE modes appear in
the expression for the tensorial Green's function, Eq.
(\ref{Gfex}), through the combination $\vec{\nabla} \times
(\psi^\mathrm{TE}_j \vec{e}_z) = \vec{e}_x~\partial_{y~}
\psi^\mathrm{TE}_j-\vec{e}_y~\partial_x \psi^\mathrm{TE}_j$. Then,
TE modes contribute in $\partial'_\beta ~G^\alpha_\beta$ for
$\vec{\nabla} \cdot [\vec{\nabla} \times (\psi^\mathrm{TE}_j
\vec{e}_z)]$, which is the divergence of a curl and yields back
zero. From a physical viewpoint this is a sound result. In fact,
radiation is related with energy change of the particle, which
happens through the scalar product $\vec{\widetilde{E}} \cdot
\vec{v}$. Since $\vec{v}_\bot(z')=0$, TE modes cannot lead to any
energy change of the electron. As a result, they are not excited.
Note that $\vec{v}_\bot(z')=0$ means that
$\vec{r}_0(z')=\mathrm{constant}$. Since the electron beam
trajectory should be centered, one recovers $\vec{r}_0(z')=0$,
leading to extra-simplifications in Eq. (\ref{efielG2}).

\subsection{Theory of Edge Radiation in a circular waveguide}

We now want to apply Eq. (\ref{efielG2}) to study the case of the
setup in Fig. \ref{guide} in the presence of a circular waveguide
with radius $\rho$. Therefore, we fix $z=L/2$ in Eq.
(\ref{efielG2}). This allows us to calculate a field distribution
at the end of the straight section, which acts as a source to be
used for further propagation in unbounded space or through the
beamline optics.

In order to explicitly compute the tensor Green's function in Eq.
(\ref{Gfex}) we need to solve the problem in (\ref{pbeigpsiH})
with this particular choice of geometry. Because of the particular
symmetry in the case of a circular waveguide, it is convenient to
use transverse polar coordinates so that $\vec{r}$ is identified
by radial distance $r=\sqrt{x^2+y^2}$ and angle $\varphi =
\arctan(y/x)$.

Solutions of the equation set in (\ref{pbeigpsiH}) are

\begin{eqnarray}
&& \psi_{mk1}^\mathrm{TM} = A_{mk}^\mathrm{TM}
J_m\left(\nu_{mk}r~\right) \sin(m\phi)~,~~~~~~
\psi_{mk2}^\mathrm{TM}= A_{mk}^\mathrm{TM}
J_m\left(\nu_{mk}r~\right)\cos(m\phi) ~,\label{solHpolar}
\end{eqnarray}
where  $\nu_{mk}$ are defined as the roots of
$J_m\left(\nu_{mk}\right) = 0$, and the normalization coefficients
for TM modes, $A_{mk}^{\mathrm{TM}}$, are given by

\begin{eqnarray}
A_{mk}^\mathrm{TM} = \sqrt{\frac{a_m}{\pi}}\frac{1}{\nu_{mk}~
J_{m-1}(\nu_{mk})}~, \label{couplA}
\end{eqnarray}
where $a_0 = 1$ and $a_m = 2$ for $m \geq 1$.

This leads to the following final result for the tensor Green's
function components in cartesian coordinates:

\begin{eqnarray}
G&&\left(\vec{r},\vec{r'},z-z'\right) = \frac{c}{2 i \omega}
\sum_{m=0}^{\infty}\sum_{k=1}^{\infty} \left(A^{TM}_{mk}\right)^2
\left(\frac{\nu_{mk}}{2 \rho }\right)^2 \exp\left[-\frac{i c
(z-z')}{2\omega \rho^2 } \nu_{mk}^2\right] \cr && \times
\left\{\left(\begin{array}{c} J_{m-1}(\nu_{mk}r/\rho)
\sin[(m-1)\varphi]-J_{m+1}(\nu_{mk}r/\rho) \sin[(m+1)\varphi]
\\
J_{m-1}(\nu_{mk}r/\rho) \cos[(m-1)\varphi]+J_{m+1}(\nu_{mk}r/\rho)
\cos[(m+1)\varphi]
\end{array}\right) \right. \cr && \left. \otimes \left(\begin{array}{c} J_{m-1}(\nu_{mk}r'/\rho)
\sin[(m-1)\varphi']-J_{m+1}(\nu_{mk}r'/\rho) \sin[(m+1)\varphi']
\\
J_{m-1}(\nu_{mk}r'/\rho)
\cos[(m-1)\varphi']+J_{m+1}(\nu_{mk}r'/\rho) \cos[(m+1)\varphi']
\end{array}\right) \right. \cr && \left. +
\left(\begin{array}{c} J_{m-1}(\nu_{mk}r/\rho)
\cos[(m-1)\varphi]-J_{m+1}(\nu_{mk}r/\rho) \cos[(m+1)\varphi]
\\
-J_{m-1}(\nu_{mk}r/\rho)
\sin[(m-1)\varphi]-J_{m+1}(\nu_{mk}r/\rho) \sin[(m+1)\varphi]
\end{array}\right) \right. \cr && \left. \otimes
\left(\begin{array}{c} J_{m-1}(\nu_{mk}r'/\rho)
\cos[(m-1)\varphi']-J_{m+1}(\nu_{mk}r'/\rho) \cos[(m+1)\varphi']
\\
-J_{m-1}(\nu_{mk}r'/\rho)
\sin[(m-1)\varphi']-J_{m+1}(\nu_{mk}r'/\rho) \sin[(m+1)\varphi']
\end{array}\right)\right\}~,
\label{Gfexcir}
\end{eqnarray}
where we already dropped TE modes. Eq. (\ref{Gfexcir}) is valid in
the case of a circular waveguide.  These facts are demonstrated in
\cite{OURW}. As a result we have

\begin{eqnarray}
&&\partial'_\beta ~G^1_\beta= -\frac{c}{4 i \omega}
\sum_{m=0}^{\infty}\sum_{k=1}^{\infty} \left(A^{TM}_{mk}\right)^2
\left(\frac{\nu_{mk}}{\rho }\right)^3 \exp\left[-\frac{i c
(z-z')}{2\omega \rho^2 } \nu_{mk}^2\right]
J_{m}\left(\frac{\nu_{mk}r'}{\rho}\right)\cr && \times
\left\{\left[J_{m-1}\left(\frac{\nu_{mk}r}{\rho}\right)
\sin[(m-1)\varphi]-J_{m+1}\left(\frac{\nu_{mk}r}{\rho}\right)
\sin[(m+1)\varphi] \right]\sin(m\varphi')\right.\cr &&\left.
~+\left[J_{m-1}\left(\frac{\nu_{mk}r}{\rho}\right)
\cos[(m-1)\varphi]-J_{m+1}\left(\frac{\nu_{mk}r}{\rho}\right)
\cos[(m+1)\varphi] \right]\cos(m\varphi')\right\} \cr &&
\label{Gfexcir2x}
\end{eqnarray}
along $\vec{e}_x$ and

\begin{eqnarray}
&&\partial'_\beta ~G^2_\beta= -\frac{c}{4 i \omega}
\sum_{m=0}^{\infty}\sum_{k=1}^{\infty} \left(A^{TM}_{mk}\right)^2
\left(\frac{\nu_{mk}}{\rho }\right)^3 \exp\left[-\frac{i c
(z-z')}{2\omega \rho^2 } \nu_{mk}^2\right]
J_{m}\left(\frac{\nu_{mk}r'}{\rho}\right)\cr && \times
\left\{\left[J_{m-1}\left(\frac{\nu_{mk}r}{\rho}\right)
\cos[(m-1)\varphi]+J_{m+1}\left(\frac{\nu_{mk}r}{\rho}\right)
\cos[(m+1)\varphi] \right]\sin(m\varphi')\right.\cr &&\left.
~-\left[J_{m-1}\left(\frac{\nu_{mk}r}{\rho}\right)
\sin[(m-1)\varphi]+J_{m+1}\left(\frac{\nu_{mk}r}{\rho}\right)
\sin[(m+1)\varphi] \right]\cos(m\varphi')\right\}
\label{Gfexcir2y}
\end{eqnarray}
along $\vec{e}_y$. A major simplification arises since
$\vec{r}_0(z') = 0$. Because of this, the only non-zero
contribution to Eq. (\ref{Gfexcir2x}) and Eq. (\ref{Gfexcir2y}) is
for $m = 0$, otherwise $J_m(\nu_{mk} r'/\rho)$ yields zero.

Substitution in Eq. (\ref{efielG2}) yields, therefore, the
following result for the field at $z=L/2$:

\begin{eqnarray}
\vec{\widetilde{E}} =  -{2 i e} \sqrt{\frac{\omega}{L c^3}}
\sum_{k=1}^{\infty} \mathcal{A}_k^\nu\left(\frac{L}{2}\right)
J_1\left(\frac{\nu_{0k} r}{\rho}\right)
\vec{e}_r\label{guidefinal}
\end{eqnarray}
where

\begin{eqnarray}
\mathcal{A}_k^\nu({L}/{2}) = \sqrt{\left(\frac{L
c}{\omega}\right)^3}\frac{ \nu_{0k} \exp\left[-i C_k^\nu
{L}/{2}\right]}{\rho^3 J_1^2(\nu_{0k})}
\mathrm{sinc}\left[\frac{L}{2}
C_k^\nu+\frac{\phi}{4}\right]~.\label{coeffguide}
\end{eqnarray}
Here we defined $C_k^\nu = \nu_{0k}^2 c/(2\omega \rho^2)$ and
$\vec{e}_r = \sin(\varphi) \vec{e}_x + \cos(\varphi)\vec{e}_y$.
Eq. (\ref{guidefinal}) and Eq. (\ref{coeffguide}) solve the
problem of characterization of ER in the setup of Fig. \ref{guide}
for $\delta \ll 1$.

It is interesting to show that the treatment of edge radiation in
terms of virtual sources has a straightforward generalization for
the case of a waveguide.

It is convenient to introduce the subject by first noting that Eq.
(\ref{pb1}), the non-homogenous Maxwell's equation  with boundary
conditions at the waveguide walls, admits a self-reproducing
solution. Because of the circular cross section of the waveguide,
there is no preferential direction on the transverse plane. As a
result, the transverse electric field must be radially polarized,
and axially symmetric. We make the ansatz (i) $\vec{\widetilde{E}}
= E_r(r) E_z(z) \vec{e}_r$, and (ii) $E_z(z) = \exp[i z/(2
\gamma^2 \lambdabar)]$. Note that (i) is already equivalent to
postulating a self-reproducing field, meaning that the transverse
shape of the field is independent of the longitudinal coordinate
$z$ down the waveguide. Substituting (i) and (ii) in Eq.
(\ref{pb1}) and shifting to polar coordinates we obtain the
following problem for $E_r(\tilde{r})$:

\begin{eqnarray}
&& \tilde{r} \frac{d^2 E_r}{d \tilde{r}^2} + \frac{d E_r}{d
\tilde{r}} - \left(\tilde{r}+\frac{1}{\tilde{r}}\right) E_r =
\frac{4 \omega e}{c^2 \gamma} \tilde{r} \frac{d}{d\tilde{r}}
\left[\frac{\delta(\tilde{r})}{\tilde{r}}\right]~,\cr &&
\frac{d}{d\tilde{r}}\left(\tilde{r}
E_r\right)_{|_{\tilde{r}=\tilde{\rho}}}=0~, \label{pbb1}
\end{eqnarray}
where we introduced $\tilde{r} = r/(\gamma \lambdabar)$ and
$\tilde{\rho} = \rho/(\gamma\lambdabar)$ ad hoc to simplify
notations in Eq. (\ref{pbb1}). Solution of Eq. (\ref{pbb1}) yields
the final result

\begin{eqnarray}
\vec{\widetilde{E}}_\mathrm{s-rep} = - \frac{2 \omega e}{c^2
\gamma} \exp\left[\frac{i \omega  z}{2 c \gamma^2}\right]
\vec{e}_r \left[K_1\left(\frac{\omega r}{c \gamma}\right)+
\frac{K_0\left(\frac{\omega \rho}{c \gamma}\right)}
{I_0\left(\frac{\omega \rho}{c \gamma}\right)}
I_1\left(\frac{\omega r}{c \gamma}\right)\right]~.\label{vettoE}
\end{eqnarray}
Eq. (\ref{vettoE}) is well-known (see, e.g. \cite{HEIF}), and
explicitly describes the self-reproducing nature of the field,
because the transverse structure of the radiation field is fixed
at any value $z$.

At each longitudinal coordinate $z_i$, the radiation field can be
expanded in a series of modes of the passive (empty) waveguide.
Although we have not calculated explicitly the coefficients in
this expansion, these are independent of $z$, as $z=z_i$ has been
fixed. This leads to a paradox. In fact, one can propagate each
waveguide mode till another coordinate $z_f$. However, each mode
of the passive waveguide has its own phase velocity. Therefore,
one will not recover Eq. (\ref{vettoE}) at position $z_f$, because
relative phases of modes have changed, at that position. The
paradox is solved noting that the field in Eq. (\ref{vettoE})
represents a result for an active (loaded) waveguide. In other
words, to obtain back Eq. (\ref{vettoE}), one should add to the
propagated field, the additional field radiated by the electron
beam between the points $z_i$ and $z_f$, which is nothing but the
edge radiation contribution from the setup in Fig. (\ref{guide})
with $A=z_i$ and $B=z_f$. In other words, in order to recover the
self-reproducing field structure, we should add to the right hand
side of the propagation equation

\begin{eqnarray}
\widetilde{E}^\alpha (z_f, \vec{r}_f) = -\frac{2i \omega }{c} \int
d \vec{r'} \widetilde{E}^\alpha (z_i, \vec{r'}) G^\alpha_\beta
(\vec{r}_f,\vec{r'}, z_f-z_i)~, \label{propaaa}
\end{eqnarray}
where $G$ is given in Eq. (\ref{Gfexcir}). Note that Eq.
(\ref{propaaa}) generalizes Eq. (\ref{fieldpropback}). It follows
that

\begin{eqnarray}
\widetilde{E}_\mathrm{s-rep}^\alpha (z_f, \vec{r}_f) &=& -\frac{2i
\omega }{c} \int d \vec{r'} \widetilde{E}_\mathrm{s-rep}^\alpha
(z_i, \vec{r'}) G^\alpha_\beta (\vec{r}_f,\vec{r'}, z_f-z_i) \cr
&&+ \frac{4 \pi e}{c} \int_{z_i}^{z_f} dz' \left[\partial'_\beta
G^\alpha_\beta(\vec{r}_f,\vec{r'},z_f-z')\right]_{\vec{r'}=0}
\exp\left[\frac{i \omega  z'}{2 c \gamma^2}\right]~.
\label{propaaa2}
\end{eqnarray}
Note that the second term on the right hand side of Eq.
(\ref{propaaa2}) is nothing but the edge radiation contribution
from $z_i$ to $z_f$ (for $z_i=-L/2$ and $z_f=L/2$ one may compare
with Eq. (\ref{efielG2})). Eq. (\ref{propaaa2}) can be rewritten
as

\begin{eqnarray}
\vec{\widetilde{E}}_\mathrm{ER}(L/2,\vec{r}) &&= \frac{4 \pi e}{c}
\int_{z_i}^{z_f} dz' \left[\partial'_\beta
G^\alpha_\beta(\vec{r}_f,\vec{r'},z_f-z')\right]_{\vec{r'}=0}
\exp\left[\frac{i \omega z'}{2 c\gamma^2}\right] \cr && =
\widetilde{E}_{s2,\vec{r}}^\alpha (L/2) - \frac{2i\omega}{c} \int
d \vec{r'} \widetilde{E}_{s1}^\alpha (-L/2, \vec{r'})
G^\alpha_\beta (\vec{r},\vec{r'}, L)~, \label{propaaaf}
\end{eqnarray}
where we defined $\widetilde{\vec{E}}_{s1} (-L/2)= -
\widetilde{\vec{E}}_\mathrm{s-rep} (-L/2)$ and
$\widetilde{\vec{E}}_{s2} (L/2) =
\widetilde{\vec{E}}_\mathrm{s-rep} (L/2)$. Having introduced this
notation we are in the position to see that the treatment of edge
radiation in terms of virtual sources has a straightforward
generalization for the case of a waveguide. In fact, Eq.
(\ref{propaaaf}) means that the edge radiation field can be
interpreted as the sum of a given field distribution
$\widetilde{E}_{s2,\vec{r}}^\alpha (L/2)$ at $z=L/2$ plus the
field obtained propagating another field distribution,
$\widetilde{E}_{s1}^\alpha (-L/2, \vec{r'})$, from $z=-L/2$ up to
$z=L/2$ according to Eq. (\ref{propaaa}). We thus
straightforwardly interpret $\widetilde{\vec{E}}_{s1,s2}$ as the
generalized virtual sources at $z=-L/2$ and $z=L/2$ respectively.
Note that in the limit for $\rho \longrightarrow \infty$, the
self-reproducing field solution in Eq. (\ref{vettoE}) is related
to the virtual source fields in Eq. (\ref{virpm05}) consistently
with what has just been stated.

Also note that Eq. (\ref{propaaaf}) and Eq. (\ref{guidefinal})
present the same field, and must coincide. This hints to the fact
that the sum in Eq. (\ref{guidefinal}) can be calculated
analytically. Since the use of virtual sources provides conceptual
insight and should facilitate the design of ER setups, it is
instructive to explicitly demonstrate the equivalence of Eq.
(\ref{propaaaf}) and Eq. (\ref{guidefinal}). This is demonstrated
in Appendix A.

\subsection{Analysis of results (exemplifications)}

We now analyze our main results, Eq. (\ref{guidefinal}) and Eq.
(\ref{coeffguide}). To this purpose, it is convenient to re-write
these equations in normalized units, so that:

\begin{eqnarray}
\vec{\hat{E}} =  - 2 i \sum_{k=1}^{\infty}
\mathcal{A}_k^\nu(\hat{z}) J_1\left(\frac{\nu_{0k}
\hat{r}}{\sqrt{\Omega}}\right) \vec{e}_r\label{guidenorm}
\end{eqnarray}
with

\begin{eqnarray}
\mathcal{A}_k^\nu(\hat{z}) = \frac{\nu_{0k} \exp[-i \hat{C}_k^\nu
\hat{z}]}{\Omega^{3/2} J_1^2(\nu_{0k})}
\mathrm{sinc}\left[\frac{1}{2}
\hat{C}_k^\nu+\frac{\phi}{4}\right]~,\label{coeffnorm}
\end{eqnarray}
where we defined $\hat{C}_k^\nu = {C}_k^\nu L =
\nu_{0k}^2/(2\Omega)$, we reminded $\Omega = \rho^2/(\lambdabar
L)$, and normalized quantities $\vec{\hat{E}}$, ${\hat{r}}$ and
$\hat{z}$ are defined as in unbounded space.

Note that, with reference to Fig. \ref{guide}, we are assuming
$\hat{z}<1/2$, because for $\hat{z}>1/2$ we are already in
unbounded space. Using the Green's function $G^\alpha_\beta$ we
can formally propagate the field at $\hat{z}=1/2$, i.e. Eq.
(\ref{guidefinal}), up to positions $\hat{z}<1/2$. Note that,
since Eq. (\ref{guidefinal}) is presented as a sum of
empty-waveguide modes, this propagation just modifies the
exponential (phase) factor, so that instead of $\exp[-i
\hat{C}_k^\nu /2]$ we find $\exp[-i \hat{C}_k^\nu \hat{z}]$ in Eq.
(\ref{guidenorm}). Then, Eq. (\ref{guidenorm}) assumes the meaning
of a virtual field distribution in analogy with the
unbounded-space case.

Also note that the $\mathrm{sinc}(\cdot)$ function in the
expression for $\mathcal{A}^{\nu}_k$ is a direct consequence of
our model of the setup of sharp-edges at positions $\hat{z}=-1/2$
and $\hat{z}=1/2$ respectively. In fact, the
$\mathrm{sinc}({\cdot})$ function is the Fourier transform of a
rectangular function with respect to $\hat{C}^\nu_{k} +\phi/2$,
modelling the  sharp switch on and switch off of the bunch
harmonics. The presence of high frequency components in the
rectangular function implies the presence of contributions with
high values of $k$. In its turn, each contribution with large
value of $k$ in Eq. (\ref{guidenorm}) can be interpreted as a
superposition of plane waves propagating at angles
$\nu_{0k}/\sqrt{\Omega}$ in units of the diffraction angle
$\sqrt{\lambdabar/L}$. Thus, higher values of $k$ correspond to
the introduction of high spatial frequency components in the
field.

However, we should account for the fact that our theory applies
with a finite accuracy related to the value of the parameter
$\delta$. When using this approximation we neglect contributions
to the field with an accuracy related with $\delta$.  This means
that, in the analysis of our results, it does not make sense to
consider high spatial frequency contributions due to abrupt
switching of the bending magnet fields on a scale shorter than
$\sqrt[3]{R^2 \lambdabar}$, because these are outside of the
accuracy of the sharp-edge approximation. We may then introduce a
spatial frequency filter in our expression for the field by
replacing Eq. (\ref{efielG2}) with

\begin{eqnarray}
\widetilde{E}^\alpha &=& \frac{4\pi e}{c} \int_{-\infty}^{z} dz'
S(z') \left[\partial'_\beta
G^\alpha_\beta(\vec{r},\vec{r'},z-z')\right]_{\vec{r'}=0}
\exp\left[\frac{i \omega z'}{2 c
\gamma^2}\right]~,\label{efielG2s}
\end{eqnarray}

where the function $S(z')$ introduces some smoothing of the
rectangular profile on a scale $\sqrt[3]{R^2 \lambdabar}$. In
normalized units Eq. (\ref{efielG2s}) reads

\begin{eqnarray}
\hat{E}^\alpha &=& {4\pi} \int_{-\infty}^{\hat{z}} d\hat{z}'
S(\hat{z}') \left[\partial'_\beta
\hat{G}^\alpha_\beta\left(\vec{\hat{r}},\vec{\hat{r}'},
\hat{z}-\hat{z}'\right)\right]_{|_{\vec{\hat{r}'}=0}}
\exp\left[\frac{i \phi \hat{z}'}{2}\right]~,\label{efielG2sn}
\end{eqnarray}
where derivatives are now calculated with respect to normalized
quantities.

We model $S(\hat{z}')$ as a constant function along the straight
section length with exponentially decaying
edges\footnote{Formally, such model is in contrast with the
previous statement that for $\hat{z}>1/2$ we are in unbounded
space. However, we introduce $\Delta$ to account for a finite
accuracy of the sharp-edge approximation. Thus, it does not make
sense to specify where the unbounded space begins within a
distance $\Delta$.} on a typical (normalized) distance $\Delta$:

\begin{eqnarray}
S(\hat{z}') =  \left\{\begin{array}{cl}
\exp\left[-(\hat{z}'+1/2)^2/(2\Delta^2)\right] &~~~~\mathrm{for}~
\hat{z}'<-1/2
\\ 1&~~~~\mathrm{for}~ -1/2<\hat{z}'<1/2
\\ \exp\left[-(\hat{z}'-1/2)^2/(2
\Delta^2)\right] &~~~~\mathrm{for}~ \hat{z}'>1/2
\end{array}\right.~.\label{S1n}
\end{eqnarray}
It follows that Eq. (\ref{coeffnorm}) should be
replaced by

\begin{figure}
\begin{center}
\includegraphics*[width=120mm]{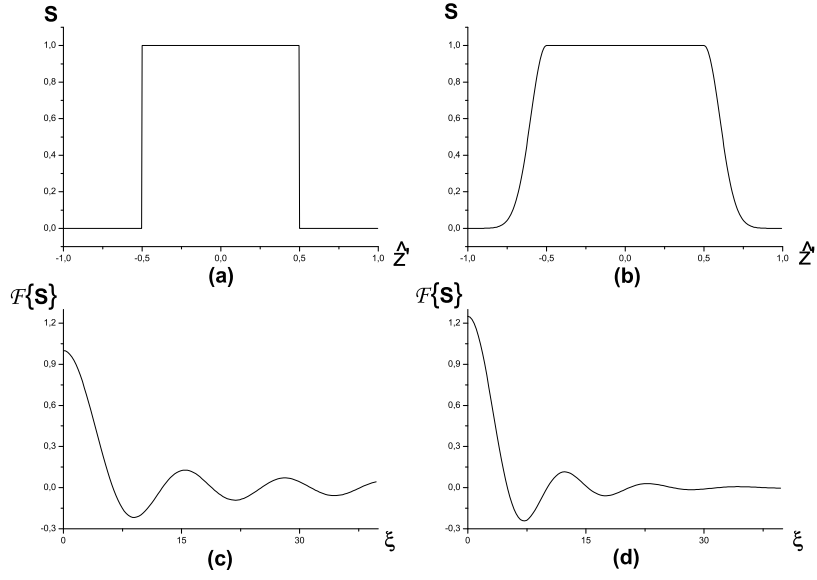}
\caption{Comparison between hard-edge case and high-frequency
filtering. Function $S(\hat{z}')$ in the hard-edge case with
${\Delta}=0$ (a) and in the case when filtering is applied with
${\Delta}=0.1$ (b). Their Fourier transforms $\mathcal{F}\{S\}$
with respect to $\xi = \hat{C}^{\mu,\nu}_k +\hat{C}>0$ are plotted
in (c) for the hard edge case and in (d) when filtering is
present. \label{SFTS}}
\end{center}
\end{figure}
\begin{figure}
\begin{center}
\includegraphics*[width=120mm]{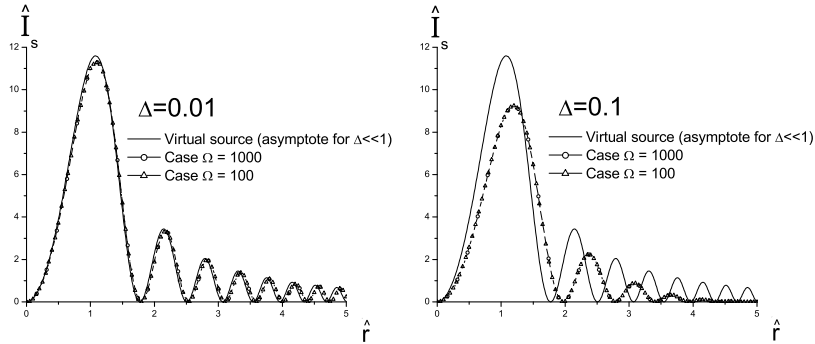}
\caption{Intensity profiles of the virtual source ($\hat{z}=0$) at
large values of $\Omega = R^2/(\lambdabar L_w)$ for different
values of $\Delta$ in the limit for $\hat{\phi}\ll 1$. These plots
are obtained from Eq. (\ref{guidenorm}) and Eq. (\ref{Anun}),
while the asymptotic limit for $\Delta \ll 1$ is found with the
help of Eq. (\ref{vir4ei}). \label{Ohigh}}
\end{center}
\end{figure}

\begin{eqnarray}
\mathcal{A}^{\nu}_k(\hat{z}) = \frac{\nu_{0k} \exp[-i \hat{C}_k^\nu
\hat{z}]}{\Omega^{3/2} J_1^2(\nu_{0k})}
\mathcal{F}\left\{S(\hat{z}'),\left(\hat{C}^\nu_k
+\phi/2\right)\right\}~, \label{Anun}
\end{eqnarray}
where $\mathcal{F}\{S(\hat{z}'),(\hat{C}^{\mu,\nu}_k +\phi/2)\}$
is the Fourier transform of the function $S$ with respect to
$(\hat{C}^{\mu,\nu}_k +\phi/2)$. The introduction of the function
$S(\hat{z}')$ introduces a suppression of higher frequency
components of $\mathcal{F}\{S\}$, and corresponds to a suppression
of higher spatial frequencies in the field distribution, i.e. to a
smoother field distribution. For a fixed value of $\Omega$ and
$\hat{z}$ we qualitatively expect finite and smooth field
distribution for finite values of ${\Delta}$. On the contrary, in
the sharp-edge limit for ${\Delta} \longrightarrow 0$, the field
must diverge for some value of $\hat{r}$, as a result of the
already discussed divergence of the integrated flux and of the
fact that the radius of the vacuum chamber is finite. Note that
the level of high spatial frequencies depends in a complicated way
on $\Omega$, because the perfectly metallic waveguide effectively
acts as a mirror. The value of $\Delta$ should be actually chosen
to cut off high spatial frequencies that are outside the region of
applicability of the sharp-edge approximation, i.e. $\Delta \sim
\delta$. Thus, the correct value of $\Delta$ depends, case by
case, on the type of setup considered. The Fourier transform of
the function $S(\hat{z}')$ in Eq. (\ref{S1n}) to be inserted into
Eq. (\ref{Anun}) reads

\begin{eqnarray}
\mathcal{F}&&\left\{S(\hat{z}'),\left(\hat{C}^{\nu,\mu}_k
+\frac{\phi}{2}\right)\right\} =
\mathrm{sinc}\left[\frac{\hat{C}^{\nu,\mu}_k}{2}+\frac{\phi}{4}\right]+
\sqrt{2\pi}~ {\Delta}
\exp\left[-\frac{{\Delta}^2}{2}\left({\hat{C}^{\nu,\mu}_k}+\frac{\phi}{2}\right)^2\right]
\cr && \times
\left\{\cos\left[\frac{\hat{C}^{\nu,\mu}_k}{2}+\frac{\phi}{4}\right]-\sin\left[\frac{\hat{C}^{\nu,\mu}_k}{2}+\frac{\phi}{4}\right]
\mathrm{erfi}\left[\frac{{\Delta}}{\sqrt{2}}\left({\hat{C}^{\nu,\mu}_k}+\frac{\phi}{2}\right)\right]\right\}
~,\label{FTSn}
\end{eqnarray}
where the imaginary error function $\mathrm{erfi}(\cdot)$ is
defined as

\begin{eqnarray}
\mathrm{erfi}\left(z\right) = \frac{1}{i} \mathrm{erf}(i z) =
\frac{2}{i\sqrt{\pi}} \int_{0}^{i z} \exp{[-t^2]} dt
~.\label{erfidef}
\end{eqnarray}

Fig. \ref{SFTS} presents a comparison between functions $S$ and
$\mathcal{F}\{S\}$ for ${\Delta}=0.1$ and ${\Delta}=0$. A filtering effect
can be clearly seen, suppressing higher frequency components of
$\mathcal{F}\{S\}$.

Eq. (\ref{guidenorm}) and Eq. (\ref{coeffnorm}) completely
describe ER in the presence of a circular waveguide. Position
$\hat{z}=0$ corresponds to the single virtual source description
discussed before. In fact, as we have seen, the concept of
laser-like radiation beam can be naturally extended to the case a
waveguide is present. As said before one must carefully select the
number of modes used for computation, according to $k \gg
\sqrt{\Omega}$.

\begin{figure}
\begin{center}
\includegraphics*[width=120mm]{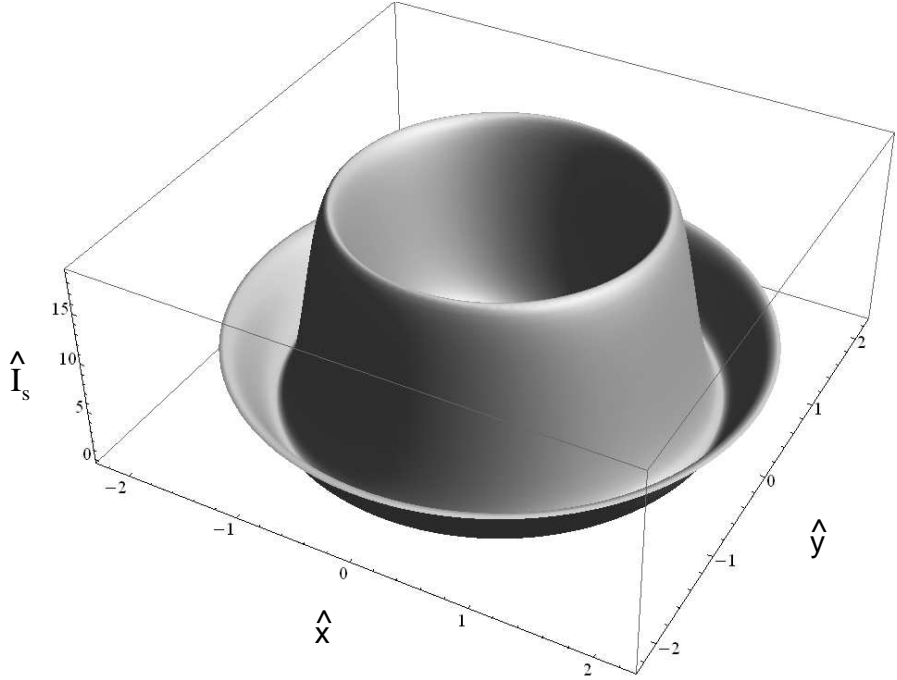}
\caption{Three-dimensional plot of the intensity profile of the
virtual source ($\hat{z}=0$) at $\Omega = R^2/(\lambdabar L_w) =
5$ for $\Delta=0.1$ in the limit for $\hat{\phi}\ll 1$.
\label{Nuova3D}}
\end{center}
\end{figure}

\begin{figure}
\begin{center}
\includegraphics*[width=100mm]{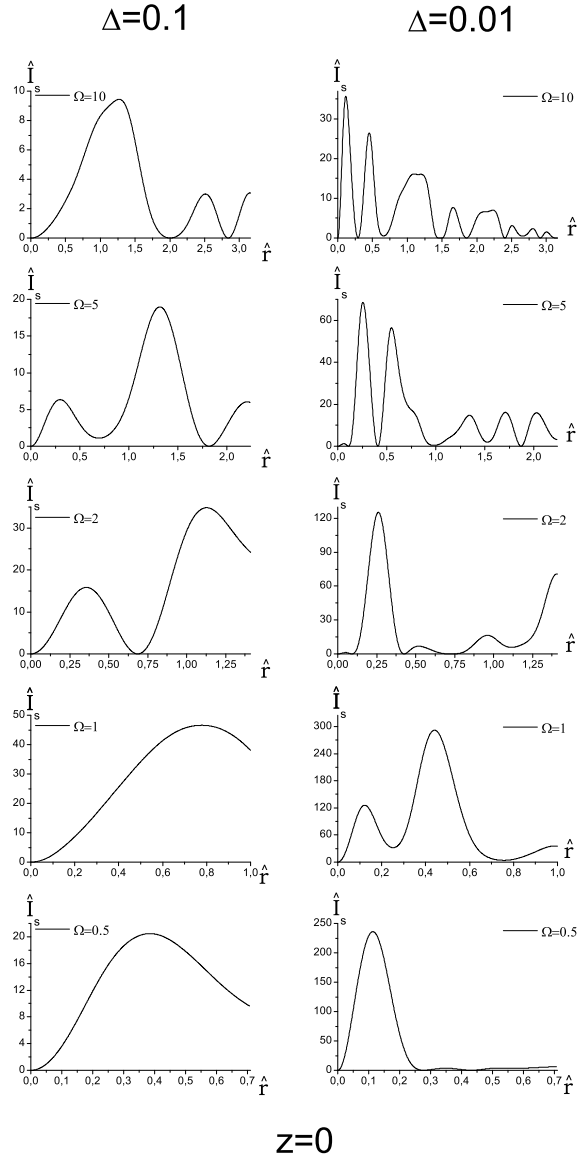}
\caption{Intensity profiles of the virtual source ($\hat{z}=0$) at
different values of $\Omega=R^2/(\lambdabar L_w)$ for
$\hat{\phi}\ll 1$ at two values $\Delta = 0.1$ and $\Delta =
0.01$. \label{z0p0}}
\end{center}
\end{figure}
\begin{figure}
\begin{center}
\includegraphics*[width=100mm]{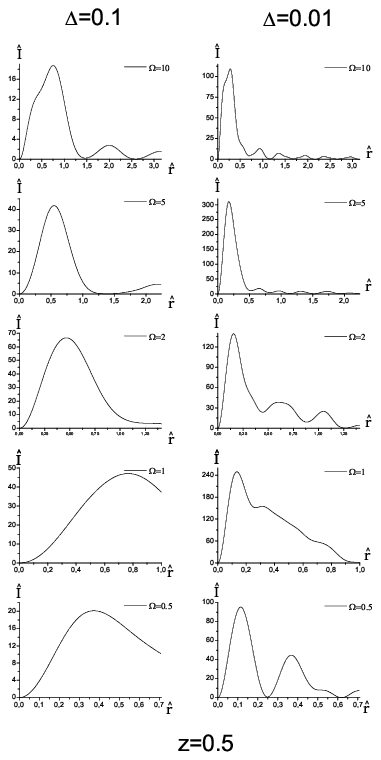}
\caption{Intensity profiles at $\hat{z}=1/2$ at different values
of $\Omega=R^2/(\lambdabar L_w)$ for $\hat{\phi}\ll 1$ at two
values $\Delta = 0.1$ and $\Delta = 0.01$. \label{z0p5}}
\end{center}
\end{figure}
We further consider, as an example, the limit for $\phi \ll 1$. In
Fig. \ref{Ohigh} we show a comparison between the analytic
expression for the intensity distribution $\hat{I}$ at the virtual
source position, and numerical expressions obtained through Eq.
(\ref{guidenorm}) for different values of ${\Delta}$ at
$\Omega=1000$ and $\Omega=100$.  Differences between results for
${\Delta}=0.1$ and Eq. (\ref{vir4ei}) should be taken as
exemplification of the filtering process.  A three-dimensional
view of the intensity profile of the virtual source for $\Omega=5$
and $\Delta = 0.1$ is shown in Fig. \ref{Nuova3D}, to be compared
with Fig. \ref{viredge} for the unbounded-space limit. The number
of modes used in all plots are in all cases much larger than
$\sqrt{\Omega}$. Furthermore, we verified that results do not
change by changing the number of modes used in the computation,
provided that condition $k_\mathrm{max} \gg \sqrt{\Omega}$ is
fulfilled. We also checked that the same results obtained with the
help of Eq. (\ref{guidenorm}) can also be recovered by introducing
apodization directly in Eq. (\ref{ABcontre}). This confirms that
numerical evaluation of Eq. (\ref{guidenorm}) is sound.

The behavior of the intensity profile at the virtual source when
waveguide effects are important is shown in Fig. \ref{z0p0} for
different values of $\Omega$ at $\Delta = 0.01$ $\Delta = 0.1$.
Further on, in Fig. \ref{z0p5} we plot the intensity distribution
$\hat{I} =|\hat{E}|^2$ as a function of the reduced radius
$\hat{r}$ for the same values of $\Delta$  and $\Omega$ at
$\hat{z}=1/2$. All plots are evaluated with the help of Eq.
(\ref{guidenorm}). It should be remarked that the wavefront at
$\hat{z}=1/2$ is not plane (as at $\hat{z}=0$) and the phase
distribution carries important information when one needs to
propagate the field from $\hat{z}=1/2$ to any other point. The
derived analytical algorithm in Eq. (\ref{guidenorm}) enables one
to calculate all these informations.

Comparison of the cases for $\Delta=0.01$ and $\Delta = 0.1$
clearly shows the effect of different edge dimensions,
corresponding to filtering of high spatial frequencies. It should
be reminded that when $\Delta = 0$ the angle-integrated spectral
flux of ER is divergent in the unbounded space, as well as in a
waveguide. Such divergency was motivated as an effect of the
sharp-edge approximation, and depends logarithmically on $\Delta$.

To sum up, knowledge of the field at $\hat{z}=1/2$ fulfill our
goal for this Section. Expression for the field at $\hat{z}=1/2$
can be interpreted as a virtual-field distribution. From this
viewpoint, the presence of a finite parameter $\Delta$ corresponds
to the presence of an angular filter, related to the violation of
ER expression for angles comparable with the bending magnet
radiation angle. The virtual-field distribution at $\hat{z}=1/2$
should be further propagated either in the unbounded space or
through any other kind of vacuum chamber up to the experimental
station. In other words, the knowledge of the field at the
straight section exit constitutes the starting point for further
propagation. Typically, a user-beamline vacuum chamber will be
connected with the electron vacuum chamber (that follows the
bending magnet after position $\hat{z}=1/2$). The user-beamline
vacuum chamber has a typical characteristic size which is larger
than the electron vacuum chamber. It follows that, after position
$\hat{z}=1/2$, the field can usually
be propagated as in unbounded space. 

\section{\label{sec:bunch} Coherent edge radiation}

The physics of radiation processes from electron beams critically
depends on the ratio of the bunch length to the wavelength of the
emitted radiation. We can thus define two opposite asymptotes for
the electron bunch length, corresponding to very different
characteristics of radiation.

In the first limit, when the electron bunch is long compared to
the radiation wavelength, one deals with the conventional case of
spontaneous radiation. In this regime the electron phases are not
correlated, i.e. distances between electrons are randomly
distributed on a characteristic scale much longer than the
radiation wavelength. Incoherent superposition of each electron
contribution to the radiation field yields low-power emission,
proportional to the charge in the bunch, with the usual incoherent
phase-noise statistics.

In the second limit, when the electron bunch is short compared to
the radiation wavelength, one deals with the case of coherent
radiation. Electron phases are correlated, in the sense that
random variations of distances between electrons are distributed
on a characteristic scale shorter than the radiation wavelength.
The bunch essentially behaves as a single point-charge, which
coherently emits high-power radiation. In this case, the radiation
power scales quadratically with the total electric charge in the
bunch, rather than linearly as in the other limit. It is because
of this quadratic effect that the radiation power can be greatly
enhanced, because the bunch population is usually of order
$10^{10}$ particles.

Coherent radiation is usually not emitted in electron storage
rings, because the bunch length is of order of a centimeter.
However, in linear accelerators, much shorter electron bunches of
the order of $100 \mu$m can be produced in magnetic bunch
compressors. In this Section we will discuss the physics of
coherent ER distinguishing two cases, when coherent emission is
due to an overall temporal bunch profile shorter than the
wavelength or to short structures (microbunches of order of the
radiation wavelength) superimposed to the temporal profile of the
bunch.

When it comes to the use of ER as a diagnostic tool, the
availability of a much larger number of photons constitutes an
important advantage of coherent ER compared to the incoherent
case. By detecting coherent ER, characteristics of the electron
bunch such as their three-dimensional distribution and divergence
can be measured. Moreover, coherent radiation presents the unique
feature that electron-beam microbunching can be investigated too.

\subsection{Review of known methods}

There are two equivalent approaches to describe coherent SR in
general: a Lagrangian approach and a Eulerian approach.

In the Lagrangian approach particles are labelled with a given
index, and the motion of individual charges is followed through
space. One follows the evolution of each particle as a function of
energy deviation, angular direction, position and arrival time at
a given reference (longitudinal) position. Knowing the evolution
of each particle, individual contributions to the field are
separately calculated and summed up.

In the Eulerian approach instead, one begins defining charge
density $\rho(\vec{r},z,t)$ and current density
$\vec{j}(\vec{r},z,t)$, i.e. particular field quantities, as a
function of coordinates and time. One is no more interested in
knowing the evolution of each single particle but only to observe
charges passing through a given longitudinal position, in order to
obtain information about $\rho$ and $\vec{j}$. Once these two
fields are determined, one can solve Maxwell's equations for the
electric field, with $\rho$ and $\vec{j}$ as macroscopic
electromagnetic sources.

\subsubsection{Lagrangian approach}

We consider an electron bunch moving along the setup specified in
Fig. \ref{geome}(a).

Given a reference point, e.g. $z=0$, along the trajectory one may
specify the phase-space coordinates of a given electron with the
help of six data: position of the electron, $\vec{l}$, angular
deflection $\vec{\eta}$, energy  $\delta \gamma$, and arrival
time, $\tau$, all relative to the position of a reference electron
with position $\vec{r}_0(0)$, transverse velocity $\vec{v}_{\bot
0}(0)$, nominal energy $\mathcal{E}_0$ and arrival time $t=0$, all
given at position $z=0$. Using the index $k$ to identify a certain
particle one finds the temporal Fourier transform of the radiation
pulse at a given frequency $\omega$ as a sum of many independent
contributions:

\begin{eqnarray}
\vec{\bar{E}}(\vec{r},z,\omega) = \sum_{k=1}^{N_e}
\vec{\bar{E}}_{(1)}\left(\vec{l}_k,\vec{\eta}_k,\delta
\gamma_k,\tau_k;\vec{r},z,\omega\right)~, \label{barE}
\end{eqnarray}
where $N_e$ is the total number of particles in the bunch, and
$\vec{\bar{E}}_{(1)}$ is the single particle field contribution
from a given particle identified by variables
$\vec{l}_k,\vec{\eta}_k,\delta \gamma_k,\tau_k$. Note that we
separated these variables from the space-frequency coordinates
where the field is calculated with a semicolon.

Eq. (\ref{barE}) forms the basis for calculating the angular
spectral flux of radiation. This is essentially done by taking the
square modulus of the field in Eq. (\ref{barE}) and averaging over
an ensemble of bunches:

\begin{eqnarray}
\frac{dW}{d\omega d \Omega} = \frac{c z^2}{4\pi^2} \left\langle
\left|\vec{\widetilde{E}}\right|^2 \right\rangle = \frac{c
z^2}{4\pi^2} \left\langle \left|\sum_{k=1}^{N_e}
\vec{\widetilde{E}}_{(1)}\left(\vec{l}_k,\vec{\eta}_k,\delta
\gamma_k,\tau_k;\vec{r},z,\omega\right) \right|^2 \right\rangle~.
\label{endenegen00}
\end{eqnarray}
Let us show that in the particular case when no focusing elements
are present, $\vec{\widetilde{E}}_{(1)}$ in the far zone depends
on $\vec{l}_k$ through a phase only. To this purpose, it is
sufficient to perform the substitution $\vec{r}_0(z')
\longrightarrow \vec{r}_0(z') + \vec{l}$ in Eq.
(\ref{generalfin}), which leads straightforwardly to the phase $-
\omega \vec{\theta}\cdot \vec{l}/c$.  Thus, in the particular case
when there are no focusing elements and electrons have random
arrival times (phases) and random offsets, but no energy spread,
nor deflections, Eq. (\ref{endenegen00}) yields

\begin{eqnarray}
&& \frac{dW}{d\omega d \Omega} =  \frac{c z^2}{4\pi^2}
\left\langle \sum_{m,n=1}^{N_e}
\left|\vec{\widetilde{E}}_{(1)}\left(\vec{r},z,\omega\right)\right|^2
\exp\left[i \omega (\tau_m-\tau_n)\right] \exp\left[-\frac{i
\omega}{c z} \vec{r}\cdot (\vec{l}_m-\vec{l}_n)\right]
\right\rangle \cr && = \frac{c z^2}{4\pi^2}
\left|\vec{\widetilde{E}}_{(1)}\left(\vec{r},z,\omega\right)\right|^2
\left\{N_e + \sum_{m \neq n}  \left \langle \exp\left[i \omega
(\tau_m-\tau_n)\right] \right\rangle \left\langle
\exp\left[-\frac{i \omega}{c z} \vec{r}\cdot
(\vec{l}_m-\vec{l}_n)\right] \right\rangle \right\} \cr && =
\frac{c z^2}{4\pi^2}
\left|\vec{\widetilde{E}}_{(1)}\left(\vec{r},z,\omega\right)\right|^2
\cdot N_e
\left[1+(N_e-1)\left|\bar{f}_t(\omega)\right|^2\left|\bar{f}_l\left(\frac{\omega
\vec{r}}{c z} \right)\right|^2\right]~, \label{endenegen0}
\end{eqnarray}
having recognized that if $f_t$ is the longitudinal bunch profile
and $f_{l}$ the offset distribution at the reference position
$z=0$, one has:

\begin{eqnarray}
\langle\exp(i\omega \tau_k)\rangle = \int_{-\infty}^{\infty} d
\tau_k ~f_t(\tau_k) \exp(i\omega \tau_k) \equiv
\bar{f}_t(\omega)\label{fttra}
\end{eqnarray}

and

\begin{eqnarray}
\left\langle\exp\left(-\frac{i \omega}{c z} \vec{r}\cdot
\vec{l}_k\right)\right\rangle = \int_{-\infty}^{\infty} d
\vec{l}_k ~f_l\left(\vec{l}_k\right) \exp\left(-\frac{i \omega}{c
z} \vec{r}\cdot \vec{l}_k\right) \equiv
\bar{f}_l\left(-\frac{\omega \vec{r}}{c z} \right) ~.\label{fltra}
\end{eqnarray}
In the rest of this Section we will mainly concentrate on the
second term in Eq. (\ref{endenegen0}), scaling as $\sim
N_e(N_e-1)$, i.e. the coherent term. Although Eq.
(\ref{endenegen0}) is known (see e.g. \cite{SHIB, ZHEN,SCHR}), we
choose to shortly report it, because it is still the object of
revisions in recently published articles \cite{ORL1,ORL2,ORL3},
where the angular spectral flux of spontaneous Transition
Radiation is found to obey a modified angular distribution
depending on the transverse distribution of electrons (see e.g.
Eq. (25) of \cite{ORL3}, dealing with the case of an infinitely
large mirror). In the next Section we will demonstrate that setups
in Fig. \ref{geome}(a) and Fig. \ref{geome}(c) can be treated with
the help of the same mathematical machinery, i.e. a Transition
Radiation setup can be formally dealt with as if it were an edge
radiation setup. While an exponential cutoff is to be expected in
the coherent angular spectral flux, results in
\cite{ORL1,ORL2,ORL3} indicating a dependence of the incoherent
angular spectral flux on the transverse distribution of electrons
is counterintuitive. In fact, as we have seen, the spectrum is
calculated by first summing single electron contributions to the
field and, second, by averaging the square modulus of the field
over an ensemble of bunches. As we have discussed above, a given
offset modifies the far-field contribution from a single electron
by a phase factor only. Our conclusion, in contrast with
\cite{ORL1,ORL2,ORL3}, is that the final incoherent spectrum
cannot depend on the transverse distribution of electrons. The
incorrectness of \cite{ORL1,ORL2,ORL3} is due to a calculus
mistake, which can be found e.g. in Eq. (19) of \cite{ORL3}: the
average of the square modulus of a quantity, in fact, is not equal
to the square modulus of the average.


Let us go back to the most general situation described in Eq.
(\ref{barE}). One can represent the sum in Eq. (\ref{barE}) in
terms of an integral over a continuous phase-space distribution of
electrons specified at a given position $z$. Let us assume for
simplicity that this reference position is $z=0$, i.e. in the
middle of the straight section. We will indicate the phase-space
distribution at $z=0$ with $f(\vec{l}, \vec{\eta}, \delta
\gamma,\tau;0)$. Eq. (\ref{barE}) can then be substituted with an
average of the single particle field over $f$, yielding

\begin{eqnarray}
\vec{\widetilde{E}}\left(\vec{r},z,\omega\right) &&=  N_e \int
d\vec{l} d\vec{\eta} d(\delta\gamma) d \tau~
f\left(\vec{l},\vec{\eta},{\delta \gamma},\tau\right)
\vec{\widetilde{E}}_{(1)}\left(\vec{l},\vec{\eta},{\delta
\gamma},\tau;\vec{r},z,\omega\right).\cr && \label{farangletot}
\end{eqnarray}
Note that substitution of Eq. (\ref{barE}) with Eq.
(\ref{farangletot}) rules out the possibility of recovering the
incoherent term in Eq. (\ref{endenegen0}). In fact, the continuous
phase-space distribution $f$ already includes an average over an
ensemble of bunches. In other words, Eq. (\ref{farangletot}) has
the meaning of a mean field, whose square modulus is essentially
the second term in Eq. (\ref{endenegen0}), i.e. the coherent term.

\subsubsection{Eulerian approach}

Up to now we have been using a microscopic approach, where the
single-electron field is averaged over the six-dimensional
phase-space distribution of electrons. It is possible to obtain
the same results using a macroscopic viewpoint, where solution of
Maxwell's equations is found with given macroscopic charge density
$\rho(\vec{r},z,t)$ and current density $\vec{j}(\vec{r},z,t)$. In
this case, one can integrate Maxwell's equations with the help of
the usual Green's function.

We begin by writing a general expression for the temporal Fourier
transform of charge and current density in terms of the
phase-space distribution at a given position down the beamline
$z$. Such phase space-distribution, $f(\vec{r},
\vec{v}_\bot/c,\delta\gamma, t,z)$ can always be written as
$f(\vec{l}+\vec{r}_0(z),\vec{\eta}+\vec{v}_{\bot 0}(z)/c,\delta
\gamma,\tau+s(z)/v,z)$ where, as we have seen before, the
subscript "0" referring to the constrained motion of an electron
with $\vec{r}=0$, $\vec{v}_\bot = 0$, $\delta \gamma = 0$ and
$t=0$ at $z=0$. The meaning of this writing is that here we
consider the evolution of $f$ along $z$ as the result of two
combined effects. On the one hand we have a z-dependent shift of
the $f$-distribution as a whole due to the reference motion
$\vec{r}_0(z)$, $\vec{v}_{\bot 0}(z)$ and $s(z)$, while on the
other hand we have a more generic evolution in $z$ due to
self-interactions and external fields. The distribution $f$ is
obviously of the form $f(\vec{l},\vec{\eta},\delta
\gamma,\tau,z)$. It follows that the Fourier transform of $f$,
$\bar{f}$, can be written as

\begin{eqnarray}
\bar{f}(\vec{l},\vec{\eta},\delta
\gamma,\omega,z)&&=\int_{-\infty}^{\infty} dt \exp[i\omega t]
f(\vec{l},\vec{\eta},\delta \gamma,\tau,z) \cr && =
\exp\left[\frac{i \omega s(z)}{v}\right]\int_{-\infty}^{\infty}
d\tau \exp[i\omega \tau] f(\vec{l},\vec{\eta},\delta
\gamma,\tau,z) \cr && \equiv \exp\left[\frac{i \omega
s(z)}{v}\right] \widetilde{f}(\vec{l},\vec{\eta},\delta
\gamma,\omega,z)~, \label{widetf}
\end{eqnarray}
where the usefulness in the definition of $\widetilde{f}$ is that
$\widetilde{f}$ is a slowly function of $z$ on the scale of
$\lambdabar$. One has therefore

\begin{eqnarray}
\bar{\rho}(\vec{r},z,\omega) && = \frac{(-e) N_e}{c} \int d\vec{l}
d \vec{\eta} d(\delta \gamma) \bar{f}(\vec{l},\vec{\eta},\delta
\gamma,\omega,z) \delta(\vec{l}+\vec{r}_0(z)-\vec{r}~)\cr &&
=\frac{(-e) N_e}{c} \int  d \vec{\eta} d(\delta \gamma)
\widetilde{f}(\vec{r}-\vec{r}_0(z),\vec{\eta},\delta
\gamma,\omega,z) \exp\left[\frac{i  \omega s(z)}{v}\right]\cr &&
\equiv \widetilde{\rho}(\vec{r}-\vec{r}_0(z),z,\omega)
\exp\left[\frac{i \omega s(z)}{v}\right]\label{chargetr}
\end{eqnarray}
and

\begin{eqnarray}
\vec{\bar{j}}(\vec{r},z,\omega) &&= \frac{(-e) N_e}{c} \int
d\vec{l} d \vec{\eta} d(\delta \gamma)  \left(c
\vec{\eta}+\vec{v}_{\bot}(z)\right)\bar{f}(\vec{l},\vec{\eta},\delta
\gamma,\omega,z) \delta(\vec{l}+\vec{r}_0(z)-\vec{r}~)\cr && =
\frac{(-e) N_e}{c} \int  d \vec{\eta} d(\delta \gamma) \left(c
\vec{\eta}+\vec{v}_{\bot}(z)\right)\widetilde{f}(\vec{r}-\vec{r}_0(z),\vec{\eta},\delta
\gamma,\omega,z) \exp\left[\frac{i \omega s(z)}{v}\right]\cr &&
\equiv \vec{\widetilde{j}}(\vec{r}-\vec{r}_0(z),z,\omega)
\exp\left[\frac{i \omega s(z)}{v}\right]~, \label{currtr}
\end{eqnarray}
where we introduced, together with $\widetilde{f}$ and with the
usual field envelope $\vec{\widetilde{E}} = \vec{\bar{E}}  \exp[-i
\omega z/c]$, also quantities $\widetilde{\rho} = \bar{\rho}
\exp[- i \omega s(z)/v]$ and $\vec{\widetilde{j}} = \vec{\bar{j}}
\exp[- i \omega s(z)/v]$. Note that $\widetilde{\rho} =
\widetilde{\rho}(\vec{r}-\vec{r}_0(z),z,\omega)$ (the same applies
to $\vec{\widetilde{j}}$). One writes the paraxial equation for
the field as

\begin{eqnarray}
\left({\nabla_\bot}^2 + {2 i \omega \over{c}}
{\partial\over{\partial z}}\right) \vec{\widetilde{E}} &=& {4 \pi}
\exp\left\{{i \omega
\left({s(z)\over{v}}-{z\over{c}}\right)}\right\}\cr && \times
\left[{i\omega\over{c^2}}\vec{\widetilde{j}}\left(\vec{r}-\vec{r}_0(z),z,\omega\right)
-\vec{\nabla}_\bot
\widetilde{\rho}\left(\vec{r}-\vec{r}_0(z),z,\omega\right)\right]
\label{incipit3}
\end{eqnarray}
in analogy with Eq. (\ref{incipit2}) for a single particle. We
solve Eq. (\ref{incipit3})  with the help of the Green's function
in Eq. (\ref{green}), similarly as in Eq. (\ref{efielGfree}), but
with the important difference that, now, we cannot carry out the
integration over transverse coordinates. We therefore obtain

\begin{eqnarray}
\vec{\widetilde{E}} &=& {4\pi} \int_{-\infty}^{z} dz' \int d
\vec{r'} \left\{ \frac{i\omega}{c^2}
\vec{\widetilde{j}}(\vec{r'}-\vec{r}_0(z'),z',\omega)
G\left(\vec{r}-\vec{r'}, z-z' \right)\right.\cr
&&\left.+\widetilde{\rho}(\vec{r'}-\vec{r}_0(z'),z',\omega)\vec{\nabla'}_\bot
G\left(\vec{r}-\vec{r'}, z-z'\right) \right\} \exp\left\{{i \omega
\left({s(z)\over{v}}-{z\over{c}}\right)}\right\}~,\label{efielGfreeg}
\end{eqnarray}
Explicit substitution of Eq. (\ref{green}) in Eq.
(\ref{efielGfreeg}) finally yields

\begin{eqnarray}
\vec{\widetilde{E}}(z, \vec{r}) &=& -\frac{i \omega}{c}
\int_{-\infty}^{z} dz' \int d \vec{r'} \frac{1}{z-z'}
\exp\left\{i\omega\left[{\mid \vec{r}-\vec{r'} \mid^2\over{2c
(z-z')}}+ \int_{0}^{z'}  \frac{d \bar{z}}{2 \gamma_{z}^2(\bar{z})
c}\right] \right\} \cr && \times
\left[\frac{\vec{\widetilde{j}}(\vec{r'}-\vec{r}_0(z'),z',\omega)}{c}
-{\vec{r}-\vec{r'}\over{z-z'}}\widetilde{\rho}(\vec{r'}-\vec{r}_0(z'),z',\omega)\right]
~. \label{generalfin2g}
\end{eqnarray}

Lagrangian and Eulerian approaches are equivalent. In simple
cases, e.g. when trajectories of electrons are a simple function
like a straight motion and Eq. (\ref{generalfin}) can be
calculated analytically, the Lagrangian method looks more
transparent. However, in the most general case, one needs to
account for the presence of focusing elements and
self-interactions. The evolution of the six-dimensional phase
space becomes complicated and numerical methods should be used.
Naturally, the Lagrangian approach can be applied when the
evolution of the six-dimensional phase-space is known via
simulations. Then, Eq. (\ref{generalfin}) has to be calculated
numerically for each (macro)particle. The Eulerian approach can be
more advantageous. Computer codes can calculate charge and current
density distribution as a function of $z$ accounting for focusing
elements and self-interactions, and the macroscopic method may be
more practical, because one integrates Eq. (\ref{generalfin2g})
once and for all. One does not need to have any explicit knowledge
of the angle distribution, or of the energy spread distribution to
calculate the field. Once charge and current density are known,
one only needs to (numerically) integrate Maxwell's equations.
This is because we are dealing with a macroscopic system, where
only macroscopic current and charge density are important. In
other words, the behavior of Maxwell's equations is influenced by
projections of the 6D phase-space on the real space ($\rho$ and
$\vec{j}$), but not by the 6D phase-space itself.

\subsection{Near Field Coherent Edge Radiation}

Let us consider the usual setup in Fig. \ref{geome}(a) and assume,
till the end of the Section, that there are no focusing elements
nor self-interactions along the straight section. By this, we
assume that electrons move along the straight line between two
bending magnets. We will adopt the Lagrangian approach to
characterize coherent ER emission due to overall temporal bunch
profile, without considering, for now, microbunching.

An algorithm for describing radiation emission is straightforward.
One begins with the expression for the field generated in the far
zone by a single electron occupying a certain position in
phase-space at a fixed reference longitudinal position $z$. Then,
one integrates the single-particle expression over the phase-space
distribution, thus getting the equivalent of Eq.
(\ref{farangletot}) in the far zone. Finally, when the far-zone
field (averaged over a given electron distribution) is known, the
field in the near-zone can be calculated as has been done before,
by back-propagating the far-zone expression.

In the case of edge radiation, the far-zone expression for the
field of a single-particle having phase-space coordinates
$(\vec{l},\vec{\eta},\delta \gamma,\tau)$ at $z=0$ is given by:

\begin{eqnarray}
\vec{\widetilde{E}}_{(1)}\left(\vec{l},\vec{\eta},{\delta
\gamma},\tau;\vec{\theta},z,\omega\right)&=&\frac{ i \omega  e
L}{c^2 z} \exp\left[{i \omega \tau}\right] \exp\left[\frac{i
\omega \theta^2 z}{2 c}\right] \exp\left[- \frac{i \omega
\vec{\theta}\cdot\vec{l}}{c}\right] \cr && \times
\left(\vec{\theta}-\vec{\eta}\right)
\mathrm{sinc}\left\{\frac{\omega L}{4
c}\left[\left|\vec{\theta}-\vec{\eta}\right|^2+\frac{1}{\gamma^2}-
\frac{2\delta\gamma}{\gamma^3}\right]\right\}~. \label{faranglee}
\end{eqnarray}
Note that Eq. (\ref{faranglee}) is a straightforward
generalization of Eq. (\ref{ABcontrint4e}), obtained by performing
a rigid rotation of the reference system and a translation of the
observation point \textit{without} changing the observation plane
of interest, which remains perpendicular to the $z$ axis of the
initial reference system. This explains why the combination
$\vec{\theta} - \vec{\eta}$ does not enter the phase in Eq.
(\ref{faranglee}). This solves the problem for the field in the
case when trajectories are straight lines between two bending
magnets without focusing elements, and self-interactions are
negligible.

The field in the far zone should be subsequently averaged over
$f$. This is accomplished by substituting Eq. (\ref{faranglee})
into Eq. (\ref{farangletot}).
Simplifications arise when the phase-space distribution can be
separated in product of different factors depending only on
offset, deflection, energy spread and arrival time for some choice
of the reference position $z$. Such separability is not always
granted. However,  the choice of some particular reference point
may help simplifying calculations. For example, the distribution
of offsets and deflections factorizes when the reference position
in $z$ is fixed at the minimal value of the betatron function. In
the following we will consider two different study cases of
interest.

\paragraph*{I. Case $f=f_l(\vec{l})\delta(\vec{\eta})
\delta(\delta\gamma/\gamma)f_\tau(\tau) $.} This is one of the
simplest situations, when both angular distribution and energy
distribution can be neglected, typical of well-collimated,
monochromatic high-quality beams produced in conventional
accelerators. Simpler subcases for $f_l = \delta(\vec{l})$ or
$f_\tau = \delta(\tau)$ (or both) can be easily inferred from this
more general model.

As we discussed before, the presence of an offset modifies the
single-particle field by a phase factor only. In case of ER,  Eq.
(\ref{farangletot}) simplifies to

%

\begin{eqnarray}
\vec{\widetilde{{E}}}(z, \vec{\theta}, \omega)&& = {N_e
F\left(\omega ,\vec{\theta}\right)} \widetilde{E}_{(1)} \cr && =
\frac{ i \omega e N_e F(\omega ,\vec{\theta}) L}{c^2 z}
\exp\left[\frac{i \omega \theta^2 z}{2 c}\right] \vec{\theta}~
\mathrm{sinc}\left\{\frac{\omega L}{4
c}\left[\frac{1}{\gamma^2}+{\theta}^2\right]\right\}~,\cr &&
\label{generalfingen}
\end{eqnarray}
where

\begin{eqnarray}
F\left(\omega ,\vec{\theta}\right) = \int dt f_\tau(\tau)
\exp\left[i {\omega \tau}\right] \int d\vec{l} f_l(\vec{l})
\exp\left[-\frac{i \omega }{c} \vec{\theta}\cdot
\vec{l}\right]~.\label{bigf}
\end{eqnarray}
Thus, in this case, we obtain a simple result amounting to a
multiplication of the structure factor $F$ defined in Eq.
(\ref{bigf}) by the single-particle field. The structure factor
$F$ is the product of the Fourier transform of $f_\tau$ (the bunch
longitudinal profile), $\bar{f}_\tau$, and the Fourier transform
of the bunch transverse profile.

To sum up, if one knows results for a single electron, like those
calculated above for ER, TUR, or any other kind of radiation, and
one needs to calculate radiation from an electron bunch, all is
needed to do is to multiply the result for a single electron by
$F$; results in the near zone are obtained remembering that
$\vec{\theta} = \vec{r}/z$. Note that in the case of a single
electron with zero offset and arrival time $\tau_k=0$,
$\bar{f}_\tau = 1$ and $f_l(\vec{l}) = \delta(\vec{l})$, so that
$F= 1$ and one recovers Eq. (\ref{generalfin}).

In the present case I, the form factor $F$ does not change during
the motion of electrons along the $z$ axis. This is in agreement
with our previous assumption about energy spread and angular
distribution of the bunch. In fact, as said before, we assume that
$f_\tau(\tau)$ remains unchanged during the motion along the
beamline. Concerning the transverse structure, we are assuming
that the transverse size and shape of the bunch do not vary
either.

With the help of Eq. (\ref{endene}) and Eq. (\ref{generalfingen})
we obtain the angular spectral flux

\begin{eqnarray}
\frac{dW}{d\omega d \Omega} = \frac{c N_e^2 z^2}{4\pi^2}
\left|\vec{\widetilde{E}}\right|^2 =
\left|F(\omega,\theta)\right|^2 \frac{c N_e^2 z^2}{4\pi^2}
\left|\vec{\widetilde{E}}_{(1)}\right|^2~. \label{endenegen}
\end{eqnarray}
Note that Eq. (\ref{endenegen}) is just the second (coherent) term
in Eq. (\ref{endenegen0}). We should stress that the near-zone
problem for coherent ER can be solved, similarly as in the single
particle case, starting from the far-zone data. Therefore, it can
be solved using Eq. (\ref{generalfingen}).

Consider first the case $\phi \ll 1$. The problem of calculation
of the virtual source distribution in the middle of the straight
section  can be basically reduced to a Fourier transform of the
far-field distribution. Therefore, the specification of virtual
sources also simplifies to a convolution between the single
particle virtual source and a spatial Fourier Transform of the
form factor $F$. Remembering Eq. (\ref{virfiemody}), and with the
help of Eq. (\ref{ABcontrint4e}), Eq. (\ref{vir3e}) and Eq.
(\ref{generalfingen}) one explicitly obtains

\begin{eqnarray}
\vec{\widetilde{E}}\left(0,\vec{r}\right)  && = N_e
\bar{f}_\tau(\omega) \vec{\widetilde{E}}_{(1)}(0,\vec{r}) \ast
f_l(\vec{r}) \cr && = N_e \bar{f}_\tau(\omega) \int d \vec{l}
~\left[ -\frac{4 i \omega e}{c^2 L} \left(\vec{r}-\vec{l}~\right)
\mathrm{sinc}\left(\frac{\omega \left|\vec{r}-\vec{l}~\right|^2}{c
L}\right)\right]f_l(\vec{l}) ~,\label{sorge}
\end{eqnarray}
where $\vec{\widetilde{E}}_{(1)}(0,\vec{r})$ indicates the
single-particle field distribution at the virtual source in Eq.
(\ref{vir3e}), and the symbol $"\ast"$ denotes the convolution
product as defined in the second line of Eq. (\ref{sorge}). Note
that the virtual source in Eq. (\ref{sorge}) can be imaged by a
lens when the object plane corresponds to position $z = 0$.

Similar remarks hold for the case of a generic value of $\phi$. In
fact, as is well-known from Eq. (\ref{Efarsum2a}), the expression
for the single-particle field in the far-zone can be split in the
sum of two spherical-wave contributions, and both terms are
subsequently multiplied by the same form factor. Thus, analogously
to Eq. (\ref{sorge}), one obtains

\begin{eqnarray}
\vec{\widetilde{E}}\left(\pm L/2,\vec{r}\right)  = N_e
\bar{f}_\tau(\omega) \vec{\widetilde{E}}_{(1)}(\pm L/2,\vec{r})
\ast f_l(\vec{r}) ~,\label{sorge2}
\end{eqnarray}
where now $\vec{\widetilde{E}}_{(1)}(\pm L/2,\vec{r})$ indicate
the single-particle field distribution at the virtual source
positions $\pm L/2$ in Eq. (\ref{virpm05}).

\paragraph*{II. Case $f=f_{l,\eta}(\vec{l},\vec{\eta})\delta(\delta\gamma/\gamma)
f_\tau(\tau)$.} In this situation we still neglect the
energy-spread distribution, but we account for a given
distribution in angles in addition to the case treated in I. In
other words, the bunch has a finite emittance. Factorization of
$f_l$ and $f_\eta$ at the reference point ($z=0$) indicates that
the betatron functions are minimal at that position. In practice,
angular distribution is important for the case depicted in Fig.
\ref{geome}(d), i.e. for a laser-plasma accelerator device. In
this case, it is more natural to assume that the minimal betatron
function is located at $z=-L/2$, rather than at $z=0$. Then, the
product $f_l(\vec{l})f_\eta(\vec{\eta})$ must be consistently
substituted by a non-factorized distribution
$f_{l,\eta}(\vec{l},\vec{\eta})$. We will consider a Gaussian
electron bunch with rms size $\sigma_l$. Additionally, we
introduce a Gaussian angular distribution with rms $\sigma_\eta$
and assume, for simplicity, that these rms quantities apply both
in the horizontal and in the vertical plane at $z=-L/2$. As a
result, one has the following distribution at $z=0$ (which may be
alternatively rewritten in terms of Twiss parameters):

\begin{eqnarray}
f_{l,\eta}\left({l},{\eta}\right) = \frac{1}{4\pi^2 \sigma_l^2
\sigma_\eta^2}
\exp\left[-\frac{(l-L\eta/2)^2}{\sigma_l^2}-\frac{\eta^2}{\sigma_\eta^2}\right]
. \label{ftlzeta}
\end{eqnarray}
With this in mind, Eq. (\ref{farangletot}) simplifies to

\begin{eqnarray}
\vec{\widetilde{E}}\left(z,\vec{\theta},\omega\right) && =\frac{ i
\omega e N_e L \bar{f}_\tau(\omega)}{c^2 z}\int d\vec{l}
d\vec{\eta} ~ f_{l,\eta}\left({l},{\eta}\right) \exp\left[\frac{i
\omega \theta^2 z}{2 c}\right] \cr && \times \exp\left[-
\frac{i\omega \vec{\theta}\cdot\vec{l}}{c}\right]
\left(\vec{\theta}-\vec{\eta}\right)
\mathrm{sinc}\left\{\frac{\omega L}{4
c}\left[\left|\vec{\theta}-\vec{\eta}\right|^2+\frac{1}{\gamma^2}\right]\right\}~,
\cr &&\label{farangletot2b}
\end{eqnarray}
Generalization of this study case to the situation with a finite
energy spread, still relevant e.g. for laser-plasma accelerator
devices can be performed straightforwardly.

\subsection{Microbunching}

We will now study the particular case of an electron bunch
modulated in density at a given wavelength. In this case, coherent
emission is not due to the overall bunch profile, but to the
presence of a modulation superimposed to such profile.

In some cases modulation of the bunch can be set up on purpose,
while in others it can be a detrimental effect.

Typical examples of parasitic microbunching mechanisms at FEL
facilities are driven by Coherent Synchrotron Radiation (CSR) in
the bending magnet of a bunch compressor chicane
\cite{CSR1,CSR2,CSR3} or Longitudinal Space Charge force (LSC)
\cite{LSC1}. Initial shot-noise density modulation at a given
frequency induces energy modulation at the same frequency due to
CSR or LSC impedances. Such energy modulation is subsequently
transformed back into density modulation through a dispersive
section, eventually leading to a beam instability. The dispersive
section is typically the magnetic compression chicane itself. In
this case, microbunching is usually an unwanted phenomenon, which
tends to spoil the high-quality electron beam needed for the FEL
process.

However, in other cases, a modulation at optical wavelengths can
be printed on purpose onto the electron bunch with the help of a
(quantum) laser and a modulator undulator followed by a dispersive
section. Optically modulated electron bunches can be used for
diagnostic purposes and in pump-probe techniques as well
\cite{REPL,OURM}.

If the bunch is modulated at a given wavelength $\lambdabar_m$,
one has $f_\tau(\tau) = f_{\tau 0}(\tau)[1 + a_0 \cos(c
\tau/\lambdabar_m )]$, where $f_{\tau 0}$ is a given longitudinal
profile and $a_0$ is the modulation level.

Let us consider a typical Gaussian model for the beam, specified
by

\begin{eqnarray}
f_l\left(\vec{l} ~\right) = \frac{1}{2\pi \sigma_l^2}
\exp\left[-\frac{l^{2}}{2 \sigma_l^2}\right]~, \label{rhor}
\end{eqnarray}
$\sigma_l$ being the $rms$ beam transverse dimension and by

\begin{eqnarray}
f_{\tau 0}(\tau) = \frac{1}{\sqrt{2\pi} \sigma_T}
\exp\left[-\frac{\tau^2}{2 \sigma_T^2}\right]~. \label{ft}
\end{eqnarray}
Here we are assuming that the wavefront of the density
distribution of the modulation is perpendicular to the velocity of
electrons, and is uniform in the transverse direction as $a_0$ is
not a function of $\vec{r}$. The latter assumption is close to
reality, because the waist of the seed laser is much larger than
the size of the electron bunch in the modulator undulator, due to
the extreme high-quality of the electron bunch in XFELs.

Near the modulation frequency $\omega_m = c/\lambdabar_m$, a bunch
with modulated Gaussian temporal profile and rms duration
$\sigma_T$ yields

\begin{eqnarray}
\bar{f}_\tau(\omega )=\frac{
a_0}{2}\left\{\exp\left[-\frac{\sigma_T^2}{2} \left(\omega -\omega
_m\right)^2\right] +\exp\left[-\frac{\sigma_T^2}{2} \left(\omega
+\omega_m\right)^2\right]\right\}~, \label{fbarrrr}
\end{eqnarray}
where $\bar{f}_\tau(\omega )$ is, as before, the Fourier transform
of the temporal profile of the bunch. In deriving Eq.
(\ref{fbarrrr}) we used an adiabatic approximation that can be
taken advantage of in practical situations involving XFELs, where
the optical modulation wavelength is much shorter than the bunch
length, i.e. $c \sigma_T/\lambdabar_m \gg 1$. In this case we have
another large parameter in the problem, which considerably
simplifies the treatment. In fact, since we are interested in
coherent emission around optical wavelengths, in calculating Eq.
(\ref{fbarrrr}) we neglected the Fourier transform of the first
term of $f_\tau(\tau)$, i.e. $f_{\tau 0}(\tau)$. Obviously, this
would not be possible for frequencies in the range $1/\sigma_T$.
From Eq. (\ref{fbarrrr}) follows that radiation is exponentially
suppressed for frequencies outside the bandwidth $1/\sigma_T$
centered at the modulation frequency $\omega_m$.

The exponential suppression in Eq. (\ref{fbarrrr}) has its
analogous in the transverse direction too. With the help of Eq.
(\ref{rhor}) one finds

\begin{eqnarray}
F\left(\omega ,\vec{\theta}\right) = \bar{f}_\tau(\omega )
\exp\left[-\frac{\omega^2 \theta^2 \sigma_l^2}{2
c^2}\right]~.\label{bigf22}
\end{eqnarray}
Note that $\theta$ is usually normalized to a characteristic angle
that depends on the system considered. In our ER case, such angle
is $\sqrt{\lambdabar/L_f}$, with $L_f =
\min{[L,\gamma^2\lambdabar]}$. With this in mind, the exponential
suppression in Eq. (\ref{bigf22}) can be written in normalized
units as

\begin{eqnarray}
F\left(\omega ,\vec{\hat{\theta}}\right) = \bar{f}_\tau(\omega )
\exp\left[-\frac{\hat{\theta}^2 N}{2}\right]~,\label{bigf33}
\end{eqnarray}
where we defined

\begin{eqnarray}
N= \frac{\sigma_l^2}{\lambdabar L_f}~. \label{NN}
\end{eqnarray}
The parameter $N$ is analogous to a Fresnel number in diffraction
theory, and is the only (dimensionless) transverse parameter
related to the radiation emission in the model defined by Eq.
(\ref{rhor}) and Eq. (\ref{ft}). When $N \ll 1$ the exponential
factor in Eq. (\ref{bigf33}) returns unity, and one gets back the
single-electron radiation profile, whereas the flux is possibly
modified by the presence of $\bar{f}_\tau$. When $N \gtrsim 1$ the
intensity profile as a function of angles is exponentially
suppressed.

Integration of Eq. (\ref{endenegen}) in $d \omega$ and in $d
\Omega$ gives the total energy radiated. Since we are dealing with
an electron bunch modulated at optical wavelengths we are
interested in narrow-bandwidth emission.

Within the above-mentioned adiabatic approximation we can consider
the wavelength in $\vec{\widetilde{E}}_{(1)}$ appearing in Eq.
(\ref{endenegen}) fixed, i.e. $\lambdabar = \lambdabar_m$. Then,
integration in $d\omega$ can be easily performed using the fact
that $\int_0^\infty d\omega
\left|\bar{f}_\tau(\omega)\right|^2={\sqrt{\pi} a_0^2}/({4
\sigma_T})$, leading to

\begin{eqnarray}
\frac{dW}{d \Omega} = \frac{\sqrt{\pi} c z^2 N_e^2 a_0^2}{16 \pi^2
\sigma_T} \exp\left[-\frac{\omega_m^2 \theta^2
\sigma_l^2}{c^2}\right] \left|\vec{\widetilde{E}}_{(1)}\right|^2~.
\label{endenegen2}
\end{eqnarray}
Integration over angles can be performed analytically in
particular limiting cases.

In the following we will discuss the setup in Fig. \ref{geome}(a).
However, our result can be easily extended to the case of Fig.
\ref{geome}(b) and Fig. \ref{geome}(c). Let us first consider the
case $\phi \ll 1$. One has

\begin{eqnarray}
W =  \frac{2  N_e^2 e^2 a_0^2}{\sqrt{\pi} \sigma_T  c }
\int_0^\infty \frac{d\theta}{\theta} \sin^2\left[\frac{\omega_m
L\theta^2}{4 c}\right] \exp\left[-\frac{\omega_m^2 \theta^2
\sigma_l^2}{c^2}\right]~. \label{endenegen3}
\end{eqnarray}
Introducing dimensionless units allows one to write the total
number of photons emitted at $\omega_m$ as

\begin{eqnarray}
N_{ph} =  \frac{2  N_e^2 \alpha a_0^2}{N_m \sqrt{\pi}}
\int_0^\infty \frac{d \hat{\theta}}{\hat{\theta}}
\sin^2\left[\frac{\hat{\theta}^2}{4}\right]
\exp\left[-N\hat{\theta}^2\right]=\frac{N_e^2 \alpha a_0^2}{4 N_m
\sqrt{\pi}} \ln\left[1+\frac{1}{4 N^2}\right]~, \label{endenegen4}
\end{eqnarray}
where $\alpha \equiv e^2/(\hbar c) = 1/137$ is the fine structure
constant and  $N_m = {\sigma_T \omega_m }$ is the number of
modulation wavelengths included in the bunch.  It should be noted
that Eq. (\ref{endenegen4}) is logarithmically divergent for
$N\longrightarrow 0$. It is therefore interesting to discuss the
region of applicability of Eq. (\ref{endenegen4}) in the limit for
small values of $N$. Such asymptote can usually be exploited when
dealing with XFEL setups, because of the extremely high-quality of
the electron bunch (small emittance). The expression used for
$\vec{\widetilde{E}}_{(1)}$ assumes that the sharp-edge
approximation holds, i.e. $\delta \ll 1$. In this case, the
formation length of the bending magnet at the end of the straight
section $L_{fb}$ is much smaller than the straight section length
$L$. The transverse spot-size of the radiation associated with
bending magnet emission, and is of order $\lambdabar L_{fb}$ is
then the smallest characteristic transverse size of the problem.
If the electron bunch size becomes of order $\lambdabar L_{fb}$,
our electrodynamic description cannot distinguish anymore between
a bunch with finite transverse size and a point. In this case one
should substitute $\sigma_l$ with $\lambdabar L_{fb}$. This
amounts to the substitution $N \longrightarrow \delta$. Thus in
the limit $N \ll \delta$ we prescribe the substitution of $N$ with
$\delta$ for estimations.

%

Finally, since $c \sigma_T/\lambdabar_m \gg 1$, it makes sense to
introduce an expression for the instantaneous power as a function
of the instantaneous (peak) current $I(\tau)$ and modulation
$a_0(\tau)$, if the seed laser duration is not too long compared
to the bunch duration. To this purpose, we consider a
stepped-profile model for the bunch $f_{\tau 0}(\tau) = 1/T$ for
$-T/2<\tau<T/2$ and zero elsewhere. It follows that

\begin{eqnarray}
\bar{f}_\tau(\omega ) &=& \frac{a_0 c}{\omega \omega_m}
\left\{\frac{\omega -\omega_m}{c}\sin\left[\frac{T  \omega
\omega_m }{2
(\omega-\omega_m)}\right]+\frac{\omega+\omega_m}{c}\sin\left[\frac{T
\omega\omega_m}{2 (\omega+\omega_m)}\right]\right\}~.\cr &&
\label{fbarrrr2}
\end{eqnarray}
Then, in the limit for $T \gg \lambdabar_m/c$ one has
$\int_0^\infty d\omega \left|\bar{f}_\tau(\omega)\right|^2 \simeq
a_0^2 \pi T$, and

\begin{eqnarray}
\frac{dP}{d\Omega} &=& \frac{1}{T} \frac{dW}{d \Omega} = \frac{c
a_0^2 I_0^2 \pi z^2}{4 \pi^2 e^2} \exp\left[-\frac{\omega_m^2
\theta^2 \sigma_l^2}{c^2}\right]
\left|\vec{\widetilde{E}}_{(1)}\right|^2 ~, \label{Pinst0}
\end{eqnarray}
where $dP/d\Omega$ is the angular distribution of power. If now
the current $I=I(\tau)$ and the modulation level $a=a(\tau)$ are
slowly varying functions of time on the scale $\lambdabar_m/c$,
one obtains for $\phi \ll 1$

\begin{eqnarray}
\frac{dP}{d\Omega} &=&  \frac{ \omega_m^2 L^2 a(\tau)^2
I(\tau)^2}{4 \pi c^3 } \exp\left[-\frac{\omega_m^2 \theta^2
\sigma_l^2}{c^2}\right] \theta^2 \mathrm{sinc}\left[\frac{\omega_m
L\theta^2}{4 c}\right] ~. \label{Pinst}
\end{eqnarray}
Eq. (\ref{Pinst}) can be used in order to study the general case
of an electron bunch with arbitrary gradient profile and amplitude
of modulation.

Let us now turn to the case when $\phi \gg 1$ for the single-edge
radiation ($\gamma^2 \lambdabar \ll d_2 \ll L$ and $r \ll
L/\gamma$). Introducing apodization of the detector on a
characteristic angular scale $\sigma_a$ and using Eq.
(\ref{Isefar0}) we obtain\footnote{In particular, Eq.
(\ref{endenegen30}) applies to the case when coherent TR due to
microbunching is outcoupled by a mirror and postmirror optics can
be modelled with a Gaussian pupil profile. In this case, $z-L/2$
indicates the distance from the mirror.}

\begin{eqnarray}
W =  \frac{N_e^2 e^2 a_0^2}{2 \sqrt{\pi} \sigma_T  c}
\int_0^\infty {d\xi} \frac{\gamma^4 \xi^3}{ (\gamma^2 \xi^2 +
1)^2} \exp\left[-\frac{\omega_m^2 \xi^2
\sigma_l^2}{c^2}-\frac{\xi^2}{2\sigma_a^2}\right]~.\cr &&
\label{endenegen30}
\end{eqnarray}
Note that apodization (through $\sigma_a$) and transverse bunch
length (through $\lambdabar_m/\sigma_l$) play the same role from a
mathematical viewpoint. Also note that, in order to use Eq.
(\ref{Isefar0}), we must require that we are within its limits of
applicability, i.e. $d_2 \cdot \min(\sigma_a,
\lambdabar_m/\sigma_l) \ll L/\gamma$ (always with $\gamma^2
\lambdabar_m \ll d_2 \ll L$).

Taking advantage of dimensionless units one can write the total
number of photons emitted at $\omega_m$ as

\begin{eqnarray}
N_{ph} &=&  \frac{N_e^2 \alpha a_0^2}{2 N_m \sqrt{\pi} }
\int_0^\infty {d\hat{\xi}} \frac{ \hat{\xi}^3
}{(\hat{\xi}^2+\phi)^2} \exp\left[-(N+A) \hat{\xi}^2 \right] \cr
&=&  \frac{N_e^2 \alpha a_0^2}{4 N_m \sqrt{\pi} }
\left\{-1+\left[1+(A+N)\phi\right]\Gamma[0,(A+N)\phi]\exp[(A+N)\phi]\right\}~,
\label{endenegen31}
\end{eqnarray}
where $A \equiv \lambdabar /(2 L \sigma_a^2)$ is analogous to the
$N$ parameter in Eq. (\ref{NN}) and accounts for apodization
effects, and $\Gamma(a,z)=\int_z^\infty t^{a-1} \exp(-t) dt$ is
the incomplete gamma function. Discussion of asymptotic cases for
different values of $N$ and $A$ is possible on the basis of Eq.
(\ref{endenegen31}).

\section{\label{sec:mirror} Extraction of edge radiation by a
mirror}

\subsection{\label{sub:backt} Backward Transition and Diffraction
radiation}


A setup where edge radiation from an upstream bending magnet is
extracted by a mirror is shown in Fig. \ref{geome}(c). In the
particular case depicted in that figure, radiation is reflected by
a hole mirror and sent to a far-infrared spectrometer. This
example refers to a method \cite{LOOS} where coherent radiation is
used to monitor the bunch length at XFEL setup, but similar setups
can also be used to collect incoherent radiation in the optical
range. The size and shape of the mirror may vary. A hole may be
present in the mirror or not, and the geometry of the experimental
arrangement depends on the particular setup considered. Anyway, in
its basic lines, a standard edge-radiation diagnostic setup
consists, similarly as in Fig. \ref{geome}(c), of a mirror
positioned at some distance after a bend and rotated by an angle
$\pi/4$ in the direction of the electron bunch to allow
extraction of radiation through a vacuum window. 
Such setup can be used for longitudinal bunch-length measurements
in the mm and sub-mm range as well as for transverse electron
bunch diagnostics in the optical wavelength range
\cite{LOOS,LUM1,LUM2}.

\begin{figure}
\begin{center}
\includegraphics*[width=100mm]{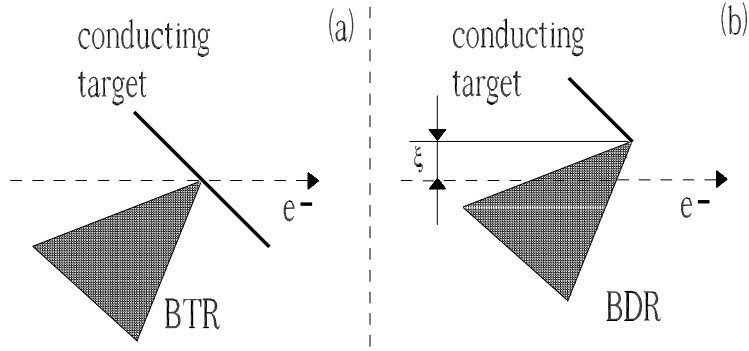}
\caption{\label{ODROTR} Geometry of backward transition radiation
(BTR) (a), backward diffraction radiation (BDR) (b).}
\end{center}
\end{figure}
Consider the straight section after the bend. If this is long
enough so that $\phi \gg 1$ (and $\delta \cdot \phi \ll 1$, i.e.
$\lambdabar \gg \lambdabar_c$), radiation collected by the mirror
is the usual Transition Radiation (TR) first noted by Ginzburg and
Frank \cite{GINZ} and considered in a number of publications
written during the last half century. In recent years there has
been a good deal of interest in TR, because this form of radiation
has found useful applications, e.g. in diagnostic for
ultrarelativistic beams \cite{KARL,KAR1,NAUM,KAR2,SUTT}. These
papers describe an electron crossing an interface between two
media with different dielectric constants and they refer, in
particular, to the case of the boundary between vacuum and an
ideal conductor. As a consequence of the crossing, time-varying
currents are induced at the boundary. These currents are
responsible for TR (see Fig. \ref{ODROTR} left). The metallic
mirror, that is treated as the source of TR, is usually modelled
with the help of a Physical Optics approach. This is a well-known
high-frequency approximation technique, often used in the analysis
of electromagnetic waves scattered from large metallic objects.
Surface current entering as the source term in the propagation
equations of the scattered field are calculated by assuming that
the magnetic field induced on the surface of the object can be
characterized using Geometrical Optics, i.e. assuming that the
surface is locally replaced, at each point, by its tangent plane.

As mentioned before, one typically assumes that the Ginzburg-Frank
formula can be used, i.e. the field distribution is proportional
to $K_1[\omega r/(\gamma c)]$. In terms of parameters introduced
in this paper, this approximation has sense only when $\phi \gg
1$. However, even for this case, one should always specify the
transverse region of applicability of the usual TR formulas, which
is typically neglected (as we will see, this region is for $ r \ll
\phi \lambdabar \gamma = L/\gamma$, with $\phi \gg 1$). Moreover,
mainly due to an increased electron energy, today setups often
meet, in practice, condition $\phi \lesssim 1$ even in the optical
wavelength range. In this case one should account for the whole
setup when discussing the reflected radiation, and not only for
the metallic mirror. In other words, as we will see, the problem
of extraction of edge radiation reduces to the well-studied TR
problem only asymptotically, i.e. for $\phi \gg 1$. In this paper
we will still keep on talking about TR or, with some abuse of
language, even when usual TR formulas do not apply anymore ($\phi
\lesssim 1$, or $r \gtrsim L/\gamma$).

When dealing with the above-discussed setup, one usually talks
about Backward Transition Radiation (BTR). BTR is widely used for
different purposes, because it allows for low background radiation
levels. We will consider a whole range of setups for BTR, without
treating their geometry in detail, and we will give an algorithm
to deal with these systems in all generality. In fact, in all
cases, the main problem is in the specification of the electric
field distribution at some position where a mirror is present (the
hole mirror in the example of Fig. \ref{geome}(c)). The only
element to be accounted for, aside for the mirror itself, is an
upstream dipole magnet. Note that the mirror is tilted as in Fig.
\ref{geome}(c), i.e. it is not perpendicular with respect to the
optical axis. In principle, one may account for this tilting, but
in practice the projection of the mirror on the optical axis is
negligible compared with the distance to the dipole, and one may
consider the screen as not tilted.

Note that Diffraction Radiation  (DR) appears when charged
particles move in the vicinity of a medium (e.g. a conductive
mirror) at some given impact parameter $\xi$, and has been
recently suggested as a possible tool for non-invasive bunch
diagnostics. DR is obviously related to TR and, in particular, it
can be treated starting from the knowledge of the electric field
distribution at the mirror, and subsequently applying physical
optics techniques. BDR in the optical wavelength range has been
measured and applied for transverse electron beam diagnostics (see
e.g. \cite{LUM1,LUM2}). A schematic comparison between a BDR and a
BTR emission is shown in Fig. \ref{ODROTR}.

It should be stressed that specification of the field at the
target position must be considered as the first step to the
solution of a more complicated problem, i.e. the characterization
of the field at a detector position. Such first step is considered
separately, because the field at the target plane is independent
of the type of target and detector. Once the field at the target
position is known, the full problem can be solved with the help of
Physical Optics techniques, where one should account for
diffraction effects due to the particular shape of the mirror.

In the following we will focus on the problem of field
characterization at the target position. A virtual source method
can be applied. Our algorithm consists of the following steps: (i)
propagate the field from the upstream virtual source to the
mirror, (ii) add the field from the downstream virtual source.

Note that, in contrast to the study case of a straight section
between two bends considered before, numerical methods involving
direct integration of Maxwell's equations (e.g. with the help of
SRW) down the optical axis up to the mirror will fail, because
electromagnetic sources must be propagated up to the observation
plane, and integration of the Green's function up to that position
yields a singularity (independently of the value of $\delta$). In
our virtual source approach we also deal with a singularity, but
this is isolated in a single term, and expressed by means of an
analytical function (the Bessel $K_1$ function in Eq.
(\ref{virpm05})), allowing straightforward presentation of
results. The same approach has been proposed in \cite{SLPB} to
discuss the setup in Fig. \ref{geome}(c).

Let us focus in detail on our generic BTR setup. Following the
algorithm described above, we model the system with the help of
two virtual sources at $z = \pm L/2$, i.e. at the end of the bend
and at the mirror. As said before, Eq. (\ref{virpm05}) can be used
to describe the virtual sources. Application of the propagation
formula Eq. (\ref{fieldpropback}) allows to calculate the field at
any distance ${z}$ in free-space for the source located at
$z=-L/2$. In order to simplify the presentation of the electric
field we take advantage of polar coordinates. We propagate the
upstream virtual source (located at $z= -L/2$) up to the mirror
position at $z=L/2$, and we add the field distribution for the
downstream source at $z=L/2$.

At ${z}= L/2$ (i.e. at the mirror position) we obtain

\begin{eqnarray}
\vec{\widetilde{E}}\left(\frac{L}{2},\vec{r}\right) &=&  - \frac{2
\omega e}{c^2  \gamma}  \exp\left[\frac{i \omega L}{4 c
\gamma^2}\right] \frac{\vec{r}}{r} K_1\left(\frac{\omega r}{c
\gamma }\right) - \frac{2 \omega e}{c^2 \gamma} \frac{\vec{r}}{r}
\exp\left[-\frac{i \omega L}{4 c \gamma^2}\right] \left.
\right.\cr &&\left.\times \exp\left[\frac{i \omega {r}^2 }{2 c
L}\right]\frac{\omega}{c L} \int_0^{\infty} d{r}' {r}'
K_1\left(\frac{\omega r'}{c \gamma}\right) J_1 \left(\frac{\omega
{r}{r}'}{c L}\right) \exp\left[\frac{i \omega {r}^{'2} }{2 c
L}\right] \right.~,\label{fieldtot2}
\end{eqnarray}
which is valid for any value of $\phi$. Exactly as in the case
treated in Section \ref{sub:vire}, one may deal with two
asymptotes of the theory for $\phi \gg 1$ and $\phi \ll 1$.

In the case $\phi \gg 1$ the contribution from the upstream source
in Eq. (\ref{fieldtot2}) can be calculated analytically. In fact,
the $K_1$ function under the integral sign is responsible for an
exponential suppression of the integrand in the limit for ${r'}
\gg \gamma \lambdabar$. As a result, we can integrate up to values
${r}' \lesssim \gamma \lambdabar$ without changing the integration
result. This means that the phase factor in the integrand will be
at most of order $1/\phi \ll 1$, and can be neglected. The result
of the integration then corresponds to the expression for a single
edge in the far-zone, Eq. (\ref{Efarsum3}), the only difference
being that the single-edge source is located at $z=-L/2$.
Altogether, for $\phi \gg 1$, Eq. (\ref{fieldtot2}) simplifies to:

\begin{eqnarray}
\vec{\widetilde{E}}\left(\frac{L}{2},\vec{{r}}\right) &=&  -
\frac{2  \omega e}{c^2 \gamma} \exp\left[\frac{i \omega L}{4 c
\gamma^2}\right] \frac{\vec{r}}{{r}} K_1\left(\frac{\omega{r}}{c
\gamma} \right)\cr && - \frac{2 e}{c} \frac{\gamma^2
\vec{r}}{(\gamma^2 r^2+L^2)} \exp\left[- \frac{i \omega L}{4 c
\gamma^2}\right] \exp\left[\frac{i \omega {r}^2 }{2 c L}\right]
~.\label{fieldtot3}
\end{eqnarray}
Eq. (\ref{fieldtot3}) is valid for $r \ll L
\sqrt[3]{\lambdabar/R}$, i.e. for angles (measured from the
upstream edge at $z=-L/2$) smaller than the angle of SR from the
bend. When $r \ll \gamma\lambdabar$, the $K_1$ term is obviously
dominant with respect to the second (polynomial) term in ${r}$.
When ${r} \gtrsim \gamma \lambdabar$, the $K_1$ term drops
exponentially, while the second term is zero at ${r} = 0$, grows
polynomially and reaches its maximum at  $r \sim L/\gamma$. Since
the polynomial growth is slower than the exponential decrease of
the $K_1$ term, one will have a region $\gamma \lambdabar \lesssim
{r} \ll L/\gamma$ where the field is strongly suppressed with
respect to the maximum at ${r} \sim L/\gamma$. In this region, the
$K_1$ term is still a good approximation to the total field. As a
result, the first term in Eq. (\ref{fieldtot3}) can be taken as a
good approximation for the total field when $r \ll L/\gamma$
(always assuming $\phi \gg 1$). In this asymptotic case, results
from our analysis coincide with usual results from TR theory. Note
that the question about how many photons will be collected
strongly depends on the acceptance of the optical system. The
first term in Eq. (\ref{fieldtot3}) is singular at ${r}=0$, so
that the larger the acceptance, the nearest to the singularity we
collect photons, and the largest the number of photons is.

The opposite asymptote for $\phi \ll 1$ can be described by
applying the propagation formula, Eq. (\ref{fieldpropback}), to
Eq. (\ref{vir3e}), like it was done for Eq. (\ref{vir4e}). When
${z}$ is fixed at the mirror position, i.e. at ${z} = L/2$ one
obtains \cite{BOSL}

\begin{eqnarray}
\vec{\widetilde{E}}\left(\frac{L}{2},\vec{{r}}\right) = - \frac{2
e}{c}  \frac{\vec{r}}{r^2} \exp\left[\frac{i \omega r^2}{2 c
L}\right]~.\label{vir4eb}
\end{eqnarray}
The quadratic phase factor in Eq. (\ref{vir4eb}) describes a
spherical wavefront centered on the optical axis at the upstream
edge (the initial bend). Note that this contribution scales as
$1/r$ and is singular on-axis. The screen positioned at the
downstream end of the setup will detect electric field given by
the spherical wavefront in Eq. (\ref{vir4eb}) at any value
$\sqrt[3]{\lambdabar R^2} \ll L \ll \gamma^2 \lambdabar$ from the
upstream magnet and $r \ll L \sqrt[3]{\lambdabar/R}$.

Following the specification of the field (in amplitude and phase)
at the target position, one faces the problem of propagation of
BTR in free-space and through optical elements, which has been
extensively discussed in \cite{KARL,KAR1,NAUM,KAR2,SUTT}. As
mentioned above, this reduces to a standard Physical Optics
problem, which is interesting within the field of Optical
Engineering, i.e. from the viewpoint of the design of post-mirror
collector optics. In these references, discussions are based on
the assumption that a field distribution $~K_1[\omega r/(\gamma
c)]$, i.e. the Ginzburg-Frank formula, can be used. In this case,
we deal with a laser-like beam, whose waist is located on the
target (i.e. the target exhibits a plane wavefront). The
transverse size of the waist is of order $\gamma \lambdabar \gg
\lambdabar$, so that Fourier Optics can be applied to solve the
propagation problem. The Rayleigh length of this laser-like beam
is about $R_L \simeq \gamma^2 \lambdabar$, and the Fresnel
diffraction-zone is located at distances from the mirror of order
$R_L$. When the distance from the mirror becomes much larger than
$R_L$ one has the Fraunhoffer diffraction-zone. If some linear
optical element is located at a given position after the mirror,
one can use available wavefront propagation/physical optics codes
(e.g. SRW \cite{CHUB}, ZEMAX \cite{ZEMA}, GLAD \cite{GLAD}, PHASE
\cite{PHAS}) for calculations. In other words, the propagation
problem is well-defined, and does not include any novel aspect
from a scientific viewpoint. Only, one needs to put attention to
special terminology and notations used in literature, which may
lead to some misunderstanding. For example, the Rayleigh length
$R_L \sim \gamma^2 \lambdabar$ is named "the formation length".
Since the waist is always given by $w_0 \simeq \sqrt{L_f
\lambdabar}$, it follows that $R_L \sim w_0^2/\lambdabar = L_f$.
However, it should be noted that, conceptually, the formation
length can only be discussed in connection with the particle
trajectories, and not in free-space. Similarly, the Fresnel
diffraction-zone is usually named as the "pre-wave zone", while
the Fraunhoffer diffraction-zone is called "wave-zone".
From this viewpoint, articles dealing with pre-wave-zone effects
actually treat a routine Optical Engineering problem.

It should be remarked that the problem of specifying the field at
a mirror position is fundamentally different from what we have
done in the previous Sections. Before we considered a system
composed by a straight section between two magnets, and radiation
was detected downstream of the last magnet. That setup could be
treated entirely within a SR-theory approach, and computer codes
like SRW could be used to cross-check our results. Here the
situation is different. 
In fact, Eq. (\ref{efielGfree}) must now include integration up to
a position $z$ where sources are present

\begin{eqnarray}
\vec{\widetilde{E}} &=& \frac{4\pi e}{c} \int_{-L/2}^{L/2} dz'
\left[\vec{\nabla'}_\bot G\left(\vec{r}-\vec{r'},
z-z'\right)\right]_{\vec{r'}=\vec{r}_0(z')} \exp\left[{i \omega
\left({s(z)\over{v}}-{z\over{c}}\right)}\right]~,\label{efielGfree2}
\end{eqnarray}
with $z=L/2$, and therefore includes a singularity. In analogy
with Eq. (\ref{propaaaf}) one then obtains Eq. (\ref{fieldtot2}),
where the singularity is isolated in the first field-term, which
is entangled with sources, whereas the second term is
disentangled.

Note that in the mathematical limit for $\rho \longrightarrow
\infty$, Eq. (\ref{propaaaf}) reduces to Eq. (\ref{fieldtot2}).
Similarly, Eq. (\ref{fieldtot}) reduces to Eq. (\ref{fieldtot2})
at ${z}=L/2$. However, although formally similar, Eq.
(\ref{propaaaf}) and Eq. (\ref{fieldtot}) have different physical
meaning compared to Eq. (\ref{fieldtot2}). In this Section we are
characterizing the field \textit{along the straight section}. This
gives a different status e.g. to the singularity in Eq.
(\ref{fieldtot2}) for $r=0$ and $z=L/2$ compared to the
singularity in Eq. (\ref{fieldtot}), also located at $r =0$ and
${z} = L/2$. In fact, the singularity in Eq. (\ref{fieldtot})
(which reduces to Eq. (\ref{vir4e}) in the case $\phi \ll 1$)  is
related with the sharp-edge approximation: $\delta$ has a finite
value, and going beyond the accuracy of the sharp-edge
approximation (i.e. modelling the edges and accounting for their
contribution to the radiation) the integrated flux will not
diverge, nor the reconstructed field from the far zone data will.
In contrast to this,  the singularity in the $K_1$ term in Eq.
(\ref{fieldtot2}) at $r =0$ and ${z} = L/2$ is due to the fact
that the plane at $\hat{z}=L/2$ intercepts the electron trajectory
at that point, and is fundamentally related with our model of the
electron beam as a filament without transverse dimensions.

\subsection{\label{sub:imp} Impact on setups at XFEL facilities}

During the last decade,  the electron beam energy in linac-based
facilities increased up to around $10$ GeV and novel effects,
which were considered negligibly small at lower energies, became
important. In fact, at high energies, condition $L \gg \gamma^2
\lambdabar$ becomes unpractical even for optical wavelengths, and
the effect of the presence of an upstream bending magnet on TR
emissions must be considered. Moreover, when measuring the
long-wavelength coherent BTR, i.e. when the radiation wavelength
is comparable to, or longer than the electron bunch length, the
influence of the upstream bending magnet could be significant even
at lower energies. These facts led to several misconceptions that
can be found till recently.

An example of application for diagnostic purposes in XFEL setups
is given in \cite{LOOS}. In general, monitoring methods based on
coherent radiation take advantage of the sensitivity of the
coherent spectrum on the bunch form factor, which is related to
the bunch length. Coherent radiation from a given setup is thus
collected with the help of mirrors and its energy measured, for
example with the help of pyroelectric detectors. In order to
analyze results, one has to characterize the field distribution at
the collecting mirror position.

\begin{figure}
\begin{center}
\includegraphics*[width=100mm]{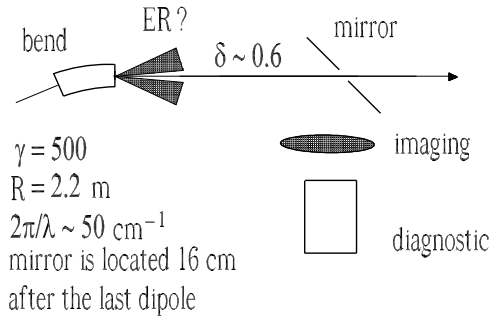}
\caption{\label{setlcls} Layout of the LCLS bunch length monitor,
located immediately after the last dipole of the first magnetic
bunch compressor.}
\end{center}
\end{figure}
The setup proposed in \cite{LOOS} uses the last bend in each
magnetic compression chicane of LCLS as depicted in Fig.
\ref{setlcls}. Radiation is collected by a mirror with a hole
(that has to be large enough to let the electron beam through) and
imaged to the diagnostic station. The hole mirror is positioned
after a short straight section, a few tens of centimeters long.
Detailed parameters can be found in \cite{LOOS}. In the case of
the BC1 chicane, for example, the length of the straight section
between magnet and mirror is $L=0.16$ m, the wavelength of
interest is of order\footnote{Since we are talking about coherent
radiation, the reduced characteristic wavelength $\lambdabar$ is
related here with the longitudinal size of the bunch
$\sigma_{rms}$, i.e. $\lambdabar \simeq \sigma_{rms}.$ }$\lambda
\sim 1$ mm, the electron energy is about $250$ MeV, and the radius
of the bend is about $R \simeq 2.2$ m.

In \cite{LOOS} one can read: "the critical wavelength of the
synchrotron radiation is orders of magnitude smaller than the
wavelength range considered here for the bunch length monitor and
the synchrotron radiation can be neglected compared to edge
radiation". Authors of \cite{LOOS} stress the fact that
$\lambdabar \gg \lambdabar_c$. Such condition is absolutely
necessary to satisfy the applicability of ER theory. However, as
we have seen, it is not sufficient, as the extra-requirement
$\delta \ll 1$ must be satisfied too. If we calculate the
parameter $\delta = {\sqrt[3]{R^2 \lambdabar}}/{L}$ for numbers
given above we obtain $\delta \simeq 0.6$. We conclude that
synchrotron radiation from the bend cannot be neglected: on the
contrary, its effect will be as important as those from the
straight section. The sharp-edge approximation fails here.

Note that there are two contributions to the field at the mirror
position. The first is the contribution due to bending magnet
radiation, while the second is the contribution due to the
straight section. The fact that $\delta \simeq 0.6$ indicates that
both contributions are of the same order of magnitude. It follows
that it is incorrect to neglect the bending magnet contribution.
Assuming that calculation of the straight section contribution is
performed correctly, this would lead to a quantitatively incorrect
numerical result from the viewpoint of practical applications.
However, authors of \cite{LOOS} calculate the straight section
contribution incorrectly as well. Let us consider the $\phi$
parameter. The reader may check that in this case the formation
length $\gamma^2 \lambdabar \simeq 50$ m is much longer than the
length of the straight section $L = 0.16$ m, so that $\phi \ll 1$.
Therefore, Eq. (\ref{vir4eb}) should be used to calculate the
straight-section contribution at the mirror position. Authors of
\cite{LOOS} proceed instead in the following way. First, they
recognize that for ER "the Coulomb field is the radiation source
due to the sudden change in acceleration". Second, they propagate
a $K_1$ source from the edge of the magnet up to the mirror "by
expressing the Coulomb field as a sum of Gauss-Laguerre modes".
This corresponds to the second term of Eq. (\ref{fieldtot2}). In
other words, the field distribution at the mirror position is
obtained by propagating a $K_1$ source located at the upstream
bending magnet. It follows that the result in \cite{LOOS} is
qualitatively incorrect.

To sum up, in \cite{LOOS}, SR from the bending magnet is
incorrectly neglected, and in the calculation of ER the downstream
source, the first term of Eq. (\ref{fieldtot2}), is missed,
qualitatively changing the result. The correct procedure to find
the field distribution at the mirror for this parameter values
($\delta \sim 1$, $\phi \ll 1$) is to use Eq. (\ref{vir4eb}) to
calculate the straight-section contribution at the mirror
position, and to sum up the SR contribution from the bending
magnet.

\begin{figure}
\begin{center}
\includegraphics*[width=100mm]{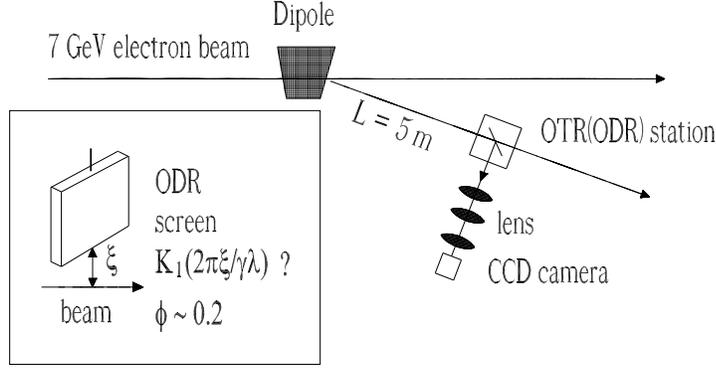}
\caption{\label{setlum} ODR setup at the APS facility. In this
experiment, ODR at visible wavelength is emitted for impact
parameter $\xi$ between $1$ mm and $2$ mm, values close to the
Lorentz factor $\gamma$ times the reduced observation wavelength.
}
\end{center}
\end{figure}
We now turn to analyze \cite{LUM1,LUM2}, where a setup similar to
that in Fig. \ref{setlum} is studied from an experimental
viewpoint. An electron bunch travels a straight section of length
$L \simeq 5$ m downstream of a bending magnet before intercepting
an offset mirror at the ODR station. The wavelength considered
here is $\lambda \simeq 800$ nm, and the electron energy is $7$
GeV. In this case $\gamma^2 \lambdabar \simeq 30$ m, i.e. $\phi
\simeq 0.17$. Authors of \cite{LUM1,LUM2} use a downstream virtual
source $\sim K_1[\xi/(\gamma\lambdabar)]$ to characterize the
field at the BDR screen, effectively describing an ER setup by
means of usual Ginzburg-Frank expression \cite{GINZ} for TR.
However, here $\delta \ll 1$ and $\phi \ll 1$. As we have seen
before, in this parameter range the correct procedure to evaluate
the field distribution at the mirror is to use Eq. (\ref{vir4eb}),
which is different from the Ginzburg-Frank expression. In other
words, in this parameter-range, the Ginzburg-Frank formula cannot
be applied anymore.

Note that in \cite{LUM1} is reported that "the ODR peak signal
intensity dependence is consistent with $\exp[-2d/(\gamma
\lambdabar)]$, where $d$ is the distance from beam center to the
screen edge, $\gamma$ is the Lorentz factor". However, in this
region of parameters the asymptotic behavior in Eq.
(\ref{vir4eb}), does not exhibit any exponential cutoff, nor
dependence on $\gamma\lambdabar$. Therefore, the interpretation of
experimental result in \cite{LUM1} cannot be accepted.

This example and that in \cite{LOOS} help to understand the
importance of edge-radiation theory in relation with future XFEL
facilities. They also demonstrate the usefulness of similarity
techniques in the efficient planning of experimens.

\section{\label{sec:conc} Conclusions}

In this article we showed how the theory of laser beams can be
used to characterize radiation field associated with any Edge
Radiation (ER) setup. In fact, in the space-frequency domain, ER
beams can be described in terms of laser-like beams, with large
transverse dimensions compared to the wavelength. Similarly to
usual laser beams, ER beams were shown to exhibit a virtual
"waist" with a plane wavefront. The field distribution of ER
across the waist turned out to be strictly related to the inverse
Fourier transform of the angular field-distribution in the
far-zone. As a result, standard Fourier Optics techniques could be
taken advantage of, and the field could be propagated to
characterize ER beams at any position down the beamline. In
particular, we reconstructed the near-field distribution from the
knowledge of the far-field ER pattern. This could be accomplished
by (i) describing the far-field pattern with known analytical
formulas, (ii) finding the virtual source(s) and (iii) propagating
the virtual source distribution in the near (and far) zone.

After a qualitative discussion in Section \ref{sec:param}, we
applied our techniques to a typical setup constituted by a
straight section between bends in Sections \ref{sec:theor},
\ref{zero} and {\ref{sub:vire}. These Sections constitute the
first comprehensive treatment of ER, in the sense that we
consistently used similarity techniques for the first time,
allowing discussion and physical understanding of many asymptotes
of the parameter space together with their region of
applicability. In particular, the main parameters of the theory
are found to be $\delta$, the ratio between the bends formation
length and the straight section length, and $\phi$, the ratio
between the length of the straight section and the maximal
formation length of ER. Note that introduction of the parameter
$\delta$ allowed us to define for the first time "how sharp" the
edges are, and to specify the region of applicability of ER
theory. A classification of regions of observation of interest,
which is regarded by us as a novel result, is presented in Section
\ref{sub:classi} with the help of dimensionless parameters.

In Section \ref{sec:five} we applied our treatment to deal with a
Transition Undulator Radiation (TUR) setup. As before, we relied
on virtual source expressions derived from the far-field pattern.
These virtual sources were propagated in free-space in the near
zone, thus providing for the first time an exact analytical
characterization of TUR in the near zone.

In the following Section \ref{sec:waveguide} we presented the
first exhaustive theory of ER within a waveguide. The analysis of
the problem was performed by introducing a tensor Green's function
technique. This complicates the mathematical structure of
equations that depends, contrarily to the unbounded-space case, on
the waveguide geometry. We outlined a solution for a homogeneous
waveguide with arbitrary cross-section, further specializing it to
the case of a circular waveguide. The electric field was found as
a superposition of waveguide modes, and was studied for different
values of parameters. The main parameter involved in the problem
(other than $\phi$, $\delta$ and $\hat{z}$) was found to be a
waveguide parameter $\Omega$, related with the strength of guiding
effects. This can be interpreted (at $\phi \ll 1$) as the ratio
between the waveguide radius and the radiation diffraction size,
and is a purely geometrical parameter.

In Section \ref{sec:bunch} we presented considerations on the
coherent emission of ER. The availability of a large number of
photons constitutes an important advantage of coherent ER compared
to the incoherent case when this type of radiation is used as a
diagnostic tool, where characteristics of the electron bunch can
be measured, and electron-beam microbunching can be investigated
too.

Finally, in Section \ref{sec:mirror} we discussed the problem of
extraction of ER by a mirror, stressing differences and analogies
that this kind of setup presents, compared to those discussed in
the previous Sections. Particular attention was given to
diagnostics setups for XFEL facilities.

As a final remark, it should be noted that in our work we
consistently exploited both theoretical and numerical results from
simulation. These approaches are complementary, and we first took
advantage of such complementarity. On the one hand, there are
situations when existing codes cannot be applied, e.g. the case of
a waveguide (Section \ref{sec:waveguide}) and the characterization
of the field distribution at a mirror position (Section
\ref{sec:mirror}), where part of the field is entangled with the
sources. In this case, analytical results can be used for
practical calculations. On the other hand, computer codes can
easily account for finite bending magnet edge length (finite value
of $\delta$) in a particular set of problem parameters. From this
viewpoint, our theory can be used to prepare, based on similarity
techniques, particular sets of problem parameters to be used as
input for computer codes, which subsequently return universal
plots presenting the accuracy of ER theory in terms of
dimensionless parameters.

\section*{\label{sec:graz} Acknowledgements}

We thank our DESY collegues Martin Dohlus, Michael Gensch and Petr
Ilinski for many useful discussions, Massimo Altarelli , Reinhard
Brinkmann and Edgar Weckert  for their interest in this work.

\section*{Appendix: Equivalence of Eq. (\ref{propaaaf}) and Eq.
(\ref{guidefinal})}

We want to show that Eq. (\ref{propaaaf}), together with Eq.
(\ref{vettoE}), is equivalent to Eq. (\ref{guidefinal}). We can
demonstrate this fact by calculating the sum in Eq.
(\ref{guidefinal}), and presenting it in terms of an integration.

We begin expanding the $\mathrm{sinc}(X)=\sin(X)/X$ function in
terms of exponential functions. Moreover, we subtract from $X$ in
the denominator of $\sin(X)/X$ an infinitesimally small quantity
$i \epsilon$, where $\epsilon>0$, to be dropped at the end of
calculations. In other words, we substitute $1/X$ with $1/(X-i
\epsilon) = i \int_0^{\infty} d \xi \exp[-i X \xi - \epsilon
\xi]$. We therefore obtain

\begin{eqnarray}
\vec{\widetilde{E}} &=& -\frac{i e L}{\omega \rho^3}
\int_0^{\infty} d \xi \exp\left[-i \frac{\omega L \xi}{4 c
\gamma^2 }-\epsilon \xi\right] \sum_{k=1}^{\infty}
\frac{\nu_{0k}}{J_1^2(\nu_{0k})} \cr && \times
\left\{\exp\left[\frac{i \omega L}{4 c
\gamma^2}\right]-\exp\left[-\frac{i \omega L}{4 c
\gamma^2}-\frac{i L \nu_{0k}^2 c}{2 \omega \rho^2}\right]\right\}
J_1\left(\frac{\nu_{0k} r}{\rho}\right)
\exp\left[-i\frac{\nu_{0k}^2 c L \xi}{4 \omega \rho^2}\right]
\vec{e}_r~. \cr && \label{new1}
\end{eqnarray}
Eq. (\ref{new1}) consists of two field terms, corresponding to the
two terms inside the $\{...\}$ parenthesis. Let us consider the
first term and demonstrate that such term is in fact a
generalization of the virtual source at $z = L/2$ (i.e. Eq.
(\ref{virpm05}) for $z_s = L/2$) accounting for the presence of
the waveguide. Using the fact that $J_1(X)=I_0'(iX)/i$, the $"'"$
symbol indicating derivative with respect to the argument, and
remembering the Wronskian relation $1/X = I_0(X) K_1(X)+I_1(X)
K_0(X)$, we have

\begin{eqnarray}
\frac{1}{J_1(\nu_{0k})} = \frac{2  \omega \rho^2}{c L} [I_1(i
\nu_{0k}) K_0(i \nu_{0k})] \left[\frac{d}{d\eta} I_0\left(
\sqrt{-i\frac{4 \omega \rho^2 \eta}{c L}}\right)
\right]^{-1}_{\eta = -i c L \nu_{0k}^2/(4 \omega
\rho^2)}~.\label{subs1}
\end{eqnarray}
Using Eq. (\ref{subs1}) and remembering $J_1(X) = I_1(i X)/i$ we
write the first term in Eq. (\ref{new1}) as

\begin{eqnarray}
\vec{\widetilde{E}}_1 &=& - i\frac{e L }{\omega \rho^3}
\int_0^{\infty} d \xi \exp\left[-i\frac{\omega L \xi}{4 c
\gamma^2}-\epsilon \xi\right] \sum_{k=1}^{\infty}
\left[\frac{d}{d\eta} I_0\left(\sqrt{-i\frac{4 \omega \rho^2
\eta}{c L}}\right) \right]^{-1}_{\eta = -i c L \nu_{0k}^2/(4
\omega \rho^2)}\cr &&\times  \frac{2 \omega \rho^2 \nu_{0k}}{c L}
K_0(i \nu_{0k}) \exp\left[\frac{i \omega L}{4\gamma^2 c}\right]
I_1\left(i\frac{\nu_{0k} r}{\rho}\right)
\exp\left[-i\frac{\nu_{0k}^2 c L \xi}{4 \omega \rho^2}\right]
\vec{e}_r~.\label{new3}
\end{eqnarray}
We now consider the integral

\begin{eqnarray}
u &=& \int_{O_1} d \eta \exp[\eta \xi] \frac{\sqrt{-i
\eta}}{I_0\left(\sqrt{-i\frac{4 \omega \rho^2 \eta}{c L}}\right)}
\left[I_1\left(\sqrt{-i\frac{4 \omega \rho^2 \eta}{c
L}}\frac{r}{\rho}\right) K_0\left( \sqrt{-i\frac{4 \omega \rho^2
\eta}{c L}}\right)\right. \cr && \left. + I_0\left(\sqrt{-i\frac{4
\omega \rho^2 \eta}{c L}}\right) K_1\left(\sqrt{-i\frac{4 \omega
\rho^2 \eta}{c L}}\frac{r}{\rho}\right) \right]~,\label{new4}
\end{eqnarray}
where $O_1$ is an integration path on the complex $\eta$ plane
going from $\gamma' - i\infty$ to $\gamma' + i \infty$,
$\gamma'>0$ being a positive number larger than the real part of
all singularities of the integrand, and closed by a semicircle of
infinite radius in the left half plane. It can be shown that the
integrand is a single valued function of $\eta$. It follows that
$u$ can be calculated as a sum of residues of first order poles
located at $\eta = \eta_k = -i \lambdabar L \nu_{0k}^2/(4
\rho^2)$. We obtain

\begin{eqnarray}
- \frac{2\rho^3 \omega^{3/2}}{\pi (c L)^{3/2}} u &=&
\sum_{k=1}^{\infty} \frac{2 \omega \rho^2 \nu_{0k}}{c L} K_0(i
\nu_{0k}) \exp\left[-i\frac{\nu_{0k}^2 c L \xi}{4 \omega
\rho^2}\right] I_1\left(i\frac{\nu_{0k} r}{\rho}\right) \cr &&
\times \left[\frac{d}{d\eta} I_0  \left( \sqrt{-i\frac{4 \omega
\rho^2 \eta}{c L}}\right) \right]^{-1}_{\eta = -i c L
\nu_{0k}^2/(4 \omega \rho^2)} ~,\label{new5}
\end{eqnarray}
that can be immediately substituted in Eq. (\ref{new3}). The
integral along the semicircle in Eq. (\ref{new4}) can be dropped
using Jordan's lemma. Therefore

\begin{eqnarray}
\vec{\widetilde{E}}_1 &=&  \frac{2 i e  \sqrt{\omega} }{\pi
\sqrt{L c}} \exp\left[\frac{i \omega L}{4\gamma^2 c}\right]
\int_0^{\infty} d \xi \exp\left[-i\frac{\omega L \xi}{4 c \gamma^2
}-\epsilon \xi\right] \int_{\gamma' -i \infty}^{\gamma' + i
\infty} d \eta \exp[\eta \xi] \cr &&
\times\frac{\sqrt{-i\eta}}{I_0\left(\sqrt{-i\frac{4 \omega \rho^2
\eta}{c L}}\right)} \left[I_1\left( \sqrt{-i\frac{4 \omega \rho^2
\eta}{c L}}\frac{r}{\rho}\right) K_0\left( \sqrt{-i\frac{4 \omega
\rho^2 \eta}{c L}}\right)\right. \cr && \left. +
I_0\left(\sqrt{-i\frac{4 \omega \rho^2 \eta}{c L}}\right)
K_1\left( \sqrt{-i\frac{4 \omega \rho^2 \eta}{c
L}}\frac{r}{\rho}\right)\right]\vec{e}_r~.\label{new6}
\end{eqnarray}
Eq. (\ref{new6}) is easily seen to be the Laplace transform of an
inverse Laplace transform. Letting $\epsilon \longrightarrow 0$,
it follows straightforwardly that

\begin{eqnarray}
\vec{\widetilde{E}}_1 &=&  - \frac{2 \omega e}{c^2 \gamma}
\exp\left[\frac{i \omega L}{4\gamma^2 c}\right] \left[
K_1\left(\frac{\omega r}{c \gamma }\right) +
\frac{K_0\left(\frac{\omega \rho}{c
\gamma}\right)}{I_0\left(\frac{\omega \rho}{c \gamma}\right)}I_1
\left(\frac{\omega r}{c\gamma}\right)\right]
\vec{e}_r~.\label{new7}
\end{eqnarray}
Eq. (\ref{new7}) is a generalization for the virtual source
located at $z=L/2$ that accounts for the presence of a waveguide
(compare with Eq. (\ref{vettoE})). 

Let us now consider the second term in Eq. (\ref{new1}), which can
also be written as

\begin{eqnarray}
\vec{\widetilde{E}}_2 &=&  \frac{e L}{\omega \rho^3}
\sum_{k=1}^{\infty} \frac{\nu_{0k}}{J_1^2(\nu_{0k})}
\exp\left[-\frac{i \omega L}{4 c \gamma^2}-\frac{i L \nu_{0k}^2
c}{2 \omega \rho^2}\right] \frac{1}{\frac{\omega L}{4 c \gamma^2
}+\frac{L\nu_{0k}^2 c}{4 \omega \rho^2}} J_1\left(\frac{\nu_{0k}
r}{\rho}\right) \vec{e}_r~.\label{newwfin}
\end{eqnarray}
Since

\begin{eqnarray}
&&\int_0^\rho dr' ~r'K_1\left(\frac{\omega r'}{c \gamma}\right)
J_1 \left(\frac{\nu_{0k} r'}{\rho}\right) = \cr && \left[\frac{c
\gamma \nu_{0k}}{\omega \rho}+\nu_{0k}
J_2(\nu_{0k})K_1\left(\frac{\omega\rho}{c\gamma}\right)-
\frac{\omega\rho}{c\gamma}J_1(\nu_{0k})K_2\left
(\frac{\omega\rho}{c\gamma}\right)\right]\Bigg/\left(\frac{\omega^2}{\gamma^2
c^2}+\frac{\nu_{0k}^2}{\rho^2}\right)~,\cr && \label{intfin2}
\end{eqnarray}
and

\begin{eqnarray}
&&\int_0^\rho dr' ~r'I_1\left(\frac{\omega r'}{c \gamma}\right)
J_1 \left(\frac{\nu_{0k} r'}{\rho}\right) = \cr && \left[
\frac{\omega\rho}{c\gamma}
I_2\left(\frac{\omega\rho}{c\gamma}\right) J_1(\nu_{0k})+\nu_{0k}
I_1\left(\frac{\omega\rho}{c\gamma}\right)
J_2(\nu_{0k})\right]\Bigg/\left(\frac{\omega^2}{\gamma^2
c^2}+\frac{\nu_{0k}^2}{\rho^2}\right)~,\cr && \label{intfin2biss}
\end{eqnarray}

%
%
%
one can write Eq. (\ref{newwfin}) as

\begin{eqnarray}
\vec{\widetilde{E}}_2\left(\frac{L}{2},\vec{r}\right) &=& \frac{4
\omega e}{\rho^2 c^2 \gamma}  \sum_{k=1}^{\infty}
\frac{1}{J_1^2(\nu_{0k})}  \exp\left[-\frac{i \omega L}{4
c\gamma^2}-\frac{i L \nu_{0k}^2 c}{2 \omega \rho^2}\right] \cr &&
\times \int_0^\rho dr' ~r' \left[ K_1\left(\frac{\omega r'}{c
\gamma}\right) + \frac{K_0\left(\frac{\omega \rho}{c
\gamma}\right)}{I_0\left(\frac{\omega \rho}{c \gamma}\right)}I_1
\left(\frac{\omega r'}{c \gamma}\right)\right]
J_1\left(\frac{\nu_{0k} r}{\rho}\right) J_1\left(\frac{\nu_{0k}
r'}{\rho}\right) \vec{e}_r~, \cr && \label{neww8}
\end{eqnarray}
because the second and the third terms on the right hand side of
Eq. (\ref{intfin2}) cancel with Eq. (\ref{intfin2biss}) multiplied
by ${K_0\left({\omega \rho}/({c \gamma})\right)}/{I_0\left({\omega
\rho}/({c \gamma})\right)}$. Remembering that an expression for
the field propagation from some initial point $z_i$ to some final
point $z_f$ is given by Eq. (\ref{propaaa}), one finds that Eq.
(\ref{neww8}) is obtained by propagating the field distribution

\begin{eqnarray}
\vec{\widetilde{E}}\left(-\frac{L}{2}\right) &=&   \frac{2 \omega
e}{c^2 \gamma} \exp\left[-\frac{i \omega L}{4 c \gamma^2}\right]
\left[ K_1\left(\frac{\omega r}{c \gamma}\right) +
\frac{K_0\left(\frac{\omega \rho}{c
\gamma}\right)}{I_0\left(\frac{\omega \rho}{c \gamma}\right)}I_1
\left(\frac{\omega r}{c
\gamma}\right)\right]\vec{e}_r~\label{neww7}
\end{eqnarray}
up to $z_f = L/2$. This can be seen with the help of Eq.
(\ref{Gfexcir}),  selecting $m=0$.

Now, Eq. (\ref{neww7}) is just a generalized expression, valid in
the presence of a waveguide, for a virtual source at $z=-L/2$,
which is described similarly to Eq. (\ref{new7}), the difference
being only its location position.

It follows that the second term in Eq. (\ref{new1}) is obtained
propagating the virtual source at $z=-L/2$ up to position $z=L/2$,
that is the second term of Eq. (\ref{propaaaf}).

\end{document}